\documentclass[
 reprint,
 amsmath,amssymb,
 aps,
]{revtex4-2}
\RequirePackage[T1]{fontenc}
\usepackage[utf8]{inputenc}
\usepackage{lmodern}

\usepackage{graphicx}% Include figure files
\usepackage{dcolumn}% Align table columns on decimal point
\usepackage{bm}
\usepackage{float}% bold math
\usepackage{braket}
\usepackage{graphicx}
\usepackage{lipsum,capt-of,graphicx}
\usepackage{hyperref}
\usepackage{todonotes}
\usepackage{multirow}
\usepackage{bigstrut}
\usepackage{setspace}
\usepackage{needspace}
\usepackage{color, colortbl}
\hypersetup{
    colorlinks=true,
    linkcolor=blue,
    filecolor=magenta,   
    citecolor=blue,
    urlcolor=cyan,
    pdftitle={Overleaf Example},
    pdfpagemode=FullScreen,
    }

\definecolor{slateblue}{rgb}{0.97, 0.96, 1.0}
\definecolor{seagreen}{rgb}{0.96, 1.0, 0.98}
\definecolor{lightblue}{rgb}{0.74, 0.83, 0.9}
\definecolor{lightcyan}{rgb}{0.94, 0.97, 1.0}
\usepackage[mathlines]{lineno}% Enable numbering of text and disp
\usepackage{subfiles}

\begin{document}

\preprint{APS/123-QED}

\title{Tomography of entangling two-qubit logic operations in exchange-coupled donor electron spin qubits}

\author{Holly G. Stemp$^{1,2}$}
\author{Serwan Asaad$^{1,2}$}%
    \altaffiliation[Currently at ]{Quantum Machines, Denmark}
\author{Mark R. van Blankenstein$^{1,2}$}
\author{Arjen Vaartjes$^{1,2}$}
\author{Mark A. I. Johnson$^{1,2}$}%
    \altaffiliation[Currently at ]{Quantum Motion, London, UK}
\author{Mateusz T. M\k{a}dzik$^{1,2}$}%
    \altaffiliation[Currently at ]{Intel Corporation Hillsboro, Oregon, United States}
\author{Amber J. A. Heskes$^{1,2}$}%
    \altaffiliation[Currently at ]{University of Twente, Enschede, The Netherlands}
\author{Hannes R. Firgau$^{1,2}$}
\author{Rocky Y. Su$^{1}$}
\author{Chih Hwan Yang$^{1, 3}$}
\author{Arne Laucht$^{1, 3}$}%
\author{Corey I. Ostrove$^{4}$}%
\author{Kenneth M. Rudinger$^{4}$}%
\author{Kevin Young$^{4}$}%
\author{Robin Blume-Kohout$^{4}$}
\author{Fay E. Hudson$^{1, 3}$}%
\author{Andrew S. Dzurak$^{1,3}$}%
\author{Kohei M. Itoh$^{5}$}%
\author{Alexander M. Jakob$^{2,6}$}%
\author{Brett C. Johnson$^{7}$}%
\author{David N. Jamieson$^{2,6}$}%
\author{Andrea Morello$^{1,2}$}%
 \email{a.morello@unsw.edu.au}

\affiliation{%
 $^{1}$ School of Electrical Engineering and Telecommunications, UNSW Sydney, Sydney, NSW 2052, Australia\\
 $^{2}$ ARC Centre of Excellence for Quantum Computation and Communication Technology\\
 $^{3}$ Diraq Pty. Ltd., Sydney, New South Wales, Australia\\
 $^{4}$ Quantum Performance Laboratory, Sandia National Laboratories, Albuquerque, NM 87185 and Livermore, CA 94550, USA \\
 $^{5}$ School of Fundamental Science and Technology, Keio University, Kohoku-ku, Yokohama, Japan \\
 $^{6}$ School of Physics, University of Melbourne, Melbourne, VIC 3010, Australia\\
 $^{7}$ School of Science, RMIT University, Melbourne, VIC, 3000, Australia
}%

\begin{abstract}

Scalable quantum processors require high-fidelity universal quantum logic operations in a manufacturable physical platform. Donors in silicon provide atomic size, excellent quantum coherence and compatibility with standard semiconductor processing, but no entanglement between donor-bound electron spins has been demonstrated to date. Here we present the experimental demonstration and tomography of universal 1- and 2-qubit gates in a system of two weakly exchange-coupled electrons, bound to single phosphorus donors introduced in silicon by ion implantation. We observe that the exchange interaction has no effect on the qubit coherence. We quantify the fidelity of the quantum operations using gate set tomography (GST), and we use the universal gate set to create entangled Bell states of the electrons spins, with fidelity $91.3 \pm 3.0 \%$, and concurrence $0.87 \pm 0.05$. These results form the necessary basis for scaling up donor-based quantum computers.

\end{abstract}

\maketitle

%% THE PAPER %%
 
\section*{Introduction}
\label{sec:intro}
\begin{figure*}[ht]
    \centering
    \includegraphics[width=0.95\textwidth]{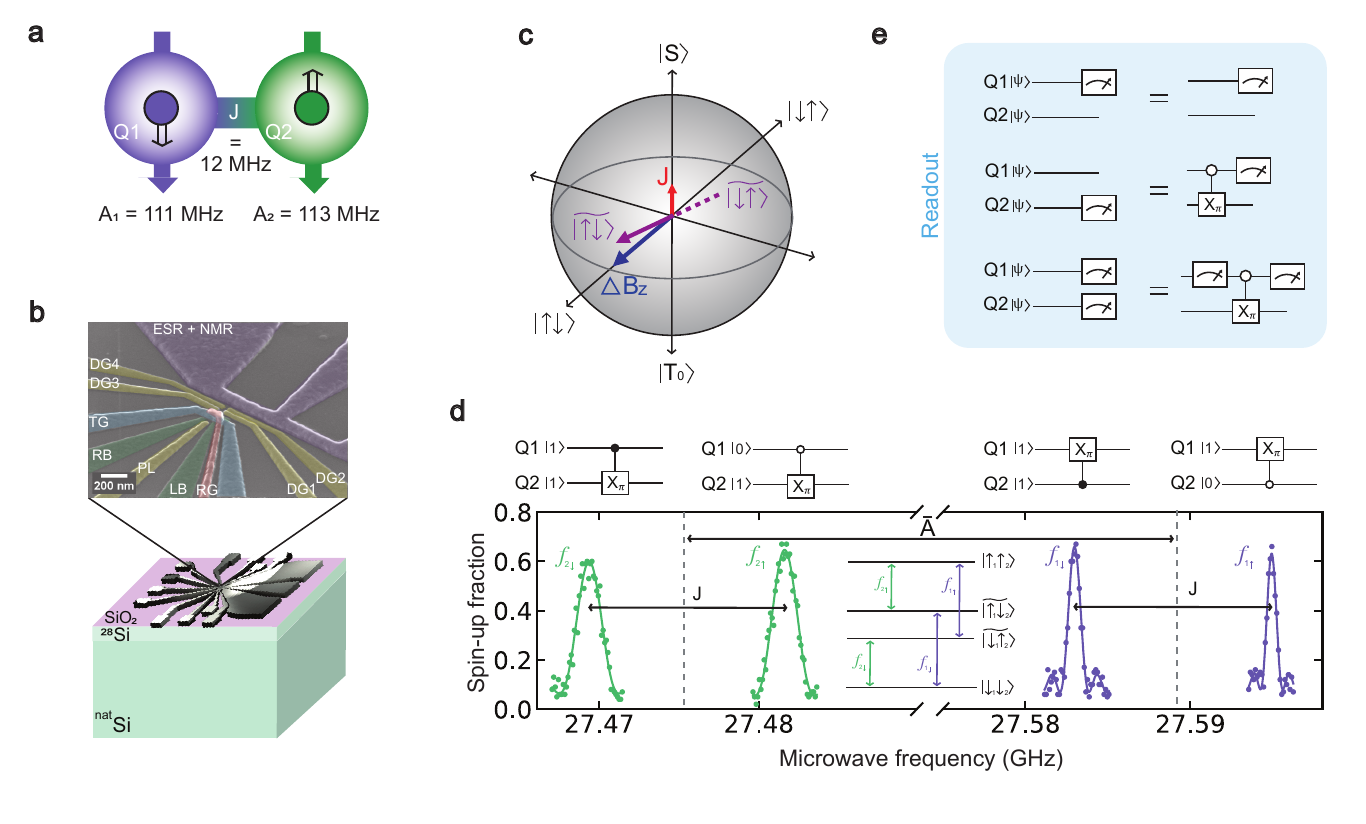}
    \caption{\textbf{Two-qubit phosphorus device operation.} \textbf{a,} Schematic of the system consisting of two phosphorus donors, each with their own single, bound electron, denoted as Q1 and Q2. Setting the nuclei in an anti-parallel configuration results in an effective $\Delta B_{z}$ between the two electrons, which is given by the average hyperfine, $\bar{A}$, between the two donors.
    \textbf{b,} Device layout showing the silicon substrate upon which Al gates are fabricated for device control and readout. The top of the figure shows an enlarged, false-coloured SEM image of the Al gate layout.
    \textbf{c,} Bloch sphere denoting the hybridised states brought about by the ratio of the energy detuning $\Delta B_{z}$ with the exchange coupling $J$, which constitute the eigenstates of the system in the weak $J$ regime. 
    \textbf{d,} ESR spectrum showing the resonant frequencies used to control Q1 (purple) and Q2 (green). These resonance frequencies are present when the nuclei are in an anti-parallel configuration of $\ket{\Uparrow_1 \Downarrow_2}$. A resonant $\pi$ pulse on any of these resonant frequencies represents a two-qubit CROT or zCROT gate, as shown in the circuit diagrams above each resonance.
    \textbf{e,} Quantum circuits denoting the readout process for Q1 and Q2. Q1 is read out via spin dependent-tunneling to an SET reservoir, while Q2 is instead read out indirectly via Q1 using quantum logic (see Supplementary Note 2).}
    
    \label{fig:1}
\end{figure*}

The exchange interaction is a fundamental form of coupling between electron spins. It stems from the Pauli exclusion principle, and its amplitude $J$ depends on the overlap between the wave functions of the electrons \cite{burkard2023semiconductor}. In the context of spin-based quantum information processing, the exchange interaction features prominently as a natural method to enable entangling operations between electron spins~\cite{loss1998quantum}. For example, preparing two spins $S=1/2$ in opposite states, $\ket{\uparrow\downarrow}$ and switching on the interaction for a time $\pi/2J$ results in the entangled state $\left(\ket{\uparrow\downarrow} - i\ket{\downarrow\uparrow} \right)/\sqrt{2}$. In early experiments on electron spin qubits in quantum dots, $J$ was controlled by detuning the two electrons' potentials with respect to each other \cite{petta2005coherent,maune2012coherent,veldhorst2015two}. Progress in device fabrication has allowed the reliable placement of thin gates between pairs of dots, through which $J$ was varied over many orders of magnitude by controlling the height of the tunnel barrier between the dots \cite{eenink2019tunable,nurizzo2022controlled}. This method also reduces the sensitivity to charge noise \cite{reed2016reduced,martins2016noise}.

The situation is more complicated in donor-based spin qubits \cite{morello2020donor}. Encoding quantum information in the nuclear spin of donor atoms in silicon was one of the earliest proposals for solid-state quantum computers \cite{kane1998silicon}. This vision has been corroborated by the experimental demonstration of exceptionally long spin coherence times, exceeding 30 seconds \cite{muhonen2014storing}, and 1- and 2-qubit gate fidelities above 99\% \cite{muhonen2015quantifying,mkadzik2022precision}. The Coulomb potential of a donor also naturally binds an electron, which is itself an excellent qubit, with demonstrated single-qubit gate fidelity up to 99.98\% \cite{dehollain2016optimization}. However, electron two-qubit logic gates based on exchange face the challenge that $J$ is both oscillating and exponentially dependent on inter-donor distance \cite{koiller2001exchange,voisin2020valley,joecker2021full}. The useful range of the exchange is only $\sim 10-20$~nm, making it difficult to fabricate and align metallic gates to control the tunnel barrier between the donors. Therefore, the only example of exchange control in donor systems to date was achieved by detuning the electrochemical potentials of a pair of multi-donor quantum dots, resulting in gate-controlled SWAP oscillations \cite{he2019two}. Since the donor-bound electrons were not operated as qubits, it was not possible to benchmark the fidelity of such operations, nor to demonstrate spin entanglement. Even in quantum computer architectures focusing on donor nuclear spins as the information carriers, electron exchange is likely to be a crucial ingredient for scale-up, since nuclear spins do not naturally interact with each other with sufficient strength.

Here we present the first experimental demonstration of exchange-based, entangling two-qubit logic gates between electrons bound to individual $^{31}$P donors in silicon. This is obtained by implementing a scheme exploiting a fixed $J$, weaker than the electron-nuclear hyperfine interaction, $A$ \cite{kalra2014robust}. In this regime, preparing the two $^{31}$P nuclei in an opposite state detunes the two electrons by $A \gg J$ and renders each electron's resonance frequency dependent on the state of the other. The native two-qubit operation is a CROT gate (equivalent to a CNOT gate, up to single-qubit rotations), implemented by electron spin resonance \cite{madzik2021conditional}. This setup is formally equivalent to that of experiments in gate-defined double quantum dots operated in the regime $J < \Delta E_z$, where $\Delta E_z$ is a Zeeman energy difference between the two electron spins, caused either by a magnetic field gradient \cite{zajac2018resonantly,noiri2022fast} or by a difference in $g$-factor across the two dots \cite{huang2019fidelity}.

Since the precise value of $J$ is irrelevant, provided it is $\ll A$ and larger than the resonance linewdith, this scheme is relatively insensitive to uncertainties in the precise location of the donors. It is thus well suited to ion-implanted donor spins \cite{morello2020donor,jakob2022deterministic,holmes2024improved,jakob2023scalable}, which retain compatibility with standard metal-oxide-semiconductor manufacturing processes \cite{pillarisetty2018qubit}. Both electron spins are operated coherently as individual qubits \cite{pla2012single} and read out in single-shot, either directly \cite{morello2010single} or via quantum logic \cite{nakajima2019quantum,xue2020repetitive}. This allows us to perform accurate tomography of the one- and two-qubit gate operations, and to show that the weak-$J$ regime does not affect the coherence of the individual spins.

\section*{Results}
\subsection*{Operation of the two-electron processor}
\label{sec:two_electron_processor}

The two-qubit processor consists of electron spins, Q1 and Q2, each with spin \textit{S}=1/2 and basis states $\ket{\downarrow}$, $\ket{\uparrow}$, bound to $^{31}$P donor nuclei with spin \textit{I}=1/2 and basis states $\ket{\Downarrow}$, $\ket{\Uparrow}$ (Fig.~\ref{fig:1}b). Denoting with $\mathbf{S_{1,2}}, \mathbf{I_{1,2}}$ the vector spin operators for each electron and nucleus, and $A_{1,2}$ the electron-nuclear hyperfine couplings on each atom, the Hamiltonian of the system (in frequency units) is:

\begin{align}
        H = &(\mu_{\text{B}}/h) B_{0}(g_{1}S_{z1} + g_{2}S_{z2})+\\
    & \gamma_\mathrm{n}B_{0}(I_{z1}+I_{z2})+ \nonumber \\
    &  A_{1}\mathbf{S_{1}}\cdot \mathbf{I_{1}} +  A_{2}\mathbf{S_{2}}\cdot \mathbf{I_{2}} + \nonumber \\
    & J(\mathbf{S_{1}\cdot S_{2}}), \nonumber
\end{align}
where $\mu_{\text{B}}$ is the Bohr magneton, $h$ is Planck's constant, $g_{1,2}\approx 1.9985$ the Land\'e  g-factors of each electron spin, $g\mu_{\text{B}}/h \approx$ 27.97 GHz/T and $\gamma_\mathrm{n} \approx$ 17.23 MHz/T is the $^{31}$P nuclear gyromagnetic ratio. 
 
We use aluminium gate electrodes, patterned on the surface of the silicon chip via electron beam lithography, to control the electrostatic environment of the donors for donor initialisation and readout \cite{pla2012single, pla2013high}. A broadband antenna delivers oscillating microwave or radio frequency (RF) magnetic fields to control the spin of the electrons and nuclei using electron spin resonance (ESR) or nuclear magnetic resonance (NMR), respectively (Fig.~\ref{fig:1}a). The donors are introduced in the silicon substrate using the industry-standard method of ion implantation. The device used here was fabricated in the same batch and with the same implantation parameters as the ones described in Ref.~\cite{madzik2021conditional}. Among the 25 electrically functional devices we fabricated in that batch, three exhibited $J\approx 10-30$~MHz, and a fourth contained very tightly spaced donors, which were used to demonstrate nuclear two-qubit gates~\cite{mkadzik2022precision}. From the modelling of the implantation parameters~\cite{madzik2021conditional} and the dependence of $J$ on donor spacing~\cite{joecker2021full}, it is likely that many more devices may have contained exchange-coupled donor pairs with $J$ in the desired range, but such pairs may not have been suitably located with respect to the SET readout device.

The Q1 electron is read out and initialised directly via the standard method of energy-dependent tunnelling to a nearby single-electron transistor (SET) island \cite{elzerman2004single, morello2010single}. This readout also automatically initialises Q1 in the $\ket{\downarrow}$ ground state. To initialise Q2, we first prepare Q1 in the $\ket{\downarrow}$ state, before transferring the spin state from Q1 to Q2 using a process similar in nature to the well established electron-nuclear double resonance (ENDOR) technique \cite{tyryshkin2006davies} (see Supplementary Note 3, for details).\\

The Q2 electron readout involves using Q1 as ancilla in a repetitive, approximately quantum nondemolition (QND) scheme (see Supplementary Note 4) \cite{nakajima2019quantum,xue2020repetitive}. This is possible thanks to the electrons' long spin relaxation time ($T_1 \approx 1.4$~s), and the fact that $J \ll A$ renders the exchange interaction almost of Ising type, approximately fulfilling the QND condition \cite{joecker2024error}. The QND readout is performed by loading a $\ket{\downarrow}$ Q1 electron from the reservoir, rotating Q1 conditional on the state of Q2, reading out Q1, and repeating the cycle 11 times. The resulting readout contrast is greatly enhanced, from a bare 0.48 to 0.76. Secondly, the Q2 electron never leaves the donor, causing the neighbouring $^{29}$Si nuclei to `freeze' \cite{madzik2020controllable}. The sparse $^{29}$Si nuclei surrounding the donor constitute a major source of decoherence and frequency jumps. Even in our isotopically enriched material, the 800 ppm residual $^{29}$Si result in a few nuclei significantly coupled to the donor electron. Fluctuations in their orientation, caused by their internal dipole-dipole coupling, causes electron spin decoherence and discrete frequency jumps. However, such fluctuations are almost entirely suppressed while the electron is bound to the donor, since the very different hyperfine couplings at each lattice site prevent energy-conserving dipolar flip-flops between the $^{29}$Si nuclei.

In a similar fashion to the readout of electron Q2, the donor nuclei are read out indirectly via the Q1 electron. In order to read out the state of the nucleus of either donor 1 or 2, Q1 is initialised in the $\ket{\downarrow}$ state and then rotated conditional on the state of the nucleus, before being read out. As with electron Q2, the nuclei can hence be read out using QND readout, resulting in high nuclear readout fidelities, exceeding 99$\%$ \cite{pla2013high}. The nuclei are initialised using an ENDOR technique \cite{tyryshkin2006davies} (see Supplementary Note 3).

\subsection*{Effect of weak exchange on qubit coherence}
\label{sec:coherence_and_exchange}

\begin{figure}[h]
    \centering
    \includegraphics[width=0.5 \textwidth]{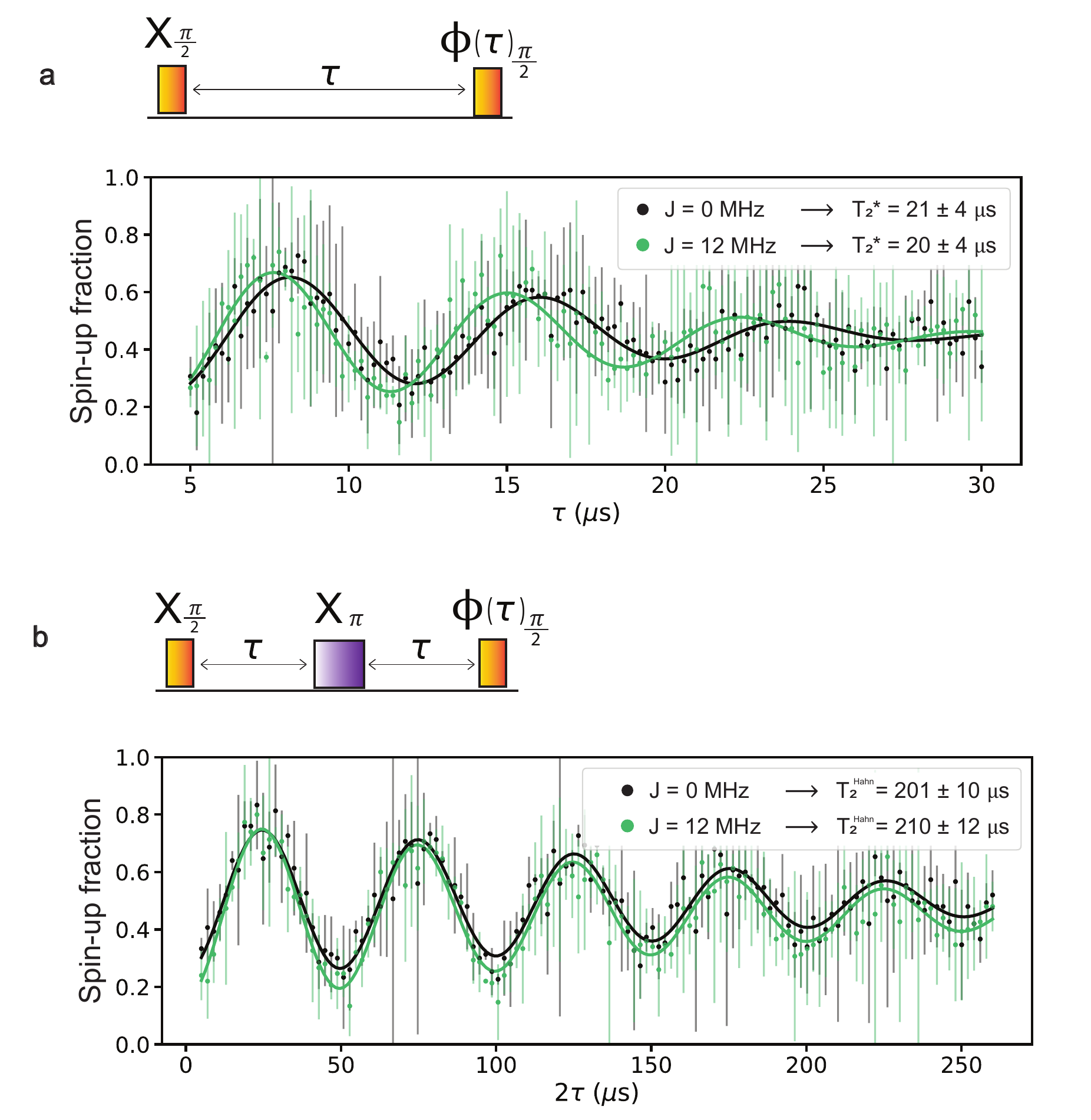}
    \caption{\textbf{Effect of weak exchange on electron coherence time.} \textbf{a,} Electron spin dephasing measurements by Ramsey experiment performed on Q2 with (green) and without (black) the presence of an exchange interaction $J$ with Q1. The $J=0$ regime was created by ionising nucleus 1, i.e. by completely removing the Q1 electron. \textbf{b,} Hahn echo experiment performed on Q2 with and without the presence of the exchange coupling. The oscillations are artificially introduced by a wait time dependent phase shift added to the final $\frac{\pi}{2}$ pulse. In all cases, the exchange coupling produces no detectable reduction in spin coherence times. In both the Ramsey and the Hahn echo experiments, each individual experiment [preparation, pulses, wait time, electron readout] was repeated 50 times. The entire Ramsey and Hahn echo experiments were then repeated multiple times, with the mean of the first three repetitions of the experiment plotted. The error bars represent 2$\sigma$ of these three repetitions.}
    \label{fig:ramsey_hahn}
\end{figure}

\begin{figure*}[ht]
    \centering
    \includegraphics[width=1\textwidth]{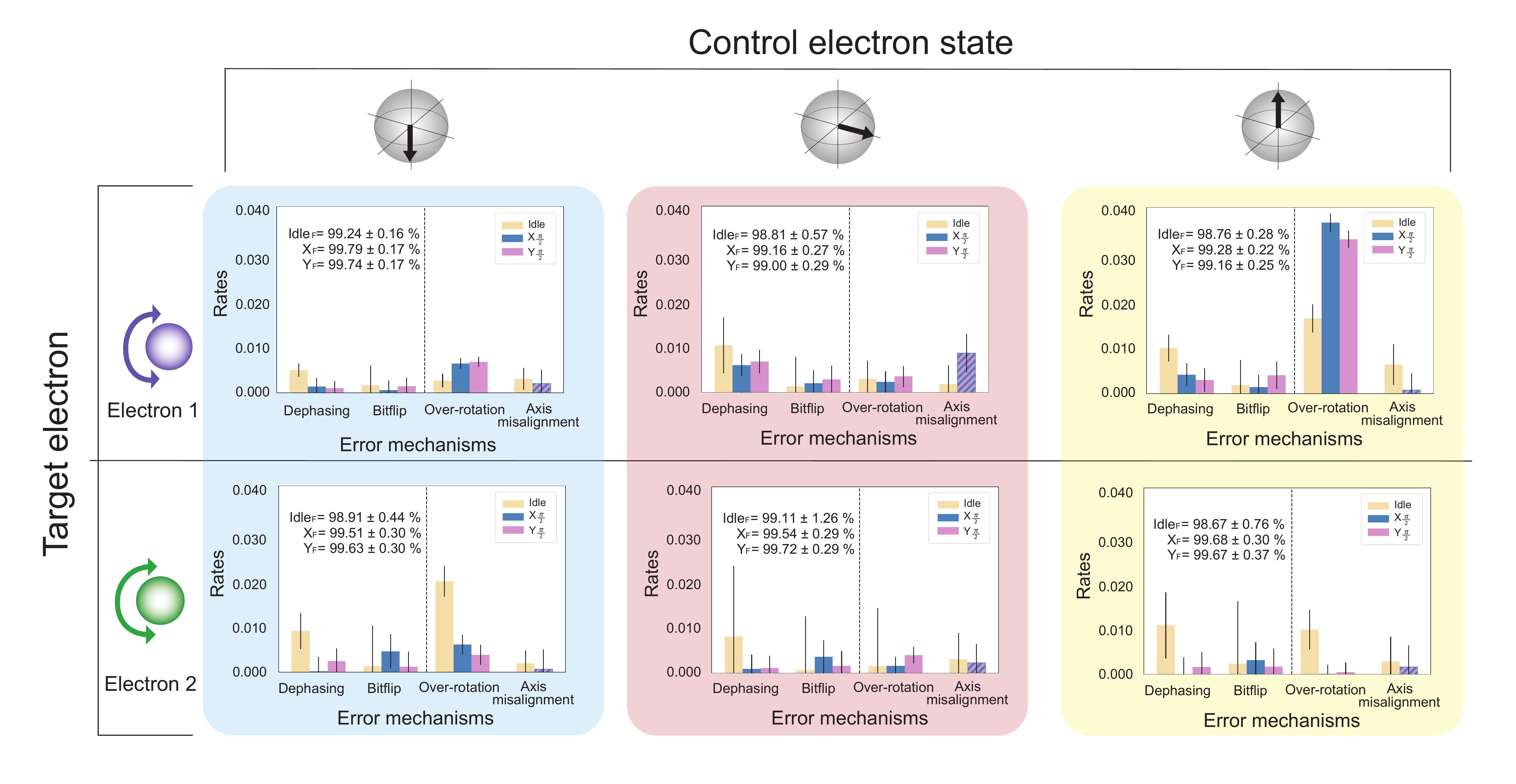}
    \caption{\textbf{Unconditional single-qubit GST.} Characterisation of the unconditional single-qubit gates, using single-qubit GST. The unconditional gates were implemented by applying two sequential $\frac{\pi}{2}$ rotations to the target electron, one conditional on the control electron being in the spin $\ket{\downarrow}$ state and one conditional on the control electron being in the $\ket{\uparrow}$ state, in order to rotate one electron unconditional on the state of the other electron. To test the effectiveness of this unconditional gate, we performed three separate GST experiments on each electron, with the other electron initialised in either the $\ket{\downarrow}$ state, a superposition state of $\frac{1}{\sqrt{2}}(\ket{\downarrow} + \ket{\uparrow})$ or the  $\ket{\uparrow}$ state. The diagrams to the left of the rows indicate which electron is acting as the target electron in each GST experiment, while the Bloch sphere schematics above each of the columns denotes the state of the corresponding control electron throughout each GST experiment. The error rates can be broken down into the two stochastic error sources: dephasing and bitflip errors (to the left of the black dashed line) and two coherent error sources: over-rotation and axis-misalignment (to the right of the black dashed line). The axis-misalignment of the $X_{\frac{\pi}{2}}$ and $Y_{\frac{\pi}{2}}$ gates on the other hand, represents the error in the relative angle between the X and Y rotation axes of the qubit and hence is represented by a single striped column, to indicate that this error is associated with both the $X_{\frac{\pi}{2}}$ and $Y_{\frac{\pi}{2}}$ gates. For a full description of how the error rates for each physical error mechanism were calculated see Supplementary Note 8 (sub note 3). The fidelities estimated for each gate, $X_F$, $Y_F$ and $\text{Idle}_F$, are quoted in the inset of each plot. For a full breakdown of the error rates extracted from GST for the unconditional single-qubit operations, see Supplementary Note 8 (sub note 2). The error bars represent confidence intervals of $2\sigma$.}
    \label{fig:uncond_1q}
\end{figure*}

We investigated the effect of the exchange interaction on the coherence of the electrons by performing Ramsey and Hahn echo experiments on electron Q2, both with donor 1 in the neutral and ionised state. Donor 1 was ionised by tuning the device gate voltages such that electron Q1 is able to tunnel from the donor to the SET island. With the removal of Q1, the exchange coupling is no longer present in the system. The $T_{2}^{*}$ and $T_2^{\rm Hahn}$ times of Q2 did not change within the error bars for the fits (shown in Fig.~\ref{fig:ramsey_hahn}) with the removal of the exchange coupling, indicating that the noise in the exchange interaction is not a dominant source of decoherence of the qubits. This finding is consistent with results obtained in lithographic quantum dots operated in the regime $J < \Delta E_z$. In such devices, it was found that fluctuations in $J$ were an insignificant source of dephasing compared to other noise sources \cite{huang2019fidelity,noiri2022fast}. Conversely, experiments conducted in the $J > \Delta E_z$ regime typically show a deterioration of spin coherence upon increasing $J$ \cite{veldhorst2015two}. Our observation thus confirms the benefit of using the weak-exchange regime for entangling operations.

\subsection*{Exchange-based two-qubit gates}
\label{sec:two_qubit_gates}

Two-qubit controlled rotation (CROT) gates are naturally obtained by preparing the two donor nuclear spins in an anti-parallel configuration, i.e. either $\ket{\Downarrow \Uparrow}$ or $\ket{\Uparrow \Downarrow}$ (Fig.~\ref{fig:1}a). In doing so, the two electron spins are frequency-detuned by the average hyperfine coupling $\overline{A} =\frac{(A_{1}+A_{2})}{2}$. This can be thought of as the switchable, digital version of the detuning caused by a gradient in Overhauser field $\Delta B_z$ in double quantum dots \cite{petta2005coherent}, i.e. $\Delta B_z = \pm \overline{A}$ (Fig.~\ref{fig:1}c).

The two-electron spin eigenstates of the system are shown in Table \ref{tab:two_qubit_states}, where $\tan(2\theta) = J/\overline{A}$. In the present device, $J \approx 12$~MHz $\ll \overline{A} \approx 112$~MHz results in $\cos(\theta) = 0.9986$, so that the eigenstates are almost (but not exactly) the tensor products of the individual spins' basis states $\{\ket{\downarrow},\ket{\uparrow}\}$.

\begin{table}[H]
    \centering
    \renewcommand{\arraystretch}{2}
    \begin{tabular}{c|c c}
         $\ket{\rm Q1 Q2}$ & Eigenbasis & $S_z$ basis  \\
         \hline
         $\ket{11}$ & $\ket{\downarrow \downarrow}$ & $\ket{\downarrow \downarrow}$ \\
         $\ket{10}$ & $\widetilde{\ket{\downarrow \uparrow}}$ & $\cos(\theta)\ket{\downarrow \uparrow} - \sin(\theta)\ket{\uparrow \downarrow}$\\
        $\ket{01}$  & $\widetilde{\ket{\uparrow \downarrow}}$ & $\cos(\theta)\ket{\uparrow \downarrow} + \sin(\theta)\ket{\downarrow \uparrow}$\\
        $\ket{00}$ & $\ket{\uparrow \uparrow}$ & $\ket{\uparrow \uparrow}$
    \end{tabular}
    \caption{Two-qubit electron spin states.}
    \label{tab:two_qubit_states}
\end{table}

\begin{figure*}[ht]
    \centering
    \includegraphics[width=1\textwidth]{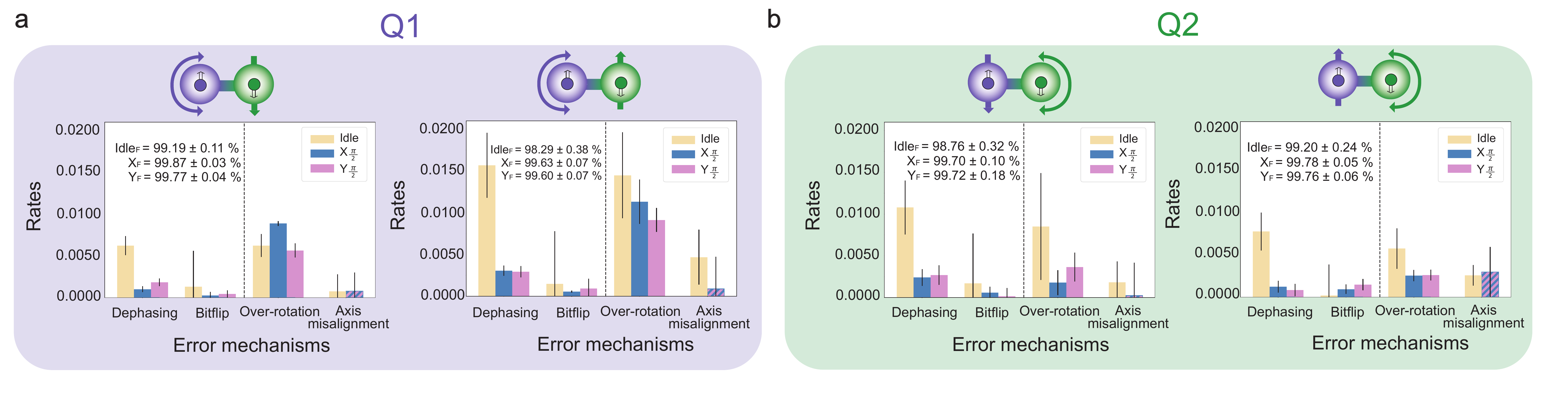}
    \caption{\textbf{Conditional single-qubit GST on both electrons.} \textbf{a,} Gate characterisation of the native, conditional gates performed on electron Q1, estimated using single-qubit GST. The schematics above each plot depict the state in which electron Q2 was initialised and upon which each gate on Q1 was conditioned, for the two separate GST experiments. \textbf{b,} Error rates for the native, conditional gates performed on electron Q2, estimated using single-qubit GST. Stochastic error mechanisms are shown to the left of the black dashed line on each plot, while coherent error mechanisms are shown to the right of the black dashed line. For more information regarding how each physical error mechanism was extracted, see Supplementary Note 8 (sub note 3). The fidelities estimated for each gate are quoted in the inset of each plot. For a full breakdown of the error rates for the conditional single-qubit operations, see Supplementary Note 8 (sub note 1). The error bars represent confidence intervals of $2\sigma$.}
    \label{fig:1q}
\end{figure*}

For each anti-parallel nuclear state, there exists two ESR frequencies for electron Q1, which are separated by $J$ and represent the following transitions: $\ket{\downarrow \downarrow} \leftrightarrow \widetilde{\ket{\uparrow \downarrow}}$ and $\widetilde{\ket{\downarrow \uparrow}} \leftrightarrow \ket{\uparrow \uparrow}$. Similarly, for electron Q2 two ESR resonances exist for each anti-parallel nuclear state, which represent the transitions: $\ket{\downarrow \downarrow} \leftrightarrow \widetilde{\ket{\downarrow \uparrow}}$ and $\widetilde{\ket{\uparrow \downarrow}} \leftrightarrow \ket{\uparrow \uparrow}$. A $\pi$ pulse applied at any of these resonance frequencies represents the inversion of one electron, conditional on the state of the other electron. By defining $\ket{\downarrow}$ as the computational $\ket{1}$ state and  $\ket{\uparrow}$ as the computational $\ket{0}$ state, a CROT gate is therefore implemented by applying a $\pi$ pulse at the resonance to flip the target electron conditional on the control electron being in the $\ket{1}$ state, while a zCROT operation is implemented by applying a $\pi$ pulse at the resonance to flip the target electron conditional on the control electron being in the $\ket{0}$ state.  For the entirety of this work we therefore set the nuclei into the anti-parallel spin configuration $\ket{\Uparrow_{1}\Downarrow_{2}}$. The native CROT gates obtained in this way are equivalent to the resonant gates in double quantum dots with fixed exchange coupling \cite{zajac2018resonantly,huang2019fidelity}, where the detuning between Q1 and Q2 is created either by a $g$-factor difference, or by a magnetic field gradient. This resonant CROT only requires $J$ to be larger than the ESR linewidth ($\sim 100$~kHz for donor electrons in $^{28}$Si) and smaller than the detuning (here given by $\bar{A}\sim 100$~MHz), making this gate robust against donor placement uncertainties.

\subsection*{Characterising gate performance}
\label{sec:fidelity}
In order to fully characterise this $J$-coupled system, we performed one-qubit GST in two separate ways.  First, we characterised the single-qubit, unconditional gates. Since the native gates in this $J$-coupled system are two-qubit CROT operations, an unconditional one-qubit gate is implemented by two sequential CROT gates, in order to rotate one electron unconditional on the state of the other electron \cite{huang2019fidelity}. We performed GST on a gate set consisting of a $\pi/2$ rotation about the $X$ axis of the Bloch sphere ($X_{\frac{\pi}{2}}$), a $\pi/2$ rotation about the $Y$ axis ($Y_{\frac{\pi}{2}}$), and an idle operation (I). This gate set was characterised for both Q1 and Q2, with the idle electron initialised in either the $\ket{\downarrow}$, $\frac{1}{\sqrt{2}}(\ket{\downarrow} + \ket{\uparrow})$ or $\ket{\uparrow}$ state.  Notwithstanding the longer gate times compared to the native conditional operations, fidelities of both $X_{\frac{\pi}{2}}$ and  $Y_{\frac{\pi}{2}}$ exceeded 99.00$\pm0.29\%$. Here and elsewhere, error bars indicate 2$\sigma$ confidence intervals. \\

We then performed one-qubit GST on the conditional single-qubit gates, in all four contexts: acting on Q1 and Q2, conditional on the control electron being initialised in either the $\ket{\downarrow}$ or $\ket{\uparrow}$ state. As with the case of the unconditional gates, we analysed a gate set consisting of $X_{\frac{\pi}{2}}$, $Y_{\frac{\pi}{2}}$. Each gate was iteratively calibrated using the one-qubit GST analysis results. The value of coherent over- or under-rotation error extracted by GST was used to  update the rotation angle of the electron in the next round of experiments, until the coherent errors were smaller than their own error bars. Fig.~\ref{fig:1q} shows the estimated error budgets for each of these four native conditional operations as extracted from the estimated error rates. The $X_{\frac{\pi}{2}}$ and $Y_{\frac{\pi}{2}}$ gates both have fidelity exceeding 99.63$\pm$0.07$\%$ for all four of the native conditional rotations.\\

\begin{figure*}[ht]
    \centering
    \includegraphics[width=1\textwidth]{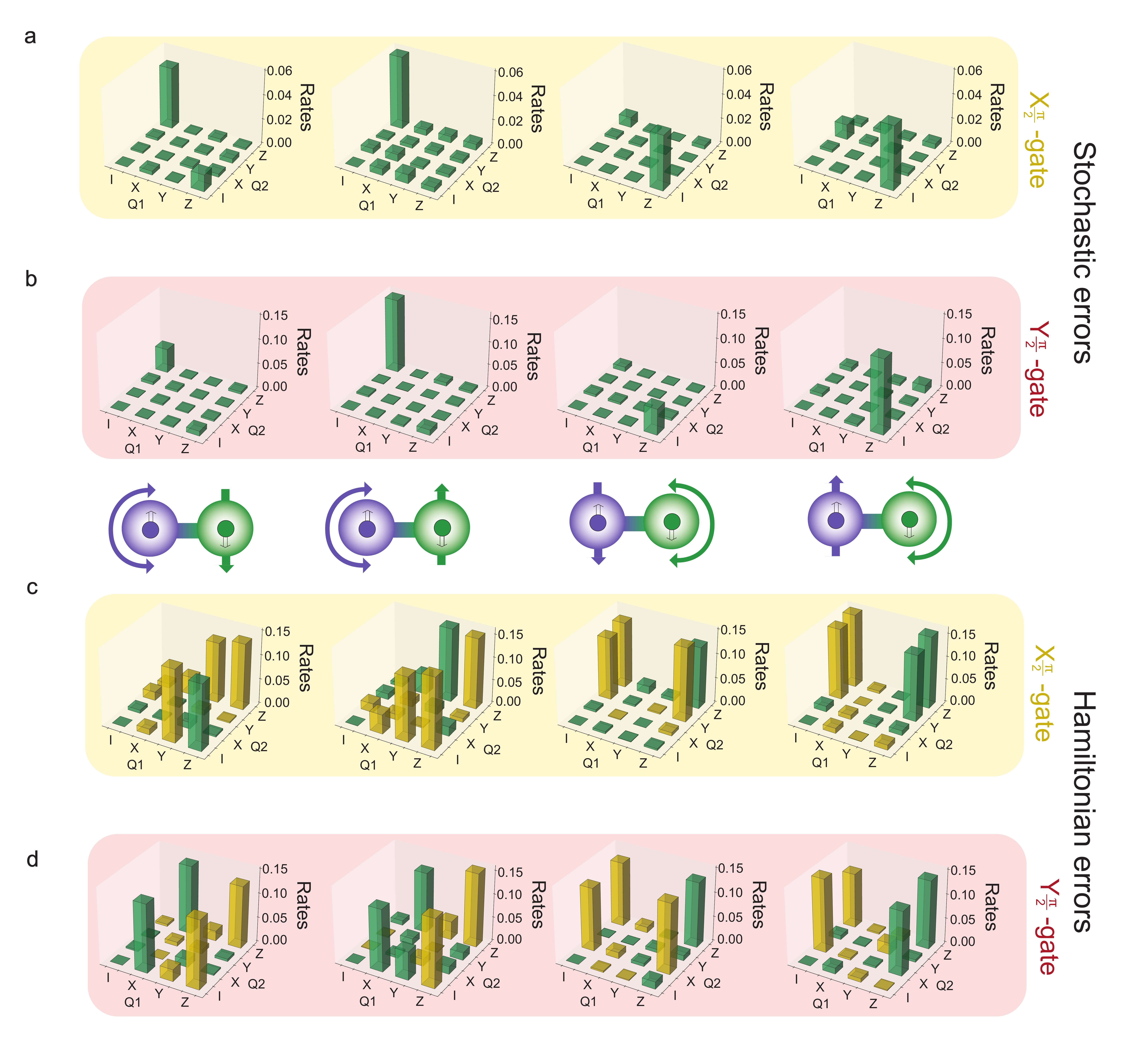}
    \caption{\textbf{Two-qubit GST on the electrons.} Stochastic and Hamiltonian error rates estimated using two-qubit GST for both the $X_{\frac{\pi}{2}}$  (\textbf{a,c}) and $Y_{\frac{\pi}{2}}$ (\textbf{b,d}) gates. From left to right in \textbf{a,c}(\textbf{b,d}) these $X_{\frac{\pi}{2}}$($Y_{\frac{\pi}{2}}$) rotations were performed on: Q1, conditional on Q2 being in the $\ket{\downarrow}$ state, Q1 conditional on Q2 being in the $\ket{\uparrow}$ state, Q2 conditional on Q1 being in the $\ket{\downarrow}$ state and Q2 conditional on Q1 being in the $\ket{\uparrow}$ state. These eight gates were all tested in the same two-qubit GST experiment. The full error rates estimated from this two-qubit GST experiment are shown in the Supplementary Note 9.}
    \label{fig:2q_GST}
\end{figure*}

Next, we performed full two-qubit GST on a gate set consisting of conditional $X_{\frac{\pi}{2}}$ and $Y_{\frac{\pi}{2}}$ rotations for each of the four configurations of the native conditional two-qubit operation (Fig. \ref{fig:2q_GST}), together with a global idle giving $9$ gates in total. The analysis of the two-qubit GST results tells a more complicated story, revealing error mechanisms that were otherwise invisible when operating the device as an effective one-qubit system as above. The error budget in the two-qubit case is dominated by three main sources. For the two-qubit conditional rotation gates there are systematic coherent errors that correspond to relational axis misalignment errors between pairs of $X_{\frac{\pi}{2}}$ and $Y_{\frac{\pi}{2}}$ gates acting on different control and target qubits. Additionally, the two-qubit conditional rotation gates induce a significant amount of dephasing on the control qubit. For the idle gate, on the other hand, the error is predominantly dephasing. 

To help understanding the underlying physical error channels, we separate out coherent errors, mostly attributable to the control system and potentially correctable through calibration, from stochastic errors which typically arise from the device physics. We perform this partition using the generator fidelity introduced in \cite{mkadzik2022precision}, and find that the two-qubit conditional rotation gates have incoherent contributions to their infidelities ranging from 6.91$\%$ to 21.88$\%$, with an average of 12.54$\pm$6.40$\%$ (see Supplementary Note 9 for details).

\begin{figure*}[ht]
    \centering
    \includegraphics[width=0.95\textwidth]{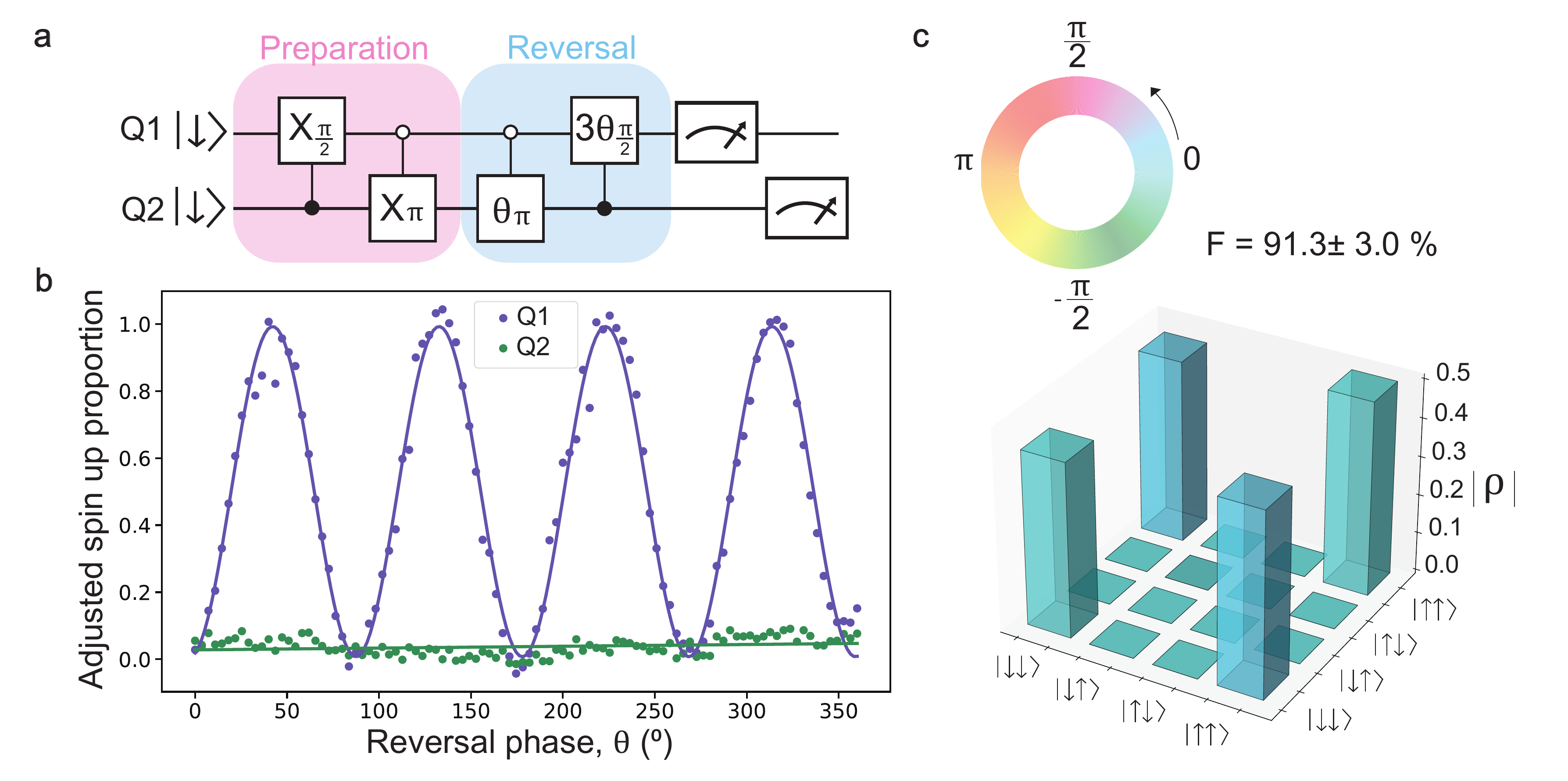}
    \caption{\textbf{Phase reversal tomography of a two-electron Bell state.} \textbf{a,} Circuit diagram associated with the preparation of the $\Phi^{+} = \frac{1}{\sqrt{2}}(\ket{\downarrow \downarrow} + \ket{\uparrow \uparrow})$ Bell state. \textbf{b,} Oscillation in the spin up proportion of electron Q1 and Q2 as a function of the phase of the reversal pulses, with the amplitude of the oscillations adjusted to account for state preparation and measurement (SPAM) error. \textbf{c,} Reconstructed density matrix of the $\Phi^{+}$ state using phase reversal tomography. The magnitude of the density matrix elements are represented by the height of the respective bars, while the phase of each element is denoted by the associated color on the color wheel. From this density matrix a Bell state fidelity of F = $91.3 \pm 3.0 \%$ was extracted. }
    \label{fig:phase_reversal}
\end{figure*}

The much lower gate fidelities obtained from two-qubit GST compared to one-qubit GST are a remarkable observation, given that all operations in a fixed-$J$ system are natively two-qubit conditional gates. This suggests that device-specific error mechanisms may have a more severe effect than previously recognised. We conjecture, for example, that much of the observed increase in dephasing on the control qubit can be attributed to jumps in the resonance frequency of the target qubit, induced by changes in the orientation of a few weakly coupled $^{29}$Si nuclei, as discussed in the sections above. These frequency jumps are small ($\sim 10$~kHz) in absolute terms, but their impact may be magnified by the geometric phase $\phi_{\rm c}$ induced on the control qubit upon performing a rotation of the target qubit. Since $\phi_{\rm c}$ is half the solid angle traversed by the target qubit on the Bloch sphere \cite{mkadzik2022precision}, its value can depend very sensitively on the detuning $\Delta \nu_{\rm t}$ between the instantaneous resonance frequency of the target qubit and the frequency of the microwave driving field. Even a small shift $\Delta \nu_{\rm t} \ll 2\nu_{\text{Rabi}}$, where $\nu_{\text{Rabi}}$ is the Rabi frequency of the target electron, can alter the path traversed by the target qubit in the Bloch sphere of its rotating frame, and cause a stochastic error on the phase of the control qubit. Frequency jumps caused by more strongly coupled $^{29}$Si nuclei, with $\Delta \nu_{\rm t} > 100$~kHz, can be mitigated by checking (e.g. with a quick Ramsey measurement) and re-calibrating the resonance frequency. This strategy becomes increasingly difficult if $\Delta \nu_{\rm t}$ approaches the intrinsic linewidth of the target qubit, proportional to $1/T_{2}^{*}$ ($\approx$ 20 kHz). In these cases, pulse engineering could be utilised to design a drive that is more robust against these small shifts in the resonance frequency. Additional work is currently underway validating this conjecture and exploring other possible decoherence mechanisms which could be partially responsible for the observed dephasing rates.\\

\subsection*{Electron spin Bell state tomography}

We demonstrated the entangling nature of the two-qubit CROT operations by creating and tomographing electron Bell states. We created a two-electron Bell state by initialising the spins in the state $\ket{\textnormal{Q1Q2}} =\ket{\downarrow \downarrow}$, performing an X$\frac{\pi}{2}$ gate on Q1, and then an entangling CROT operation on Q2. The fidelity of the resulting state was benchmarked using phase reversal tomography \cite{mehring2003entanglement, sackett2000experimental, wei2020verifying, dehollain2016bell}, whereby the pulses used to create the Bell state are applied in reverse, but with a progressively increasing phase (Fig.~\ref{fig:phase_reversal}a). Accordingly, the ability to reverse the electrons back to their initial state oscillates according to the phase of the reversal pulses. This results in the spin up proportion $P_{\uparrow 1}(\theta)$ of the Q1 electron oscillating as a function of reversal phase $\theta$, where $P_{\uparrow 1}(0)=0$ indicates that the electrons have been perfectly reversed to their initial state. The amplitude and phase of these oscillations allows for the reconstruction of the off-diagonal corner elements of the two-qubit density matrix, $\rho_{14}, \rho_{41}$. In order to obtain the diagonal corner elements, $\rho_{11}, \rho_{44}$, we directly measured the Bell state populations in the Z-basis immediately following the state's preparation. By choosing to prepare the $\Phi^{+}=\frac{1}{\sqrt{2}}(\ket{\downarrow \downarrow} + \ket{\uparrow \uparrow})$ Bell state, which only has non-zero elements in each of the four corners of the density matrix, the fidelity of this Bell state can be ascertained using these four matrix elements. Using this method we obtained a Bell state fidelity of $91.3 \pm 3.0 \%$, with state preparation and measurement (SPAM) error removed, and $72.5\pm 2.0$\% without SPAM removal (Supplementary Note 10). An optimisation algorithm was used to find the closest physical density matrix to the measured density matrix, $\rho$. A concurrence value of $0.87\pm0.05$ was obtained from the SPAM-removed density matrix (Supplementary Note 10 and 11), indicating a high degree of entanglement between Q1 and Q2.

\section*{Discussion}
\label{sec:conclusion}

We demonstrated an entangling two-qubit gate between two exchange-coupled electrons bound to ion-implanted $^{31}$P donors in silicon, by exploiting the hyperfine interaction to provide a natural source of detuning $\Delta B_{z}$ between the two qubits. 
This method works for any value of $J$ between the ESR linewidth ($\sim 100$~kHz) and the hyperfine coupling ($\sim 100$~MHz), making it robust against the precise placement of the donors. 
Furthermore, we showed that the small value of \textit{J} results in the qubit coherence being unaffected by the coupling.  

Gate-set tomography experiments showed that both conditional and unconditional single-qubit gate fidelities consistently surpass 99\%. However, the full two-qubit GST showed an increased error rate, likely caused by the coupling to nearby $^{29}$Si nuclei (Supplementary Note 12). This observation motivates the adoption of $^{28}$Si isotopic enrichment far beyond the 800 ppm residual $^{29}$Si used in this experiment \cite{madzik2020controllable,witzel2010electron}. Promising results, with residual $^{29}$Si concentrations of $2.4 \pm 0.7$~ppm have been recently obtained using focused ion beam implantation of $^{28}$Si \cite{acharya2024highly}.

The value of exchange coupling $J \approx 12$~MHz in the present device corresponds to an expected inter-donor distance in the range 15-22~nm \cite{joecker2021full}. This distance is comparable to the smallest feature sizes of ultra-scaled semiconductor devices that use industrial CMOS fabrication \cite{cao2023future}. Therefore, 2-qubit logic gates based on weak exchange provide a plausible pathway to integrate and scale up single-donor quantum devices. 

In the medium term, scaling up beyond two qubits is likely to involve coupling donors to quantum dots at the Si/SiO$_2$ interface, and then using chains of quantum dots \cite{chan2021exchange} or `jellybean' elongated dots \cite{wang2023jellybean} to reach beyond the $\sim 20$~nm range. Small logical qubit structures could be built this way \cite{jones2018logical}, or extended to full-scale surface code architectures \cite{pica2016surface}. The donor nuclear spin can act as a `digital detuning' between the donor and dot electrons \cite{harvey2017coherent}, or be used directly as a long-lived, high-fidelity qubit where the electrons act as entanglement mediators \cite{mkadzik2022precision}. The possibility of coupling to interface quantum dots is a unique feature of the MOS-compatible, ion-implanted donor fabrication route. The demonstration of entanglement between the electrons of two ion-implanted donors provided here constitutes the key stepping stone to unlocking hybrid, scalable spin-based quantum computer architectures in silicon.

\section*{Methods}
\subsection*{Sample fabrication}
For the device fabrication, an isotopically enriched epilayer of $^{28}$Si, with a residual $^{29}$Si concentration of 730 ppm, is grown atop a silicon wafer. A 200 nm thick SiO$_{2}$ field oxide is grown via a wet oxidation process, while in the active region of the device a high-quality 8 nm thermal oxide layer is grown in dry conditions. In order to implant the devices with donor atoms, a 90 nm x 100 nm implantation window is defined using electron-beam lithography (EBL). Donor atoms are then implanted into this window at an acceleration voltage of 10 kV and implantation dose of  1.4 $\times$ 10$^{12}$ cm$^{-2}$. Using SRIM simulations we have estimated that at this implantation energy, the implantion profile of the donors should exist in a range approximately 10 nm  from the oxide interface. A rapid thermal anneal for 5 seconds at 1000 $^{o}$C is carried out following implantation in order to repair the damage inflicted to the lattice during implantation and to activate the donors. Aluminium gates are then fabricated on the surface of the chip in three separate EBL steps, with each step being followed by a thermal deposition of aluminium, followed by exposure to pure, low pressure oxygen to form an Al$_{2}$O$_{3}$ layer between each gate layer. A forming gas anneal (95$\%$ N$_{2}$, 5$\%$ H$_{2}$) for 15 minutes at 400 $^{o}$C is then performed in order to passivate any interface traps. 
\subsection*{Experimental setup}
The device is glued to a copper enclosure, through a cut out in a custom printed circuit board (PCB), that contains both microwave and low frequency lines. The device is then aluminium wire bonded to the PCB and bolted to a permanent magnet assembly, which provides a strong, static magnetic field. The permanent magnet assembly is mounted onto the mixing chamber plate of a Bluefors BF-LD400 cryogen free dilution refrigerator, which cools the device to a base temperature of approximately 20 mK.\\

DC voltages are sourced by 9 Stanford Research systems (SRS) SIM 928 DC sources, hosted in two SIM 900 mainframes. The DC lines consist of a constantan loom, which is low pass filtered with a 20 Hz cutoff at the mixing chamber stage. Three of the donor gates also have an AC input to allow for fast dynamic tuning of these gate voltages. These RF signals are provided by a Keysight M3300A arbitrary waveform generator (AWG), which is bandwidth limited to 200 MHz, and passed into the AC inputs of AC/DC combiners for these gates, to allow an RF modulation to be added on top of the static DC voltage. The RF lines are made from a flexible copper-nickel (Cu-Ni) coaxial line. These lines are graphite coated to reduce triboelectric noise and filtered to a cutoff frequency of 145 MHz at the mixing chamber plate.\\

A Keysight M3201A module hosts a field-programmable gate array (FPGA), on which we
have built an in-house direct digital synthesis (DDS) system. The DDS provides RF input signals to the I and Q inputs of a Keysight E8267D PSG vector microwave source, to allow for single and dual-sideband modulation of a microwave tone produced by the vector source, for control of the electron spin. Additionally, the DDS provides the RF signal for control of the nuclear spin through NMR. The NMR signal is attenuated by 10 dB at room temperature, to protect the antenna from any unexpected spikes in NMR amplitude, before being combined with the microwave output of the vector source using a DPX1721 diplexer, allowing both NMR and ESR signals to travel down the same line to the device’s antenna. This line consists of a semi-rigid, silver-plated Cu-Ni coaxial line. The line then passes through an inner/outer DC block at room temperature, to prevent possible voltage offsets from creating a DC current through the thin short circuit termination of the microwave antenna. Additionally, this line is attenuated by a further 10 dB, with an attenuator positioned at the 4K stage of the dilution refrigerator.\\

The SET current returning from the drain is passed through a Basel SP983c transimpedence amplifier, which converts the current into a voltage with a gain of 10$^{7}$ V/A and frequency bandwidth of 100 kHz. The voltage signal is then passed into a SIM911 bipolar junction transistor (BJT) amplifier, which is housed in the same SIM900 mainframe as the DC sources and provides an additional gain of 10$^{2}$. This amplifier also has the function of breaking the ground between the fridge and the measurement setup. As the BJT amplifier can add additional noise to the current signal, the signal is passed through a further passive low pass filter, with a cutoff at 100 kHz. The signal then enters the digitiser channel of the Keysight M3300A, with a sample rate of 100 megasamples-per-second.

\section*{Data Availability}
The source data and analysis scripts  generated in this study can be found in the following repository https://doi.org/10.5061/dryad.w3r2280zm.

\providecommand{\noopsort}[1]{}\providecommand{\singleletter}[1]{#1}%
%

% \bibliography{references}

\section*{Acknowledgements}
\label{sec:acknowledgements}
This research was funded by the Australian Research Council Centre of Excellence for Quantum Computation and Communication Technology (CE170100012) and the US Army Research Office (Contracts no. W911NF-17-1-0200 and W911NF-23-1-0113). We acknowledge the facilities, and the scientific and technical assistance provided by the UNSW node of the Australian National Fabrication Facility (ANFF), and the Heavy Ion Accelerators (HIA) nodes at the University of Melbourne and the Australian National University. ANFF and HIA are supported by the Australian Government through the National Collaborative Research Infrastructure Strategy (NCRIS) program. H.G.S., M.R.v.B., A.V. acknowledge support from the Sydney Quantum Academy. C.I.O, K.M.R, K.C.Y and R.J.B-K acknowledge funding in part by the U.S. Department of Energy, Office of Science, Office of Advanced Scientific Computing Research Quantum Testbed Pathfinder Program. The views and conclusions contained in this document are those of the authors and should not be interpreted as representing the official policies, either expressed or implied, of the Army Research Office or the U.S. Government. The U.S. Government is authorized to reproduce and distribute reprints for Government purposes notwithstanding any copyright notation herein.
Sandia National Laboratories is a multimission laboratory managed and operated by National Technology \& Engineering Solutions of Sandia, LLC, a wholly owned subsidiary of Honeywell International Inc., for the U.S. Department of Energy's National Nuclear Security Administration under contract DE-NA0003525. This paper describes objective technical results and analysis. Any subjective views or opinions that might be expressed in the paper do not necessarily represent the views of the U.S. Department of Energy or the United States Government.

\section*{Author Contributions Statement}
H.G.S., S.A., A.L. and A.M. conceived and designed the experiments. H.G.S., S.A., M.R.vB., A.V., M.A.I.J., A.J.A.H., H.R.F. performed and analysed the measurements. M.T.M. and F.E.H. fabricated the device, with A.S.D.'s supervision, on materials supplied by K.M.I.. A.M.J., B.C.J. and D.N.J. designed and performed the ion implantation. C.I.O., K.M.R., K.Y. and R.B.-K. contributed to the GST analysis. R.Y.S. and C.H.Y. contributed to the data analysis. H.G.S. and A.M. wrote the manuscript, with input from all coauthors. A.M. supervised the project.

\section*{Competing Interests Statement}
A.M. and A.L. are coauthors on a patent that describes the quantum logic operations used in this work (AU2013302299B2, US10878331B2, EP2883194B1). A.S.D. is the CEO and a director of Diraq Pty Ltd. C.H.Y., A.L., F.E.H. and A.S.D. declare equity interest in Diraq Pty Ltd.

\clearpage

\preprint{APS/123-QED}

\title{Supplementary Information: Tomography of entangling two-qubit logic operations in exchange-coupled donor electron spin qubits}

\author{Holly G. Stemp$^{1,2}$}
\author{Serwan Asaad$^{1,2}$}%
    \altaffiliation[Currently at ]{Quantum Machines, Denmark}
\author{Mark R. van Blankenstein$^{1,2}$}
\author{Arjen Vaartjes$^{1,2}$}
\author{Mark A. I. Johnson$^{1,2}$}%
    \altaffiliation[Currently at ]{Quantum Motion, London, UK}
\author{Mateusz T. M\k{a}dzik$^{1,2}$}%
    \altaffiliation[Currently at ]{Intel Corporation Hillsboro, Oregon, United States}
\author{Amber J. A. Heskes$^{1,2}$}%
    \altaffiliation[Currently at ]{University of Twente, Enschede, The Netherlands}
\author{Hannes R. Firgau$^{1,2}$}
\author{Rocky Y. Su$^{1}$}
\author{Chih Hwan Yang$^{1, 3}$}
\author{Arne Laucht$^{1, 3}$}%
\author{Corey I. Ostrove$^{4}$}%
\author{Kenneth M. Rudinger$^{4}$}%
\author{Kevin Young$^{4}$}%
\author{Robin Blume-Kohout$^{4}$}
\author{Fay E. Hudson$^{1, 3}$}%
\author{Andrew S. Dzurak$^{1,3}$}%
\author{Kohei M. Itoh$^{5}$}%
\author{Alexander M. Jakob$^{2,6}$}%
\author{Brett C. Johnson$^{7}$}%
\author{David N. Jamieson$^{2,6}$}%
\author{Andrea Morello$^{1,2}$}%
 \email{a.morello@unsw.edu.au}

\affiliation{%
 $^{1}$ School of Electrical Engineering and Telecommunications, UNSW Sydney, Sydney, NSW 2052, Australia\\
 $^{2}$ ARC Centre of Excellence for Quantum Computation and Communication Technology\\
 $^{3}$ Diraq Pty. Ltd., Sydney, New South Wales, Australia\\
 $^{4}$ Quantum Performance Laboratory, Sandia National Laboratories, Albuquerque, NM 87185 and Livermore, CA 94550, USA \\
 $^{5}$ School of Fundamental Science and Technology, Keio University, Kohoku-ku, Yokohama, Japan \\
 $^{6}$ School of Physics, University of Melbourne, Melbourne, VIC 3010, Australia\\
 $^{7}$ School of Science, RMIT University, Melbourne, VIC, 3000, Australia
}%

\maketitle
\renewcommand{\thetable}{\arabic{table}}
\renewcommand{\tablename}{Supplementary Table}
\renewcommand{\figurename}{Supplementary Fig.}
\renewcommand{\thefigure}{\arabic{figure}}

\onecolumngrid
\renewcommand\thesection{\arabic{section}}
\renewcommand\thesubsection{\arabic{subsection}} 
\setcounter{figure}{0}
\setcounter{section}{0}

\renewcommand{\arraystretch}{2}
\begin{tabular}{m{30em}m{17em}m{1.5em}}
1. Full description of an exchange-coupled two-qubit system & & 2\\
2. Initialisation of electron Q2 &  & 4 \\
3. Initialisation of the nuclei &  & 5 \\
4. Quantum non-demolition (QND) readout of the Q2 electron &  & 6 \\
5. Tuning the exchange interaction &  & 7 \\
6. CROT unitary matrix &  & 8 \\
7. Experimental GST Details & & 8 \\
8. One-Qubit GST: Extended Results &  & 10 \\
9. Two-Qubit GST: Extended Results &  & 15 \\
10. Phase reversal tomography &  & 18 \\
11. Concurrence &  & 21 \\
12. $^{29}$Si Spin bath &  &  22
\end{tabular}

% \tableofcontents

\newpage

% \section{Fabrication}
% For the device fabrication, an isotopically enriched epilayer of $^{28}$Si, with a residual $^{29}$Si concentration of 730 ppm, is grown atop a silicon wafer. A 200 nm thick SiO$_{2}$ field oxide is grown via a wet oxidation process, while in the active region of the device a high-quality 8 nm thermal oxide layer is grown in dry conditions. In order to implant the devices with donor atoms, a 90 nm x 100 nm implantation window is then defined using electron-beam lithography (EBL). Donor atoms are then implanted into this window at an acceleration voltage of 10 kV and implantation dose of  1.4 $\times$ 10$^{12}$ cm$^{-2}$. Using SRIM simulations we have estimated that at this implantation energy, the implantion profile of the donors should exist in a range approximately 10 nm  from the oxide interface. A rapid thermal anneal for 5 seconds at 1000 $^{o}$C is carried out following implantation in order to repair the damage inflicted to the lattice during implantation and to activate the donors. Aluminium gates are then fabricated on the surface of the chip in three separate EBL steps, with each step being followed by a thermal deposition of aluminium, followed by exposure to pure, low pressure oxygen to form an Al$_{2}$O$_{3}$ layer between each gate layer. A forming gas anneal (95$\%$ N$_{2}$, 5$\%$ H$_{2}$) for 15 minutes at 400 $^{o}$C is then performed in order to passivate any interface traps. 

\section{Full description of an exchange-coupled two-qubit system}\label{supp_A}

The system Hamiltonian, as described in the main text Results section, is given by the following:

\begin{align}
        H = &(\mu_{\text{B}}/h) B_{0}(g_{1}S_{z1} + g_{2}S_{z2})+
     \gamma_\mathrm{n}B_{0}(I_{z1}+I_{z2})+ 
      A_{1}\mathbf{S_{1}}\cdot \mathbf{I_{1}} +  A_{2}\mathbf{S_{2}}\cdot \mathbf{I_{2}} + 
     J(\mathbf{S_{1}\cdot S_{2}})
\end{align}

where $\mu_{\text{B}}$ is the Bohr magneton, $h$ is Planck's constant, $g_{1,2}\approx 1.9985$ the Land\'e  g-factors of each electron spin, $g\mu_{\text{B}}/h \approx$ 27.97 GHz/T and $\gamma_\mathrm{n} \approx$ 17.23 MHz/T is the $^{31}$P nuclear gyromagnetic ratio.

\begin{figure}[!h]
\begin{center}
\includegraphics[width=1\textwidth]{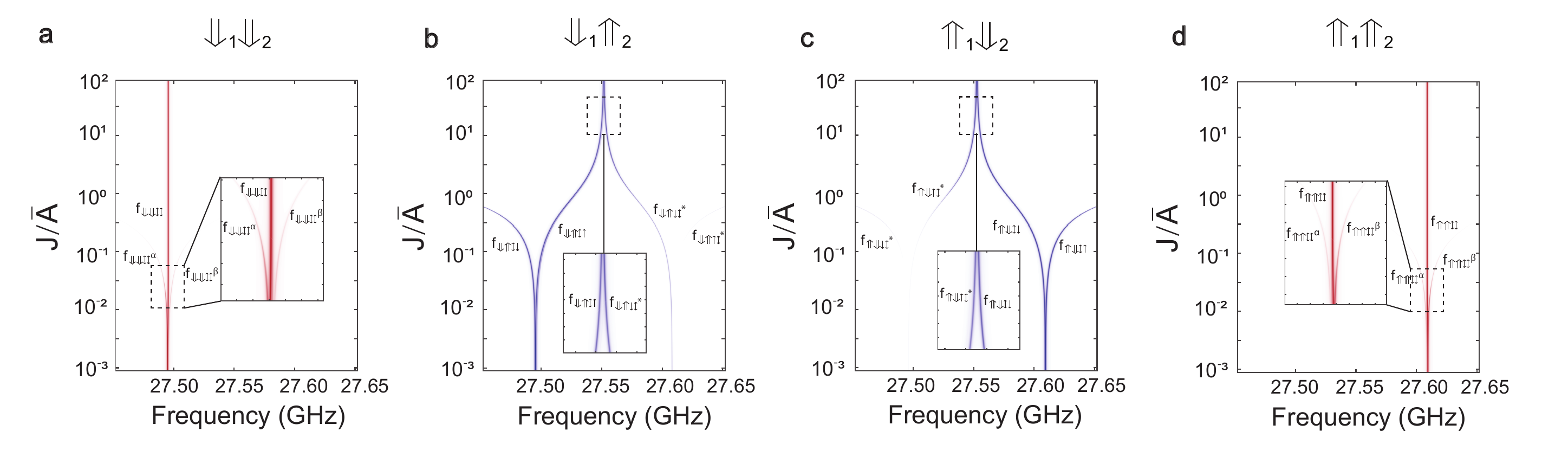}
\caption[Phase reversal tomography density matrix elements.]{Simulated ESR frequency spectrum for electron 1 for the case of the nuclei in the state $\ket{\Downarrow_{1}\Downarrow_{2}}$ (\textbf{a.}), $\ket{\Downarrow_{1}\Uparrow_{2}}$ (\textbf{b.}), $\ket{\Uparrow_{1}\Downarrow_{2}}$ (\textbf{c.}), $\ket{\Uparrow_{1}\Uparrow_{2}}$ (\textbf{d.}) plotted on a log scale of $J/\bar{A}$. Each inset shows a zoomed region of the plot for clarity. The colors of the transitions are scaled according to the probility of the transition, given by $P_{\text{ESR}}\Delta S_{1z}$ where $P_{\text{ESR}} = |\bra{\psi_{i}}(S_{1x} + S_{2x})\ket{\psi_{f}}|^{2}$ and $\Delta S_{1z} = \bra{\psi_{f}}S_{1z}\ket{\psi_{f}} - \bra{\psi_{i}}S_{1z}\ket{\psi_{i}}$ where $\ket{\psi_{i}}$ and $\ket{\psi_{f}}$ represent the initial and final state of electron 1 respectively. The term $\Delta S_{1z}$ is present in the scaling factor to mimic the readout of the electron experimentally, which occurs along the z-axis of the Bloch sphere. The simulation code used for these simulations was written by Dr Rachpon Kalra during his time at the University of New South Wales \cite{kalra2014robust}.}
\label{fig:all_J}
\end{center}
\end{figure}

Supplementary Fig.~\ref{fig:all_J} shows the simulated ESR spectrum for the electron bound to donor 1, referred to as electron 1, in an exchange-coupled donor system. The system has been simulated for each of the four possible two-nuclear states: $\ket{\Downarrow_{1}\Downarrow_{2}}, \ket{\Downarrow_{1}\Uparrow_{2}}, \ket{\Uparrow_{1}\Downarrow_{2}}$ and $\ket{\Uparrow_{1}\Uparrow_{2}}$ over a range of $J$ values spanning from $J \ll \bar{A}$ to $J \gg \bar{A}$. \\

For the case of the nuclei in the $\ket{\Downarrow_{1} \Downarrow_{2}}$ configuration (Supplementary Fig.~\ref{fig:all_J}, a) the following transition frequencies are present:

\begin{align}
    f_{\Downarrow \Downarrow \updownarrow \updownarrow \alpha} &= T_{-} \longleftrightarrow \widetilde{S},\\
    f_{\Downarrow \Downarrow \updownarrow \updownarrow} &= T_{-} \longleftrightarrow T_{0}, T_{0} \longleftrightarrow T_{+},\\
    f_{\Downarrow \Downarrow \updownarrow \updownarrow \beta} &= \widetilde{S} \longleftrightarrow T_{+},
\end{align}

where $\widetilde{S}$ represents a singlet-like state. In the limits of high $J$ this singlet-like state becomes a pure singlet, with a total spin of zero and thus can no longer be driven with ESR. For the case of $J \gg \bar{A}$, $f_{\Downarrow \Downarrow \updownarrow \updownarrow}$ represents the combined transition between $T_{-} \longleftrightarrow T_{0}$ and $T_{0} \longleftrightarrow T_{+}$. In the limit of small $J$ however, these transitions become split in frequency by $\Delta A = |A_{1} - A_{2}|$ and hence can be individually addressed.\\

For the case of the nuclei in the $\ket{\Downarrow_{1} \Uparrow_{2}}$ configuration (Supplementary Fig.~\ref{fig:all_J}, b) the transition frequencies present are the following. Electron 1 has been highlighted in blue to indicate that we are considering the frequency spectrum of this electron only, in the following discussion:

\begin{align}
    &f_{\Downarrow \Uparrow \updownarrow \downarrow} = \ket{\textcolor{blue}{\downarrow_{1}}\downarrow_{2}} \longleftrightarrow \widetilde{\ket{\textcolor{blue}{\uparrow_{1}}\downarrow_{2}}},\\
    &f_{\Downarrow \Uparrow \updownarrow \uparrow} = \widetilde{\ket{\textcolor{blue}{\downarrow_{1}}\uparrow_{2}}} \longleftrightarrow \ket{\textcolor{blue}{\uparrow_{1}}\uparrow_{2}},\\
    &f_{\Downarrow \Uparrow \downarrow \updownarrow}* = \ket{\textcolor{blue}{\downarrow_{1}}\downarrow_{2}} \longleftrightarrow \widetilde{\ket{\textcolor{blue}{\downarrow_{1}}\uparrow_{2}}},\\
    &f_{\Downarrow \Uparrow \uparrow \updownarrow}* = \widetilde{\ket{\textcolor{blue}{\uparrow_{1}}\downarrow_{2}}} \longleftrightarrow \ket{\textcolor{blue}{\uparrow_{1}}\uparrow_{2}},
\end{align}

where $\widetilde{\ket{\textcolor{blue}{\downarrow_{1}}\uparrow_{2}}} = \cos(\theta)\ket{\textcolor{blue}{\downarrow_{1}}\uparrow_{2}} + \sin(\theta)\ket{\textcolor{blue}{\uparrow_{1}}\downarrow_{2}}$, $\widetilde{\ket{\textcolor{blue}{\uparrow_{1}}\downarrow_{2}}} = \cos(\theta)\ket{\textcolor{blue}{\uparrow_{1}}\downarrow_{2}} + \sin(\theta)\ket{\textcolor{blue}{\downarrow_{1}}\uparrow_{2}}$ and $\tan(2\theta) = \frac{J}{\bar{A}}$. $f_{\Downarrow \Uparrow \downarrow \updownarrow}*$ and $f_{\Downarrow \Uparrow \uparrow \updownarrow}*$ represents transitions to flip electron 2 for the $\ket{\Downarrow_{1} \Uparrow_{2}}$ nuclear state. The reason that these transitions are seen in the ESR spectrum for electron 1 is due to the fact that for $J > 0$, the eigenstates of the two-electron system are no longer the product states $\ket{\downarrow_{1}\downarrow_{2}}, \ket{\downarrow_{1}\uparrow_{2}}, \ket{\uparrow_{1}\downarrow_{2}}$ and $\ket{\uparrow_{1}\uparrow_{2}}$ but rather the hybridised states $\ket{\downarrow_{1}\downarrow_{2}}, \widetilde{\ket{\downarrow_{1}\uparrow_{2}}}, \widetilde{\ket{\uparrow_{1}\downarrow_{2}}}$ and $\ket{\uparrow_{1}\uparrow_{2}}$. This hybridisation of states introduces an additional excitation of electron 1 upon driving electron 2 (and vice versa for driving electron 1). The magnitude of this excitation depends on the value of $\theta = \frac{1}{2}\arctan(\frac{J}{\bar{A}})$. For $J \rightarrow \infty, \theta \rightarrow \frac{\pi}{4}$ and hence $\widetilde{\ket{\downarrow_{1}\uparrow_{2}}} = \widetilde{\ket{\uparrow_{1}\downarrow_{2}}} = \frac{1}{\sqrt{2}}(\ket{\downarrow_{1}\uparrow_{2}} + \ket{\uparrow_{1}\downarrow_{2}})$. For $J \rightarrow 0$ however $\theta \rightarrow 0 $ and hence $\widetilde{\ket{\uparrow_{1}\downarrow_{2}}} = \ket{\uparrow_{1}\downarrow_{2}} $ and $\widetilde{\ket{\downarrow_{1}\uparrow_{2}}} = \ket{\downarrow_{1}\uparrow_{2}}$ resulting in an increasingly smaller excitation of electron 1 upon driving electron 2.\\

As $J$ increases, $f_{\Downarrow \Uparrow \updownarrow \uparrow}$ and $f_{\Downarrow \Uparrow \downarrow \updownarrow}*$ tend towards the transitions $T_{0} \longleftrightarrow T_{+}$ and $T_{-} \longleftrightarrow T_{0}$ respectively. In the limits of low $J$ these transitions are split by $\Delta A$ however, the two transitions become degenerate as $J \rightarrow \infty$.\\

For the case of the nuclei in the $\ket{\Uparrow_{1} \Downarrow_{2}}$ configuration (Supplementary Fig.~\ref{fig:all_J}, c) the transition frequencies present are the following:

\begin{align}
    &f_{\Uparrow \Downarrow \downarrow \updownarrow}* = \ket{\textcolor{blue}{\downarrow_{1}}\downarrow_{2}} \longleftrightarrow \widetilde{\ket{\textcolor{blue}{\downarrow_{1}}\uparrow_{2}}},\\
    &f_{\Uparrow \Downarrow \uparrow\updownarrow}* = \widetilde{\ket{\textcolor{blue}{\uparrow_{1}}\downarrow_{2}}} \longleftrightarrow \ket{\textcolor{blue}{\uparrow_{1}}\uparrow_{2}},\\
    &f_{\Uparrow \Downarrow \updownarrow \downarrow} = \ket{\textcolor{blue}{\downarrow_{1}}\downarrow_{2}} \longleftrightarrow \widetilde{\ket{\textcolor{blue}{\uparrow_{1}}\downarrow_{2}}},\\
    &f_{\Uparrow \Downarrow \updownarrow \uparrow} = \widetilde{\ket{\textcolor{blue}{\downarrow_{1}}\uparrow_{2}}} \longleftrightarrow \ket{\textcolor{blue}{\uparrow_{1}}\uparrow_{2}}.   
\end{align}

$f_{\Uparrow \Downarrow \downarrow \updownarrow}*$ and $f_{\Uparrow \Downarrow \uparrow \updownarrow}*$ represent transitions to flip electron 2 conditional on the nuclear state $\ket{\Uparrow_{1} \Downarrow_{2}}$. These transitions are observed in the ESR spectrum of electron 1 as a result of the hybridisation of the electron states for $J >0$ as explained above. As for the case of the $\ket{\Downarrow_{1} \Uparrow_{2}}$ nuclear state, $f_{\Uparrow \Downarrow \uparrow\updownarrow}*$ and $f_{\Uparrow \Downarrow \updownarrow \downarrow}$ become degenerate as $J \rightarrow \infty$.\\

For the case of the nuclei in the $\ket{\Uparrow_{1} \Uparrow_{2}}$ configuration (Supplementary Fig.~\ref{fig:all_J}, d) the transition frequencies present are:

\begin{align}
    f_{\Uparrow \Uparrow \updownarrow \updownarrow \alpha} &= T_{-} \longleftrightarrow \widetilde{S},\\
    f_{\Uparrow \Uparrow \updownarrow \updownarrow} &= T_{-} \longleftrightarrow T_{0}, T_{0} \longleftrightarrow T_{+},\\
    f_{\Uparrow \Uparrow \updownarrow \updownarrow \beta} &= \widetilde{S} \longleftrightarrow T_{+},
\end{align}

These transitions are identical in nature to the transitions present for the $\ket{\Downarrow_{1}\Downarrow_{2}}$ nuclear state. \\

\begin{figure*}[ht]
    \centering
    \includegraphics[width=1\textwidth]{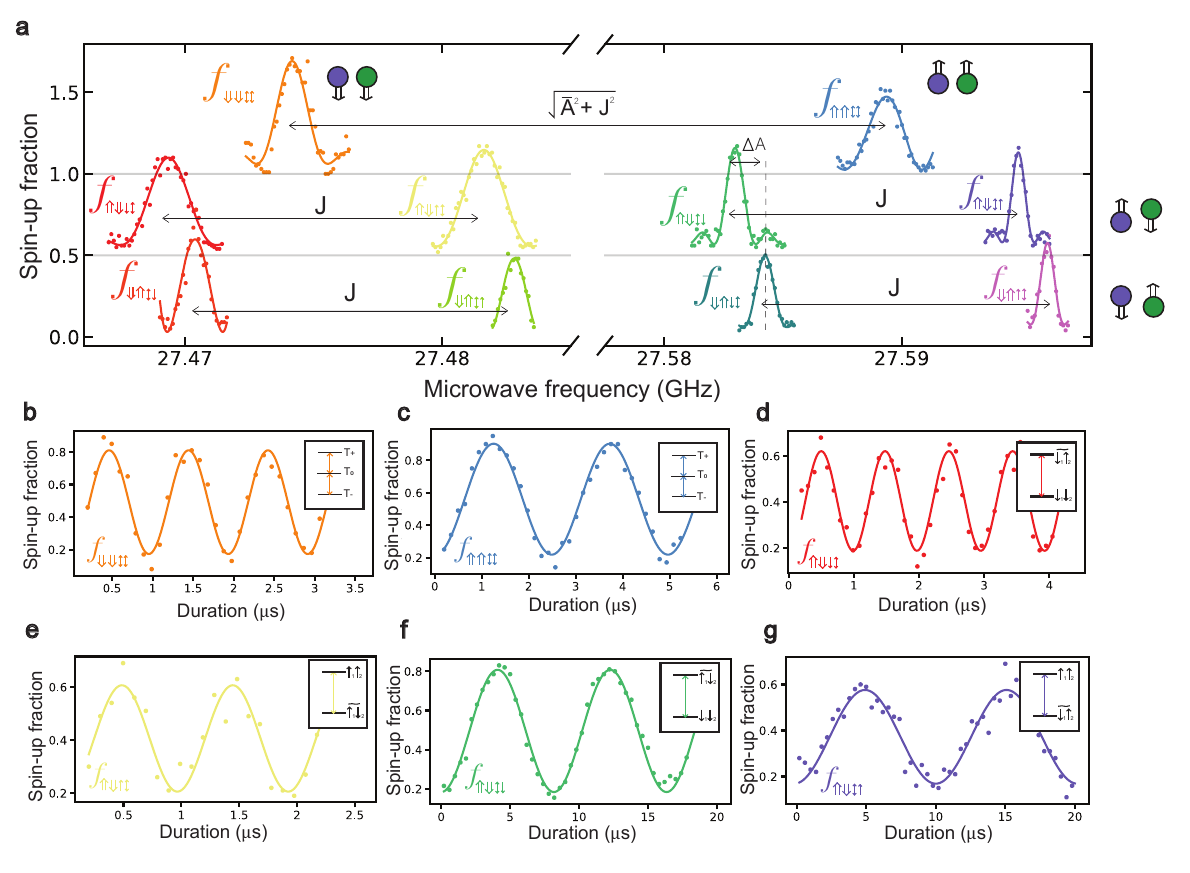}
    \caption{\textbf{ESR spectrum for different nuclear configurations. a,} ESR peaks for both electron 1 and electron 2, observed for different nuclear configurations (nuclear configurations are shown to the right of panel \textbf{a} for the case of the anti-parallel nuclear spin orientations and next to the resonance peaks for the case of the parallel nuclear spin orientations). The resonance peaks active for each nuclear configurations are offset from one another for clarity, as indicated by the grey gridlines. \textbf{b-g} Rabi oscillations performed on the ESR transitions shown in panel \textbf{a}. The inset to each figure shows the electron transitions being addressed.}
    \label{fig:whole_spectrum}
\end{figure*}
\ \\

Supplementary Fig.~\ref{fig:whole_spectrum} shows the measured electron spin resonance (ESR) transitions present for each of the four possible spin orientations of the two donor nuclei: $\ket{\Downarrow \Downarrow}$, $\ket{\Downarrow \Uparrow}$, $\ket{\Uparrow\Downarrow}$ and $\ket{\Uparrow \Uparrow}$ for the case of $\Delta A = 2$ MHz, $J = $ 12 MHz and $\bar{A} = 112$ MHz. \\

Supplementary Fig.~\ref{fig:whole_spectrum} b-g shows Rabi oscillations performed at the ESR resonance frequencies shown in Supplementary Fig.~\ref{fig:whole_spectrum}, a. The difference in Rabi frequencies observed between the different electron transitions is primarily attributed to the difference in transmission through the microwave line across the $\approx$ 100 MHz range of frequencies observed in the ESR spectrum, which results in less signal power reaching the device antenna depending on the frequency of the ESR drive applied.

\section{Initialisation of electron Q2}\label{suppB}

Supplementary Table~\ref{tab:control_electron_initialisation} outlines the initialisation procedure for electron Q2. Due to the lack of individual addressability of the electrons when the two nuclei are in a parallel spin orientation, this initialisation scheme is contingent upon the two donor nuclei being initialised in an anti-parallel spin orientation beforehand. Additionally, the scheme also relies on the electron Q1 being initialised in the $\ket{\downarrow}$ state at the beginning of the procedure. Therefore, any initialisation error on Q1 also affects the initialisation fidelity of Q2. In order to ensure we are less sensitive to small fluctuations in the resonance frequencies of the electron, adiabatic pulses are utilised, whereby the centre frequency of the pulse is swept over some frequency range \cite{laucht2014high}. Provided a large enough frequency range and that the pulse frequency is swept at a rate $\frac{\partial }{\partial t}\Delta f \ll f_{R}^{2}$, where $f_{R}^{2}$ is the Rabi frequency of the electron, the electron will be inverted adiabatically. In each step of the procedure, an adiabatic ESR pulse is performed, with a frequency span, $\Delta f > \Delta A$, chosen such that two ESR transitions can be addressed simultaneously: one to flip Q1 conditional on a given anti-parallel nuclear spin orientation, and the other to flip Q2 conditional on the opposite anti-parallel spin orientation. Q1 is re-initialised to the $\ket{\downarrow}$ state at the end of the procedure via spin-dependent tunnelling to the SET.

\begin{table}[H]
\centering
\begin{tabular}{ |>{\centering\arraybackslash}m{2cm}|>{\centering\arraybackslash}m{1.5cm}|>{\centering\arraybackslash}m{1.5cm}|>{\centering\arraybackslash}m{3cm}|>{\centering\arraybackslash}m{3cm}|>{\centering\arraybackslash}m{3cm}|>{\centering\arraybackslash}m{3cm}|} 
\hline
Nuclear state & Electron & Initial electron state & Step 1: \newline  Apply $f_{\Uparrow \Downarrow \updownarrow \uparrow}$ $\&$ \newline $f_{\Downarrow \Uparrow \uparrow \updownarrow}$ & Step 2: \newline Apply $f_{\Downarrow \Uparrow \updownarrow \uparrow}$ $\&$ \newline $f_{\Uparrow \Downarrow \uparrow \updownarrow}$ & Step 3: \newline  Apply $f_{\Uparrow \Downarrow \updownarrow \uparrow}$ $\&$ \newline $f_{\Downarrow \Uparrow \uparrow \updownarrow}$ & Step 4: \newline  Init. Q1 via readout\\[5pt]
\hline
\multirow{4}{4em}{$\Downarrow \Uparrow$} & \cellcolor{slateblue} Q1 & \cellcolor{slateblue} $\downarrow$ & \cellcolor{slateblue} $\downarrow$ & \cellcolor{slateblue} $\downarrow$ & \cellcolor{slateblue} $\downarrow$ & \cellcolor{slateblue} $\downarrow$ \\[5pt] 
& \cellcolor{seagreen}  Q2 & \cellcolor{seagreen} $\downarrow$ & \cellcolor{seagreen} $\downarrow$ & \cellcolor{seagreen} $\downarrow$ & \cellcolor{seagreen} $\downarrow$ & \cellcolor{seagreen} $\downarrow$ \\[5pt] \cline{2-7}
& \cellcolor{slateblue} Q1 & \cellcolor{slateblue} $\downarrow$ & \cellcolor{slateblue} $\downarrow$ & \cellcolor{slateblue} $\uparrow$ & \cellcolor{slateblue} $\uparrow$ & \cellcolor{slateblue} $\downarrow$ \\[5pt] 
& \cellcolor{seagreen} Q2 & \cellcolor{seagreen} $\uparrow$ & \cellcolor{seagreen} $\uparrow$ & \cellcolor{seagreen} $\uparrow$ & \cellcolor{seagreen} $\downarrow$ & \cellcolor{seagreen} $\downarrow$\\[5pt]
\hline
\multirow{4}{4em}{$\Uparrow \Downarrow$} & \cellcolor{slateblue} Q1 & \cellcolor{slateblue} $\downarrow$ & \cellcolor{slateblue} $\downarrow$ & \cellcolor{slateblue} $\downarrow$ & \cellcolor{slateblue} $\downarrow$ & \cellcolor{slateblue} $\downarrow$ \\[5pt] 
& \cellcolor{seagreen} Q2 & \cellcolor{seagreen}  $\downarrow$ & \cellcolor{seagreen} $\downarrow$ & \cellcolor{seagreen} $\downarrow$ & \cellcolor{seagreen} $\downarrow$ & \cellcolor{seagreen} $\downarrow$ \\[5pt] \cline{2-7}
& \cellcolor{slateblue} Q1 & \cellcolor{slateblue} $\downarrow$ & \cellcolor{slateblue} $\uparrow$ & \cellcolor{slateblue} $\uparrow$ & \cellcolor{slateblue} $\uparrow$ & \cellcolor{slateblue} $\downarrow$ \\[5pt] 
& \cellcolor{seagreen} Q2 & \cellcolor{seagreen}  $\uparrow$ & \cellcolor{seagreen} $\uparrow$ & \cellcolor{seagreen} $\downarrow$ & \cellcolor{seagreen} $\downarrow$ & \cellcolor{seagreen} $\downarrow$\\[5pt]
\hline
\end{tabular}
    \caption{Q2 electron initialisation procedure for different initial states of the nuclei and Q2 electron.}
    \label{tab:control_electron_initialisation}
\end{table}

\section{Initialisation of the nuclei}\label{suppC}

The two nuclei are initialised using an electron-nuclear double resonance (ENDOR)- style sequence. This sequence is outlined in Supplementary Tables \ref{tab:nuclear_initialisation_down} and \ref{tab:nuclear_initialisation_up}, for the case of the nucleus being initialised in either the spin $\ket{\Downarrow}$ state (Supplementary Table \ref{tab:nuclear_initialisation_down}) or the spin $\ket{\Uparrow}$ state (Supplementary Table \ref{tab:nuclear_initialisation_up}). The scheme relies on the electron bound to each nucleus being first initialised in the spin $\ket{\downarrow}$ state, resulting in any initialisation error associated with the electron initialisation, also affecting the nuclear initialisation fidelity. \\

The sequence to initialise the donor nucleus first involves flipping the electron conditional on the desired state of the nucleus (i.e. $\ket{\Downarrow}$ for the case of initialising the nucleus in the $\ket{\Downarrow}$ state and $\ket{\Uparrow}$ or the case of initialising the nucleus in the $\Uparrow$ state). In the J-coupled system, this is achieved by applying multiple adiabatic ESR pulses, in order to address all resonances conditional on the given state of the nucleus. A nuclear magnetic resonance (NMR) pulse then flips the nucleus, conditional on its bound electron being in the $\ket{\downarrow}$ state. The entire initialisation procedure is first carried out for donor nucleus 1 and then for donor nucleus 2, thus initialising both nuclei sequentially.

\begin{table}[H]
\renewcommand{\arraystretch}{2}
    % \caption{Nuclear initialisation protocols}
    \label{tab:nuclear_initialisation}
    \begin{minipage}{.5\linewidth}
      \caption{Initialising nuclear state $\Downarrow$}
      \label{tab:nuclear_initialisation_down}
      \centering
    \begin{tabular}{ |>{\centering\arraybackslash}m{1.7cm}|>{\centering\arraybackslash}m{1.5cm}|>{\centering\arraybackslash}m{1.5cm}|>{\centering\arraybackslash}m{1.5cm}|>{\centering\arraybackslash}m{1.5cm}|} 
    \hline
    \cellcolor{lightblue} Initial nuclear state & \cellcolor{lightblue} Spin state & \cellcolor{lightblue} Apply $f_{\Downarrow \updownarrow }$ & \cellcolor{lightblue} Apply $f_{\Updownarrow \downarrow}$  & \cellcolor{lightblue} Re-init. electron\\[10pt]
    \hline
    \cellcolor{lightcyan} \multirow{2}{4em}{$\Downarrow$} & Nucleus & $\Downarrow$ & $\Downarrow$ & $\Downarrow$ \\[10pt] \cline{2-5}
  \cellcolor{lightcyan}   & Electron & $\uparrow$ &  $\uparrow$ & $\downarrow$ \\[10pt] \cline{1-5}
    
    \cellcolor{lightcyan} \multirow{2}{4em}{$\Uparrow$} & Nucleus & $\Uparrow$ & $\Downarrow$ & $\Downarrow$ \\[10pt] \cline{2-5}
    \cellcolor{lightcyan} & Electron & $\downarrow$ &  $\downarrow$ & $\downarrow$ \\[10pt] \cline{1-5}
    \hline
    \end{tabular}
    \end{minipage}%
    \begin{minipage}{.5\linewidth}
      \centering
        \caption{Initialising nuclear state $\Uparrow$}
        \label{tab:nuclear_initialisation_up}
    \begin{tabular}{ |>{\centering\arraybackslash}m{1.7cm}|>{\centering\arraybackslash}m{1.5cm}|>{\centering\arraybackslash}m{1.5cm}|>{\centering\arraybackslash}m{1.5cm}|>{\centering\arraybackslash}m{1.5cm}|} 
    \hline
     \cellcolor{lightblue} Initial nuclear state &  \cellcolor{lightblue} Spin state &  \cellcolor{lightblue} Apply $f_{\Uparrow \updownarrow }$ &  \cellcolor{lightblue} Apply $f_{ 	\Updownarrow \downarrow}$  &  \cellcolor{lightblue} Re-init. electron\\[10pt]
    \hline
    \cellcolor{lightcyan} \multirow{2}{4em}{$\Downarrow$} & Nucleus & $\Downarrow$ & $\Uparrow$ & $\Uparrow$ \\[10pt] \cline{2-5}
  \cellcolor{lightcyan}   & Electron & $\downarrow$ &  $\downarrow$ & $\downarrow$ \\[10pt] \cline{1-5}
    
   \cellcolor{lightcyan}  \multirow{2}{4em}{$\Uparrow$} & Nucleus & $\Uparrow$ & $\Uparrow$ & $\Uparrow$ \\[10pt] \cline{2-5}
   \cellcolor{lightcyan}  & Electron & $\uparrow$ &  $\uparrow$ & $\downarrow$ \\[10pt] \cline{1-5}
    \hline
    \end{tabular}
    \end{minipage} 
\end{table}

\section{Quantum non-demolition (QND) readout of the Q2 electron}\label{suppD}

\begin{figure*}[h]
    \centering
    \includegraphics[width=1\textwidth]{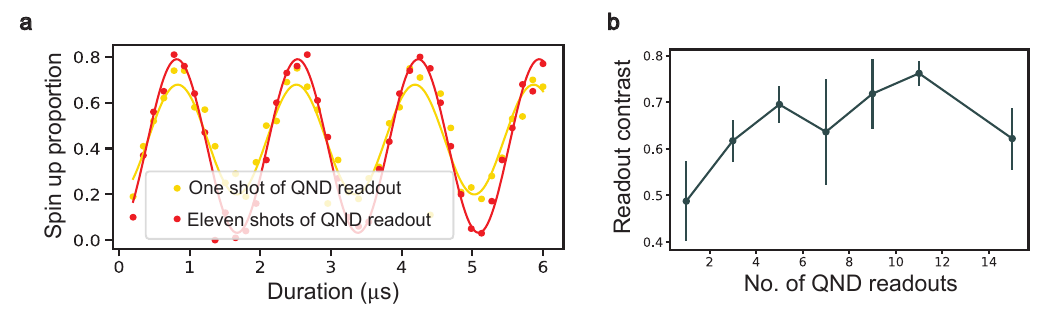}
    \caption{\textbf{Quantum non-demolition (QND) readout of the electron.} \textbf{a,} Rabi oscillation performed on electron 2 with a single repetition of QND readout, compared to 11 QND readouts, showing an increase in readout contrast. \textbf{b,} A plot of electron 2 readout contrast against number of QND readouts.}
    \label{fig:QND_readout}
\end{figure*}
Electron Q2 is read out indirectly via Q1 using the process described in the Results section of the main text. Supplementary Fig.~\ref{fig:QND_readout}a shows a Rabi oscillation performed on Q2 with either a single repetition of QND readout, or with 11 repetitions of QND readout, resulting in an increase in readout contrast from 0.48 to 0.76. Supplementary Fig.~\ref{fig:QND_readout}, b shows the readout contrast of electron Q2 as a function of number of QND readout shots, showing a maximum readout contrast at approximately 11 QND readout shots, after which the readout contrast is seen to decrease. The reason for this decrease is attributed to imperfections in the QND readout, whereby the hybridisation of electron states results in a finite probability of flipping the electron Q2 during the loading or unloading of electron Q1 onto the donor \cite{joecker2024error}.\\

In order for a measurement to be truly QND in nature, the Hamiltonian of the qubit $\hat{H}_{Q}$ must commute with the coupling term, $\hat{H}_{C}$, between the qubit and the ancilla used for readout

\begin{equation}
    [\hat{H}_{Q}, \hat{H}_{C}] = 0.
\end{equation}

For the case of $J \ll A$ the exchange interaction is only approximately of Ising type and thus for this system $[\hat{H}_{Q}, \hat{H}_{C}] \neq 0$, resulting in the readout of the electron Q2 being not fully QND in nature. As a consequence of this, the exchange interaction acts to weakly entangle the two electrons, resulting in the eigenstates of the system becoming hybridised. As discussed in Section II, the hybridised eigenstates of the two electrons in the exchange-coupled system are given by

\begin{align}
    \widetilde{\ket{\downarrow \uparrow}} = \cos(\theta)\ket{\downarrow \uparrow} + \sin(\theta)\ket{\uparrow \downarrow}\\
    \widetilde{\ket{\uparrow \downarrow}} = \cos(\theta)\ket{\uparrow \downarrow} - \sin(\theta)\ket{\downarrow \uparrow}
\end{align}

where $\tan(2\theta) = \frac{J}{\Delta}$ and $\Delta = \bar{A}$ for the case of the anti-parallel nuclei. This hybridisation of states leads to a small probability of flipping the electron Q2 during the QND readout process. This error mechanism associated with imperfections in the QND readout is the dominant source of erroneous flipping of the qubit being read out, compared to the natural relaxation time of the electron spin. 

\section{Tuning the exchange interaction}\label{suppE}
\begin{figure*}[!h]
    \centering
    \includegraphics[width=0.7\textwidth]{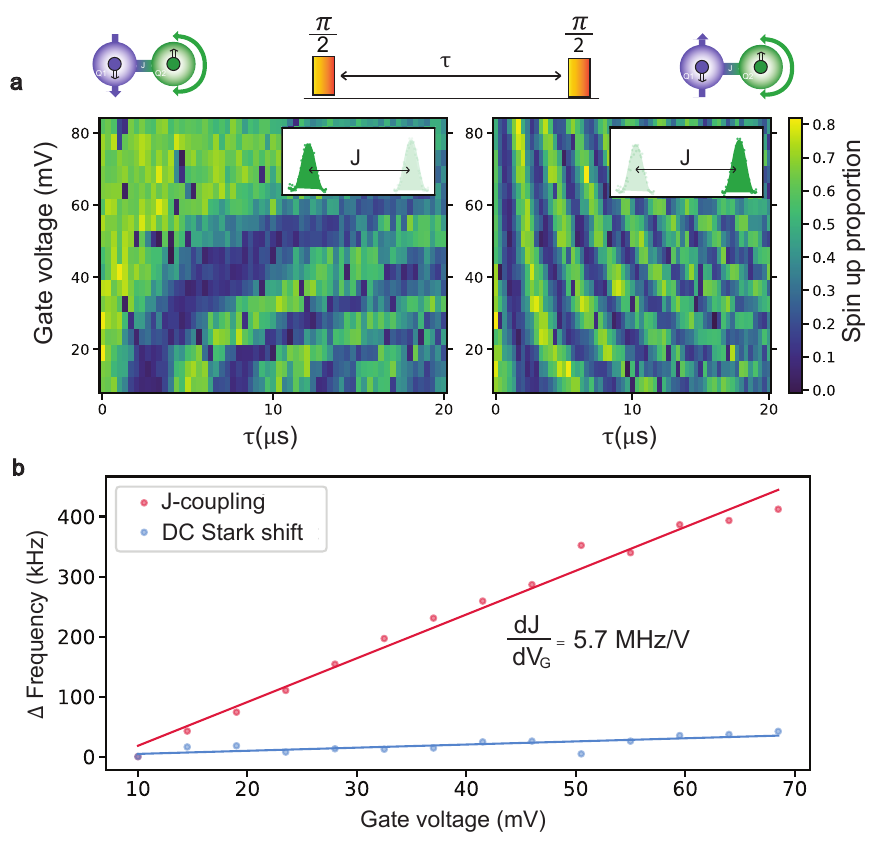}
    \caption{\textbf{Tuning the exchange interaction.} \textbf{a.} Ramsey measurements performed on the electron Q2, with electron Q1 either in the $\ket{\downarrow}$ state (left plot) or $\ket{\uparrow}$ state (right plot), as a function of gate voltage, to obtain the voltage-dependent resonance frequencies of the two qubits. \textbf{b.} DC Stark shift (combined effect of hyperfine and $g$-factor shifts) as a function of gate voltage, extracted with $\frac{\bar{f}}{\Delta V_{\rm G}}$ where $\Delta V_{\rm G}$ is the change in voltage and $\bar{f} = \frac{f_{\Uparrow \Downarrow \downarrow \updownarrow} + f_{\Uparrow \Downarrow \uparrow \updownarrow}}{2}$. $f_{\Uparrow \Downarrow \downarrow \updownarrow}$ is the frequency of the Ramsey oscillation conditional on Q1 in the  $\ket{\downarrow}$ state and $f_{\Uparrow \Downarrow \uparrow \updownarrow}$ is the frequency of the Ramsey oscillation conditional on Q1 in the $\ket{\uparrow}$ state. The change in exchange interaction as a function of gate voltage was calculated with $\frac{\Delta f}{\Delta V}$ where $\Delta f = f_{\Uparrow \Downarrow \uparrow \updownarrow} - f_{\Uparrow \Downarrow \downarrow \updownarrow}$.}
    \label{fig:J_tuning}
\end{figure*}

The exact value of the exchange interaction between two exchange-coupled electrons depends upon the overlap of their wavefunctions, which in turn can be influenced by the voltages applied to the gates fabricated on the surface of the device. In order to test the tunability of the exchange interaction with the applied gate voltages, we initialised the nuclei in an anti-parallel spin configuration of $\ket{\Uparrow \Downarrow}$, before performing two Ramsey measurements, one on Q2 conditional on Q1 being in the $\ket{\downarrow}$ state ($f_{\Uparrow \Downarrow \downarrow \updownarrow}$), and one on Q2 conditional on Q1 being in the $\ket{\uparrow}$ state ($f_{\Uparrow \Downarrow \uparrow \updownarrow}$). These Ramsey measurements were performed while sweeping the voltage applied to one of the device gates. Any frequency shift as a function of gate voltage, that shifts both $f_{\Uparrow \Downarrow \downarrow \updownarrow}$ and $f_{\Uparrow \Downarrow \uparrow \updownarrow}$ together can be explained by the combined DC Stark shift of the hyperfine interaction and the g-factor \cite{laucht2015electrically}. However, the frequency difference $|f_{\Uparrow \Downarrow \downarrow \updownarrow} - f_{\Uparrow \Downarrow \uparrow \updownarrow}|$ depends only on the exchange interaction between Q1 and Q2. Therefore, any change in $|f_{\Uparrow \Downarrow \downarrow \updownarrow} - f_{\Uparrow \Downarrow \uparrow \updownarrow}|$ represents a change in $J$. \\

For a change in voltage of $\approx$ 70 mV we observed a DC Stark shift of $\approx 40$~kHz, and a shift in exchange coupling of $\approx 400$~kHz. This translates to a tunability of the exchange coupling of $\frac{dJ}{dV_{\rm G}} = 5.7$ MHz/V. Compared to the tuneability of exchange coupling that has been reported in quantum dots, where the exchange interaction has been shown to change by over 7 orders of magnitude per volt of applied gate voltage \cite{cifuentes2024bounds}, the low tunability observed in this system indicates that we may be close to a first-order insensitive regime of $J(V_{\rm G})$. This is also corroborated by the low noise in $J$-coupling measured by comparing the coherence times of the electrons with and without the exchange coupling present, as shown in Fig.2.

\section{CROT unitary matrix}\label{suppF}
For a rotation around the X-axis of the Bloch sphere the unitary operations that define the CROT and zCROT gates for electrons 1 and 2 are therefore the following:\\

\begin{align}
\text{CROT}_{\updownarrow_{1}\downarrow_{2}} &= 
    \begin{pmatrix}
        0 & 0 & -i & 0 \\
        0 & 1 & 0 & 0 \\
        -i & 0 & 0 & 0 \\
        0 & 0 & 0 & 1 
    \end{pmatrix}, 
\text{CROT}_{\downarrow_{1}\updownarrow_{2}} = 
    \begin{pmatrix}
        0 & -i & 0 & 0 \\
        -i & 0 & 0 & 0 \\
        0 & 0 & 1 & 0 \\
        0 & 0 & 0 & 1 
    \end{pmatrix}\\
\text{zCROT}_{\updownarrow_{1}\uparrow_{2}} &= 
    \begin{pmatrix}
        1 & 0 & 0 & 0 \\
        0 & 0 & 0 & -i \\
        0 & 0 & 1 & 0 \\
        0 & -i & 0 & 0 
    \end{pmatrix}, 
\text{zCROT}_{\uparrow_{1}\updownarrow_{2}} = 
    \begin{pmatrix}
        1 & 0 & 0 & 0 \\
        0 & 1 & 0 & 0 \\
        0 & 0 & 0 & -i \\
        0 & 0 & -i & 0 
    \end{pmatrix},
\end{align}

where $\updownarrow$ represents the electron being driven during the gate implementation. The CROT gate is equivalent to a CNOT gate, but with the addition of a phase imparted on the control qubit. This phase comes about as a result of the geometric phase imparted by the target electron on the control and is equal to half the solid angle traversed on the Bloch sphere by the target electron. \\

\section{Experimental GST details} \label{supp:experimental_GST}

The time taken to perform a one-qubit GST measurement was approximately 45 minutes, when using 400 shots per circuit and a maximum circuit depth of L=64. The analysis time for one qubit GST takes approximately 15 seconds to generate a full GST report from the data. For two-qubit GST, the measurement took approximately 2 hours to perform when using 400 shots per circuit and a maximum circuit depth of L=4. The analysis time for two-qubit GST took approximately 5 hours to generate a full GST report from the data.  \\

The GST sequences for both one and two-qubit GST were performed by first uploading all of the circuit pulse sequences to the experiment instruments at once, at the start of the GST experiment. This is important, as the upload time of the pulse sequences to the instruments was the main bottleneck for GST measurement time previously. Thus, by uploading all of the sequences to the instruments at the start of the experiment, we remove the upload time of the pulses during the experiment, at the expense of spending approximately 5 minutes at the start of the experiment uploading all the pulses. We then measured each of the circuits in the entire GST circuit list with a single shot, before repeating the entire circuit list 400 times.

\subsection{Recalibration sequences}

We interleaved the GST experiment with re-calibrations, to ensure the device remained tuned throughout the experiment. Checks were integrated after each shot of the GST circuit list, to determine if any calibration was required before progressing to the next shot. More specifically, we performed the following calibrations: 

\begin{itemize}
    \item \textbf{Readout position tuning.} During a long measurement the device will undergo both sudden jumps and slow drifts in voltage that can alter the readout position of the donor. A small drift of only around 2 mV is enough to completely diminish the electron readout visibility. It is therefore important to monitor and retune the readout position during a long measurement, such as GST. We performed two adiabatic pulses after every iteration of measuring the GST circuits, one pulse which should be on resonance with the electron and thus result in a high spin up proportion, $P_{\text{high}}$, and one which should be off resonance with the electron and hence give a low spin up proportion, $P_{\text{low}}$. The readout contrast, $C$, is then given by $C =  |P_{\text{high}} - P_{\text{low}}|$. If this readout contrast fell below a pre-determined threshold, then the measurement would automatically perform an automated readout position tuning, using a Nelder Mead optimisation algorithm. This algorithm involved tuning a pair of the device gate voltages and measuring the resulting electron readout contrast, in order to determine the optimal set of gate voltages to maximise the electron readout contrast. This automated retuning algorithm took approximately 30 seconds to run.  
    \item \textbf{Ensuring we remained in the correct nuclear state.} The GST experiments were performed on the electrons, conditional on the two donor nuclei being in the state $\ket{\Uparrow_1 \Downarrow_2}$. We therefore checked the state of the nuclei, by rotating electron 1 conditionally on the state of the nuclei, after each repetition of the GST circuits, and re-initialised the nuclei into the correct state if necessary before the next shot. 
    \item \textbf{Re-calibration of the electron resonance frequency.} In order to ensure the pulses remained on resonance, we performed a low power $\pi$ pulse at the resonance frequencies of interest after each GST circuit shot. If the up proportion of this $\pi$ pulse fell below a pre-determined threshold, then we performed an automatic retuning of the relevant resonance frequencies. This was done by performing a low power frequency spectrum about the previously determined resonance frequency, automatically fitting the resulting resonance peak, extracting the centre frequency of this peak and setting the pulse frequency used for the GST circuits to the updated resonance frequency. The low power ensures that we are as sensitive as possible to any frequency drifts.

\end{itemize}

\subsection{Model violation}

For the single-qubit conditional GST presented in Fig. 4, a maximum circuit depth of L=1024 was used. Across the four conditional gates tested, the average model violation estimated by GST was $N_{\sigma}$  = 16.77, with a maximum model violation of 35.00 and a minimum model violation of 5.77. For the single-qubit unconditional GST, presented in Fig. 3. a maximum circuit depth of L=64 was used. A lower maximum circuit depth was used compared to the conditional gates, due to the fact that in this system, the unconditional gates are each composed of two conditional gates. This makes it more challenging to iteratively improve these gates using the GST results, due to the fact that the GST errors are not directly translatable to tuneable parameters in this composite pulse. Across the six unconditional GST experiments presented, the average model violation was $N_{\sigma}$ = 4.32, with a maximum model violation of 13.72 and a minimum of 1.52. We minimised the model violation by minimising the GST measurement time as much as possible and integrating the calibration schemes, such that the drift in the device parameters were minimised (see above). For two-qubit GST, a model violation of $N_{\sigma}$= 85.63 was estimated, indicating a higher level of non-Markovian noise present. For a definition of model violation in GST see \cite{nielsen2021gate}

\section{One-Qubit GST: Extended Results} \label{suppG}

In this section we present detailed results from the analysis of single-qubit GST using both native conditional two-qubit operations and composite unconditional operations, as discussed in the Results section of the main text. In Supplementary Tables \ref{tab:one_qubit_conditional_complete_generator_infidelity_table} and \ref{tab:one_qubit_unconditional_complete_generator_infidelity_table} we present a complete set of the estimated generator infidelities for the conditional and unconditional operations, respectively, as well as the contributions to this infidelity due to coherent and incoherent noise sources. The generator infidelity, $\hat{\epsilon}$, was introduced in \cite{mkadzik2022precision} and is defined as

\begin{equation}
   \hat{\epsilon} = \sum_i \epsilon_i + \sum_i \theta_i^2 
\end{equation}
\label{eqn:generator_infdl}

\noindent where $\{\epsilon_i\}$ and $\{\theta_i\}$ are the stochastic and Hamiltonian error generator rates, respectively \cite{PRXQuantum.3.020335}. To get the separate coherent and incoherent contributions to  $\hat{\epsilon}$ we simply add the sums over $\{\epsilon_i\}$ and $\{\theta_i\}$ separately. \\

\subsection{1Q Conditional operations}\label{supp:conditional}

In Supplementary Table \ref{tab:one_qubit_error_generator_rates_GST_conditional} we present the full set of estimated Hamiltonian and stochastic error generator rates for the native conditional operations. In Supplementary Tables \ref{tab:one_qubit_error_generator_rates_GST_unconditional_part_1} and \ref{tab:one_qubit_error_generator_rates_GST_unconditional_part_2} we present these rates for the composite unconditional operations.

\begin{table}[htbp]
	\centering
	\begin{tabular}{ccccc}
		\hline \hline
		\multicolumn{1}{l}{(Target, Cond. State)} & Gate  & Coherent & Incoherent & Total Generator Infidelity \bigstrut[b]\\
		\hline
		\multirow{3}[2]{*}{(0, $\ket{\downarrow}$)} & $X_{\frac{\pi}{2}}$ & $0.0080 \pm 0.0005\%$ & $0.126 \pm 0.042\%$ & $0.134 \pm 0.042\%$ \bigstrut[t]\\
		& $Y_{\frac{\pi}{2}}$ & $0.0033 \pm 0.0009\%$ & $.0.226 \pm 0.039\%$ & $0.229 \pm 0.039\%$ \\
		& Idle  & $0.068 \pm .007\%$ & $0.753 \pm 0.193 \%$ & $0.821 \pm 0.193\%$  \bigstrut[b]\\
		\hline
		\multirow{3}[2]{*}{(0, $\ket{\uparrow}$)} & $X_{\frac{\pi}{2}}$ & $0.0130 \pm 0.0056\%$ & $0.361 \pm 0.098\%$ & $0.374 \pm 0.094\%$ \bigstrut[t]\\
		& $Y_{\frac{\pi}{2}}$ & $0.0085 \pm 0.0026\%$ & $0.388 \pm 0.141\%$ & $0.397 \pm 0.140\%$ \\
		& Idle  & $0.023 \pm 0.002\%$ & $1.72 \pm 0.36\%$ & $1.74 \pm 0.36\%$ \bigstrut[b]\\
		\hline
		\multirow{3}[2]{*}{(1, $\ket{\downarrow}$)} & $X_{\frac{\pi}{2}}$ & $0.0003 \pm 0.0005\%$ & $0.30 \pm 0.16\%$ & $0.30 \pm 0.16\%$ \bigstrut[t]\\
		& $Y_{\frac{\pi}{2}}$ & $0.0013 \pm 0.0013\%$ & $0.28 \pm 0.18\%$ & $0.29 \pm 0.18\%$ \\
		& Idle  & $.008 \pm .011\%$ & $1.25 \pm 0.26\%$ & $1.26 \pm 0.27\%$ \bigstrut[b]\\
		\hline
		\multirow{3}[2]{*}{(1, $\ket{\uparrow}$)} & $X_{\frac{\pi}{2}}$ & $0.0025 \pm 0.0007\%$ & $0.22 \pm 0.12\%$ & $0.22 \pm 0.12\%$ \bigstrut[t]\\
		& $Y_{\frac{\pi}{2}}$ & $0.0025 \pm 0.0007\%$ & $0.23 \pm 0.12\%$ & $0.24 \pm 0.12\%$ \\
		& Idle  & $0.0041 \pm 0.0029\%$ & $0.81 \pm 0.31\%$ & $0.81 \pm 0.31\%$ \bigstrut[b]\\
		\hline \hline
	\end{tabular}%
	\caption{Estimated one-qubit generator infidelities for a gate set consisting of single-qubit $X_\frac{\pi}{2}$ and $Y_\frac{\pi}{2}$ rotations together with an idle operation. Here one-qubit GST is performed using the native conditional operations by first initializing the control qubit in either the $\ket{\downarrow}$ or $\ket{\uparrow}$ state and utilizing the corresponding conditional rotation gate, thereafter considering only the action on the target qubit. We report here the resulting infidelities for each of the four target qubit-conditional state pairs. In addition to the total generator infidelity for each gate we also report the separate contributions to the generator infidelity due to coherent and incoherent noise sources.}
	\label{tab:one_qubit_conditional_complete_generator_infidelity_table}%
\end{table}%

% \begin{table}[htbp]
\begin{table}[H]
	\centering
	{\setstretch{.9}
	{\renewcommand{\arraystretch}{.85}
	\begin{tabular}{ccrrrrrrr}
		\hline
		\hline\\[.05em]
		\multicolumn{1}{p{2.25cm}}{(Target, Conditional State)} & \hspace{.5em} Gate  & \multicolumn{3}{c}{Hamiltonian} &       & \multicolumn{3}{c}{Stochastic} \bigstrut[b]\\
		\hline
		\multirow{10}[20]{*}{(0, $\ket{\downarrow}$)} &       &       &       &       &       &       &       &  \bigstrut\\
		\cline{3-5}\cline{7-9}          & \multicolumn{1}{c|}{\multirow{2}[4]{*}{$X_{\frac{\pi}{2}}$}} & \multicolumn{1}{p{1.5cm}|}{$X$} & \multicolumn{1}{p{1.5cm}|}{$Y$} & \multicolumn{1}{p{1.5cm}|}{$Z$} & \multicolumn{1}{r|}{} & \multicolumn{1}{p{1.5cm}|}{$X$} & \multicolumn{1}{p{1.5cm}|}{$Y$} & \multicolumn{1}{p{1.5cm}|}{$Z$} \bigstrut\\
		\cline{3-5}\cline{7-9}          & \multicolumn{1}{c|}{} & \multicolumn{1}{p{1.5cm}|}{$0.0089\pm0.0003$} & \multicolumn{1}{p{1.5cm}|}{$0.0008\pm0.0009$} & \multicolumn{1}{p{1.5cm}|}{$0.0008\pm0.0010$} & \multicolumn{1}{r|}{} & \multicolumn{1}{p{1.5cm}|}{$0.00101\pm0.00037$} & \multicolumn{1}{p{1.5cm}|}{$0.00013\pm0.00028$} & \multicolumn{1}{p{1.5cm}|}{$0.00013\pm0.00029$} \bigstrut\\
		\cline{3-5}\cline{7-9}          &       &       &       &       &       &       &       &  \bigstrut\\
		\cline{3-5}\cline{7-9}          & \multicolumn{1}{c|}{\multirow{2}[4]{*}{$Y_{\frac{\pi}{2}}$}} & \multicolumn{1}{p{1.5cm}|}{$X$} & \multicolumn{1}{p{1.5cm}|}{$Y$} & \multicolumn{1}{p{1.5cm}|}{$Z$} & \multicolumn{1}{r|}{} & \multicolumn{1}{p{1.5cm}|}{$X$} & \multicolumn{1}{p{1.5cm}|}{$Y$} & \multicolumn{1}{p{1.5cm}|}{$Z$} \bigstrut\\
		\cline{3-5}\cline{7-9}          & \multicolumn{1}{c|}{} & \multicolumn{1}{p{1.5cm}|}{$0.0008\pm0.0013$} & \multicolumn{1}{p{1.5cm}|}{$0.0056\pm0.0008$} & \multicolumn{1}{p{1.5cm}|}{$-0.0008\pm0.0012$} & \multicolumn{1}{r|}{} & \multicolumn{1}{p{1.5cm}|}{$0.00024\pm0.00030$} & \multicolumn{1}{p{1.5cm}|}{$0.00184\pm0.00046$} & \multicolumn{1}{p{1.5cm}|}{$0.00018\pm0.00031$} \bigstrut\\
		\cline{3-5}\cline{7-9}          &       &       &       &       &       &       &       &  \bigstrut\\
		\cline{3-5}\cline{7-9}          & \multicolumn{1}{c|}{\multirow{2}[4]{*}{Idle}} & \multicolumn{1}{p{1.5cm}|}{$X$} & \multicolumn{1}{p{1.5cm}|}{$Y$} & \multicolumn{1}{p{1.5cm}|}{$Z$} & \multicolumn{1}{r|}{} & \multicolumn{1}{p{1.5cm}|}{$X$} & \multicolumn{1}{p{1.5cm}|}{$Y$} & \multicolumn{1}{p{1.5cm}|}{$Z$} \bigstrut\\
		\cline{3-5}\cline{7-9}          & \multicolumn{1}{c|}{} & \multicolumn{1}{p{1.5cm}|}{$0.0005\pm0.0014$} & \multicolumn{1}{p{1.5cm}|}{$-0.0005\pm0.0015$} & \multicolumn{1}{p{1.5cm}|}{$0.0261\pm0.0014$} & \multicolumn{1}{r|}{} & \multicolumn{1}{p{1.5cm}|}{$0.00006\pm0.00260$} & \multicolumn{1}{p{1.5cm}|}{$0.00123\pm0.00322$} & \multicolumn{1}{p{1.5cm}|}{$0.00623\pm0.00112$} \bigstrut\\
		\cline{3-5}\cline{7-9}          &       &       &       &       &       &       &       &  \bigstrut\\
		\hline
		\multirow{10}[20]{*}{(0, $\ket{\uparrow}$)} &       &       &       &       &       &       &       &  \bigstrut\\
		\cline{3-5}\cline{7-9}          & \multicolumn{1}{c|}{\multirow{2}[4]{*}{$X_{\frac{\pi}{2}}$}} & \multicolumn{1}{p{1.5cm}|}{$X$} & \multicolumn{1}{p{1.5cm}|}{$Y$} & \multicolumn{1}{p{1.5cm}|}{$Z$} & \multicolumn{1}{r|}{} & \multicolumn{1}{p{1.5cm}|}{$X$} & \multicolumn{1}{p{1.5cm}|}{$Y$} & \multicolumn{1}{p{1.5cm}|}{$Z$} \bigstrut\\
		\cline{3-5}\cline{7-9}          & \multicolumn{1}{c|}{} & \multicolumn{1}{p{1.5cm}|}{$0.0113\pm0.0027$} & \multicolumn{1}{p{1.5cm}|}{$0.0009\pm0.0018$} & \multicolumn{1}{p{1.5cm}|}{$0.0009\pm0.0018$} & \multicolumn{1}{r|}{} & \multicolumn{1}{p{1.5cm}|}{$0.0031\pm0.0006$} & \multicolumn{1}{p{1.5cm}|}{$0.0002\pm0.0002$} & \multicolumn{1}{p{1.5cm}|}{$0.0003\pm0.0003$} \bigstrut\\
		\cline{3-5}\cline{7-9}          &       &       &       &       &       &       &       &  \bigstrut\\
		\cline{3-5}\cline{7-9}          & \multicolumn{1}{c|}{\multirow{2}[4]{*}{$Y_{\frac{\pi}{2}}$}} & \multicolumn{1}{p{1.5cm}|}{$X$} & \multicolumn{1}{p{1.5cm}|}{$Y$} & \multicolumn{1}{p{1.5cm}|}{$Z$} & \multicolumn{1}{r|}{} & \multicolumn{1}{p{1.5cm}|}{$X$} & \multicolumn{1}{p{1.5cm}|}{$Y$} & \multicolumn{1}{p{1.5cm}|}{$Z$} \bigstrut\\
		\cline{3-5}\cline{7-9}          & \multicolumn{1}{c|}{} & \multicolumn{1}{p{1.5cm}|}{$0.0009\pm0.0021$} & \multicolumn{1}{p{1.5cm}|}{$0.0091\pm0.0014$} & \multicolumn{1}{p{1.5cm}|}{$-0.0009\pm0.0021$} & \multicolumn{1}{r|}{} & \multicolumn{1}{p{1.5cm}|}{$0.0005\pm0.0007$} & \multicolumn{1}{p{1.5cm}|}{$0.0029\pm0.0007$} & \multicolumn{1}{p{1.5cm}|}{$0.0004\pm0.0006$} \bigstrut\\
		\cline{3-5}\cline{7-9}          &       &       &       &       &       &       &       &  \bigstrut\\
		\cline{3-5}\cline{7-9}          & \multicolumn{1}{c|}{\multirow{2}[4]{*}{Idle}} & \multicolumn{1}{p{1.5cm}|}{$X$} & \multicolumn{1}{p{1.5cm}|}{$Y$} & \multicolumn{1}{p{1.5cm}|}{$Z$} & \multicolumn{1}{r|}{} & \multicolumn{1}{p{1.5cm}|}{$X$} & \multicolumn{1}{p{1.5cm}|}{$Y$} & \multicolumn{1}{p{1.5cm}|}{$Z$} \bigstrut\\
		\cline{3-5}\cline{7-9}          & \multicolumn{1}{c|}{} & \multicolumn{1}{p{1.5cm}|}{$0.0045\pm0.0028$} & \multicolumn{1}{p{1.5cm}|}{$0.0011\pm0.0025$} & \multicolumn{1}{p{1.5cm}|}{$0.0144\pm0.0051$} & \multicolumn{1}{r|}{} & \multicolumn{1}{p{1.5cm}|}{$0.0012\pm0.0043$} & \multicolumn{1}{p{1.5cm}|}{$0.0003\pm0.0043$} & \multicolumn{1}{p{1.5cm}|}{$0.0157\pm0.0039$} \bigstrut\\
		\cline{3-5}\cline{7-9}          &       &       &       &       &       &       &       &  \bigstrut\\
		\hline
		\multirow{10}[20]{*}{(1, $\ket{\downarrow}$)} &       &       &       &       &       &       &       &  \bigstrut\\
		\cline{3-5}\cline{7-9}          & \multicolumn{1}{c|}{\multirow{2}[4]{*}{$X_{\frac{\pi}{2}}$}} & \multicolumn{1}{p{1.5cm}|}{$X$} & \multicolumn{1}{p{1.5cm}|}{$Y$} & \multicolumn{1}{p{1.5cm}|}{$Z$} & \multicolumn{1}{r|}{} & \multicolumn{1}{p{1.5cm}|}{$X$} & \multicolumn{1}{p{1.5cm}|}{$Y$} & \multicolumn{1}{p{1.5cm}|}{$Z$} \bigstrut\\
		\cline{3-5}\cline{7-9}          & \multicolumn{1}{c|}{} & \multicolumn{1}{p{1.5cm}|}{$0.0018\pm0.0014$} & \multicolumn{1}{p{1.5cm}|}{$0.0002\pm0.0025$} & \multicolumn{1}{p{1.5cm}|}{$0.0002\pm0.0014$} & \multicolumn{1}{r|}{} & \multicolumn{1}{p{1.5cm}|}{$0.00241\pm0.00100$} & \multicolumn{1}{p{1.5cm}|}{$0.00004\pm0.00172$} & \multicolumn{1}{p{1.5cm}|}{$0.00055\pm0.00052$} \bigstrut\\
		\cline{3-5}\cline{7-9}          &       &       &       &       &       &       &       &  \bigstrut\\
		\cline{3-5}\cline{7-9}          & \multicolumn{1}{c|}{\multirow{2}[4]{*}{$Y_{\frac{\pi}{2}}$}} & \multicolumn{1}{p{1.5cm}|}{$X$} & \multicolumn{1}{p{1.5cm}|}{$Y$} & \multicolumn{1}{p{1.5cm}|}{$Z$} & \multicolumn{1}{r|}{} & \multicolumn{1}{p{1.5cm}|}{$X$} & \multicolumn{1}{p{1.5cm}|}{$Y$} & \multicolumn{1}{p{1.5cm}|}{$Z$} \bigstrut\\
		\cline{3-5}\cline{7-9}          & \multicolumn{1}{c|}{} & \multicolumn{1}{p{1.5cm}|}{$0.0002\pm0.0018$} & \multicolumn{1}{p{1.5cm}|}{$0.0036\pm0.0017$} & \multicolumn{1}{p{1.5cm}|}{$-0.0002\pm0.0022$} & \multicolumn{1}{r|}{} & \multicolumn{1}{p{1.5cm}|}{$0.00007\pm0.00074$} & \multicolumn{1}{p{1.5cm}|}{$0.00268\pm0.00117$} & \multicolumn{1}{p{1.5cm}|}{$0.00008\pm0.00068$} \bigstrut\\
		\cline{3-5}\cline{7-9}          &       &       &       &       &       &       &       &  \bigstrut\\
		\cline{3-5}\cline{7-9}          & \multicolumn{1}{c|}{\multirow{2}[4]{*}{Idle}} & \multicolumn{1}{p{1.5cm}|}{$X$} & \multicolumn{1}{p{1.5cm}|}{$Y$} & \multicolumn{1}{p{1.5cm}|}{$Z$} & \multicolumn{1}{r|}{} & \multicolumn{1}{p{1.5cm}|}{$X$} & \multicolumn{1}{p{1.5cm}|}{$Y$} & \multicolumn{1}{p{1.5cm}|}{$Z$} \bigstrut\\
		\cline{3-5}\cline{7-9}          & \multicolumn{1}{c|}{} & \multicolumn{1}{p{1.5cm}|}{$0.0018\pm0.0018$} & \multicolumn{1}{p{1.5cm}|}{$-0.0004\pm0.0033$} & \multicolumn{1}{p{1.5cm}|}{$0.0085\pm0.0064$} & \multicolumn{1}{r|}{} & \multicolumn{1}{p{1.5cm}|}{$0.00010\pm0.00364$} & \multicolumn{1}{p{1.5cm}|}{$0.00161\pm0.00439$} & \multicolumn{1}{p{1.5cm}|}{$0.01078\pm0.00321$} \bigstrut\\
		\cline{3-5}\cline{7-9}          &       &       &       &       &       &       &       &  \bigstrut\\
		\hline
		\multirow{10}[20]{*}{(1, $\ket{\uparrow}$)} &       &       &       &       &       &       &       &  \bigstrut\\
		\cline{3-5}\cline{7-9}          & \multicolumn{1}{c|}{\multirow{2}[4]{*}{$X_{\frac{\pi}{2}}$}} & \multicolumn{1}{p{1.5cm}|}{$X$} & \multicolumn{1}{p{1.5cm}|}{$Y$} & \multicolumn{1}{p{1.5cm}|}{$Z$} & \multicolumn{1}{r|}{} & \multicolumn{1}{p{1.5cm}|}{$X$} & \multicolumn{1}{p{1.5cm}|}{$Y$} & \multicolumn{1}{p{1.5cm}|}{$Z$} \bigstrut\\
		\cline{3-5}\cline{7-9}          & \multicolumn{1}{c|}{} & \multicolumn{1}{p{1.5cm}|}{$-0.0008\pm0.0010$} & \multicolumn{1}{p{1.5cm}|}{$0.0025\pm0.0010$} & \multicolumn{1}{p{1.5cm}|}{$-0.0058\pm0.0024$} & \multicolumn{1}{r|}{} & \multicolumn{1}{p{1.5cm}|}{$0.00124\pm0.00079$} & \multicolumn{1}{p{1.5cm}|}{$0.00005\pm0.00050$} & \multicolumn{1}{p{1.5cm}|}{$0.00090\pm0.00054$} \bigstrut\\
		\cline{3-5}\cline{7-9}          &       &       &       &       &       &       &       &  \bigstrut\\
		\cline{3-5}\cline{7-9}          & \multicolumn{1}{c|}{\multirow{2}[4]{*}{$Y_{\frac{\pi}{2}}$}} & \multicolumn{1}{p{1.5cm}|}{$X$} & \multicolumn{1}{p{1.5cm}|}{$Y$} & \multicolumn{1}{p{1.5cm}|}{$Z$} & \multicolumn{1}{r|}{} & \multicolumn{1}{p{1.5cm}|}{$X$} & \multicolumn{1}{p{1.5cm}|}{$Y$} & \multicolumn{1}{p{1.5cm}|}{$Z$} \bigstrut\\
		\cline{3-5}\cline{7-9}          & \multicolumn{1}{c|}{} & \multicolumn{1}{p{1.5cm}|}{$0.0030\pm0.0016$} & \multicolumn{1}{p{1.5cm}|}{$0.0026\pm0.0007$} & \multicolumn{1}{p{1.5cm}|}{$-0.0030\pm0.0017$} & \multicolumn{1}{r|}{} & \multicolumn{1}{p{1.5cm}|}{$0.00132\pm0.00071$} & \multicolumn{1}{p{1.5cm}|}{$0.00084\pm0.00075$} & \multicolumn{1}{p{1.5cm}|}{$0.00018\pm0.00052$} \bigstrut\\
		\cline{3-5}\cline{7-9}          &       &       &       &       &       &       &       &  \bigstrut\\
		\cline{3-5}\cline{7-9}          & \multicolumn{1}{c|}{\multirow{2}[4]{*}{Idle}} & \multicolumn{1}{p{1.5cm}|}{$X$} & \multicolumn{1}{p{1.5cm}|}{$Y$} & \multicolumn{1}{p{1.5cm}|}{$Z$} & \multicolumn{1}{r|}{} & \multicolumn{1}{p{1.5cm}|}{$X$} & \multicolumn{1}{p{1.5cm}|}{$Y$} & \multicolumn{1}{p{1.5cm}|}{$Z$} \bigstrut\\
		\cline{3-5}\cline{7-9}          & \multicolumn{1}{c|}{} & \multicolumn{1}{p{1.5cm}|}{$-0.0008\pm0.0010$} & \multicolumn{1}{p{1.5cm}|}{$0.0025\pm0.0010$} & \multicolumn{1}{p{1.5cm}|}{$-0.0058\pm0.0024$} & \multicolumn{1}{r|}{} & \multicolumn{1}{p{1.5cm}|}{$0.00013\pm0.00276$} & \multicolumn{1}{p{1.5cm}|}{$0.00006\pm0.00268$} & \multicolumn{1}{p{1.5cm}|}{$0.00786\pm0.00228$} \bigstrut\\
		\cline{3-5}\cline{7-9}          &       &       &       &       &       &       &       &  \bigstrut\\
		\hline
		\hline
	\end{tabular}}}%
    \caption{Estimated one-qubit error generator rates for the gate set consisting of single-qubit $X_\frac{\pi}{2}$ and $Y_\frac{\pi}{2}$ rotations together with an idle operation. One-qubit GST is performed using the native conditional operations by first initializing the control qubit in either the $\ket{\downarrow}$ or $\ket{\uparrow}$ state and utilizing the corresponding conditional rotation gate, thereafter considering only the action on the target qubit.}
	\label{tab:one_qubit_error_generator_rates_GST_conditional}%
\end{table}%

\subsection{1Q Unconditional operations}\label{supp:unconditional}

% \begin{table}[htbp]
\begin{table}[H]
	\centering
	\begin{tabular}{ccccc}
		\hline \hline
		\multicolumn{1}{l}{(Target, Control State)} & Gate  & Coherent & Incoherent & Total Generator Infidelity \bigstrut[b]\\
		\hline
		\multirow{3}[2]{*}{(0, $\ket{\downarrow}$)} & $X_{\frac{\pi}{2}}$ & $0.005 \pm 0.009\%$ & $0.21 \pm 0.20\%$ & $0.21 \pm 0.20\%$ \bigstrut[t]\\
		& $Y_{\frac{\pi}{2}}$ & $0.005 \pm 0.002\%$ & $.0.25 \pm 0.18\%$ & $0.26 \pm 0.17\%$ \\
		& Idle  & $0.079 \pm .009\%$ & $0.69 \pm 0.27 \%$ & $0.76 \pm 0.27\%$  \bigstrut[b]\\
		\hline
		\multirow{3}[2]{*}{(0, $\frac{1}{\sqrt{2}}\left(\ket{\uparrow} + \ket{\downarrow}\right)$)} & $X_{\frac{\pi}{2}}$ & $0.016 \pm 0.005\%$ & $0.83 \pm 0.32\%$ & $0.85 \pm 0.32\%$ \bigstrut[t]\\
		& $Y_{\frac{\pi}{2}}$ & $0.017 \pm 0.006\%$ & $0.99 \pm 0.35\%$ & $1.01 \pm 0.35\%$ \\
		& Idle  & $0.001 \pm 0.003\%$ & $ 1.20 \pm 0.62\%$ & $1.20 \pm 0.62\%$ \bigstrut[b]\\
		\hline
		\multirow{3}[2]{*}{(0, $\ket{\uparrow}$)} & $X_{\frac{\pi}{2}}$ & $0.150 \pm 0.016\%$ & $0.58 \pm 0.21\%$ & $0.73 \pm 0.21\%$ \bigstrut[t]\\
		& $Y_{\frac{\pi}{2}}$ & $0.121 \pm 0.013\%$ & $0.73 \pm 0.37\%$ & $0.85 \pm 0.37\%$ \\
		& Idle  & $0.033 \pm 0.010\%$ & $1.22 \pm 0.28\%$ & $1.25 \pm 0.28\%$ \bigstrut[b]\\
		\hline
		\multirow{3}[2]{*}{(1, $\ket{\downarrow}$)} & $X_{\frac{\pi}{2}}$ & $0.040 \pm 0.026\%$ & $0.49 \pm 0.28\%$ & $0.50 \pm 0.28\%$ \bigstrut[t]\\
		& $Y_{\frac{\pi}{2}}$ & $0.0016 \pm 0.0017\%$ & $0.37 \pm 0.31\%$ & $0.37 \pm 0.31\%$ \\
		& Idle  & $.042 \pm .015\%$ & $1.07 \pm 0.41\%$ & $1.11 \pm 0.40\%$ \bigstrut[b]\\
		\hline
		\multirow{3}[2]{*}{(1, $\frac{1}{\sqrt{2}}\left(\ket{\uparrow} + \ket{\downarrow}\right)$)} & $X_{\frac{\pi}{2}}$ & $0.0013 \pm 0.0013\%$ & $0.47 \pm 0.45\%$ & $0.47 \pm 0.45\%$ \bigstrut[t]\\
		& $Y_{\frac{\pi}{2}}$ & $0.0027 \pm 0.0018\%$ & $0.29 \pm 0.37\%$ & $0.29 \pm 0.37\%$ \\
		& Idle  & $0.0012 \pm 0.0058\%$ & $0.90 \pm 0.70\%$ & $0.90 \pm 0.70\%$ \bigstrut[b]\\
		\hline
		\multirow{3}[2]{*}{(1, $\ket{\uparrow}$)} & $X_{\frac{\pi}{2}}$ & $0.0006 \pm 0.0011\%$ & $0.33 \pm 0.28\%$ & $0.33 \pm 0.28\%$ \bigstrut[t]\\
		& $Y_{\frac{\pi}{2}}$ & $0.0006 \pm 0.0011\%$ & $0.34 \pm 0.36\%$ & $0.34 \pm 0.36\%$ \\
		& Idle  & $0.011 \pm 0.009\%$ & $1.34 \pm 1.04\%$ & $1.35 \pm 1.04\%$ \bigstrut[b]\\
		\hline \hline
	\end{tabular}%
	\caption{Estimated one-qubit generator infidelities for a gate set consisting of single-qubit $X_\frac{\pi}{2}$ and $Y_\frac{\pi}{2}$ rotations together with an idle operation. Here one-qubit GST is performed using unconditional operations derived from the native conditional two-qubit operations by first initializing the control qubit in either the $\ket{\downarrow}$,  $\ket{\uparrow}$, or $\frac{1}{\sqrt{2}}\left(\ket{\uparrow} + \ket{\downarrow}\right)$ and sequentially applying conditional rotations conditioned on $\ket{\downarrow}$ and $\ket{\uparrow}$ respectively. We thereafter consider only the action on the target qubit. We report here the resulting infidelities for each of the six target qubit-control state pairs. In addition to the total generator infidelity for each gate we also report the separate contributions to the generator infidelity due to coherent and incoherent noise sources.}
	\label{tab:one_qubit_unconditional_complete_generator_infidelity_table}%
\end{table}%

%\\

% \begin{table}[htbp]
\begin{table}[H]
	\centering
	{\setstretch{.9}
	{\renewcommand{\arraystretch}{.85}
	\begin{tabular}{ccrrrrrrr}
		\hline \hline\\[.05em]
		\multicolumn{1}{p{2.5cm}}{(Target, Control State)} & Gate  & \multicolumn{3}{c}{Hamiltonian} &       & \multicolumn{3}{c}{Stochastic} \bigstrut[b]\\
		\hline
		\multirow{10}[20]{*}{(0, $\ket{\downarrow}$)} &       &       &       &       &       &       &       &  \bigstrut\\
		\cline{3-5}\cline{7-9}          & \multicolumn{1}{c|}{\multirow{2}[4]{*}{$X_{\frac{\pi}{2}}$}} & \multicolumn{1}{p{1.5cm}|}{$X$} & \multicolumn{1}{p{1.5cm}|}{$Y$} & \multicolumn{1}{p{1.5cm}|}{$Z$} & \multicolumn{1}{r|}{} & \multicolumn{1}{p{1.5cm}|}{$X$} & \multicolumn{1}{p{1.5cm}|}{$Y$} & \multicolumn{1}{p{1.5cm}|}{$Z$} \bigstrut\\
		\cline{3-5}\cline{7-9}          & \multicolumn{1}{c|}{} & \multicolumn{1}{p{1.5cm}|}{$0.0066\pm0.0012$} & \multicolumn{1}{p{1.5cm}|}{$0.0021\pm0.0014$} & \multicolumn{1}{p{1.5cm}|}{$0.0021\pm0.0014$} & \multicolumn{1}{r|}{} & \multicolumn{1}{p{1.5cm}|}{$0.0014\pm0.0017$} & \multicolumn{1}{p{1.5cm}|}{$0.0003\pm0.0016$} & \multicolumn{1}{p{1.5cm}|}{$0.0003\pm0.0009$} \bigstrut\\
		\cline{3-5}\cline{7-9}          &       &       &       &       &       &       &       &  \bigstrut\\
		\cline{3-5}\cline{7-9}          & \multicolumn{1}{c|}{\multirow{2}[4]{*}{$Y_{\frac{\pi}{2}}$}} & \multicolumn{1}{p{1.5cm}|}{$X$} & \multicolumn{1}{p{1.5cm}|}{$Y$} & \multicolumn{1}{p{1.5cm}|}{$Z$} & \multicolumn{1}{r|}{} & \multicolumn{1}{p{1.5cm}|}{$X$} & \multicolumn{1}{p{1.5cm}|}{$Y$} & \multicolumn{1}{p{1.5cm}|}{$Z$} \bigstrut\\
		\cline{3-5}\cline{7-9}          & \multicolumn{1}{c|}{} & \multicolumn{1}{p{1.5cm}|}{$0.0021\pm0.0017$} & \multicolumn{1}{p{1.5cm}|}{$0.0070\pm0.0011$} & \multicolumn{1}{p{1.5cm}|}{$-0.0021\pm0.0016$} & \multicolumn{1}{r|}{} & \multicolumn{1}{p{1.5cm}|}{$0.0010\pm0.0013$} & \multicolumn{1}{p{1.5cm}|}{$0.0010\pm0.0017$} & \multicolumn{1}{p{1.5cm}|}{$0.0005\pm0.0012$} \bigstrut\\
		\cline{3-5}\cline{7-9}          &       &       &       &       &       &       &       &  \bigstrut\\
		\cline{3-5}\cline{7-9}          & \multicolumn{1}{c|}{\multirow{2}[4]{*}{Idle}} & \multicolumn{1}{p{1.5cm}|}{$X$} & \multicolumn{1}{p{1.5cm}|}{$Y$} & \multicolumn{1}{p{1.5cm}|}{$Z$} & \multicolumn{1}{r|}{} & \multicolumn{1}{p{1.5cm}|}{$X$} & \multicolumn{1}{p{1.5cm}|}{$Y$} & \multicolumn{1}{p{1.5cm}|}{$Z$} \bigstrut\\
		\cline{3-5}\cline{7-9}          & \multicolumn{1}{c|}{} & \multicolumn{1}{p{1.5cm}|}{$0.0031\pm0.0024$} & \multicolumn{1}{p{1.5cm}|}{$0\pm0.0019$} & \multicolumn{1}{p{1.5cm}|}{$0.0279\pm0.0015$} & \multicolumn{1}{r|}{} & \multicolumn{1}{p{1.5cm}|}{$0.0003\pm0.0021$} & \multicolumn{1}{p{1.5cm}|}{$0.0015\pm0.0034$} & \multicolumn{1}{p{1.5cm}|}{$0.0051\pm0.0015$} \bigstrut\\
		\cline{3-5}\cline{7-9}          &       &       &       &       &       &       &       &  \bigstrut\\
		\hline
		\multirow{10}[20]{*}{(0, $\frac{1}{\sqrt{2}}\left(\ket{\uparrow} + \ket{\downarrow}\right)$)} &       &       &       &       &       &       &       &  \bigstrut\\
		\cline{3-5}\cline{7-9}          & \multicolumn{1}{c|}{\multirow{2}[4]{*}{$X_{\frac{\pi}{2}}$}} & \multicolumn{1}{p{1.5cm}|}{$X$} & \multicolumn{1}{p{1.5cm}|}{$Y$} & \multicolumn{1}{p{1.5cm}|}{$Z$} & \multicolumn{1}{r|}{} & \multicolumn{1}{p{1.5cm}|}{$X$} & \multicolumn{1}{p{1.5cm}|}{$Y$} & \multicolumn{1}{p{1.5cm}|}{$Z$} \bigstrut\\
		\cline{3-5}\cline{7-9}          & \multicolumn{1}{c|}{} & \multicolumn{1}{p{1.5cm}|}{$0.0024\pm0.0023$} & \multicolumn{1}{p{1.5cm}|}{$0.0088\pm0.0021$} & \multicolumn{1}{p{1.5cm}|}{$0.0089\pm0.0021$} & \multicolumn{1}{r|}{} & \multicolumn{1}{p{1.5cm}|}{$0.0062\pm0.0026$} & \multicolumn{1}{p{1.5cm}|}{$0.0007\pm0.0021$} & \multicolumn{1}{p{1.5cm}|}{$0.0014\pm0.0022$} \bigstrut\\
		\cline{3-5}\cline{7-9}          &       &       &       &       &       &       &       &  \bigstrut\\
		\cline{3-5}\cline{7-9}          & \multicolumn{1}{c|}{\multirow{2}[4]{*}{$Y_{\frac{\pi}{2}}$}} & \multicolumn{1}{p{1.5cm}|}{$X$} & \multicolumn{1}{p{1.5cm}|}{$Y$} & \multicolumn{1}{p{1.5cm}|}{$Z$} & \multicolumn{1}{r|}{} & \multicolumn{1}{p{1.5cm}|}{$X$} & \multicolumn{1}{p{1.5cm}|}{$Y$} & \multicolumn{1}{p{1.5cm}|}{$Z$} \bigstrut\\
		\cline{3-5}\cline{7-9}          & \multicolumn{1}{c|}{} & \multicolumn{1}{p{1.5cm}|}{$0.0089\pm0.0023$} & \multicolumn{1}{p{1.5cm}|}{$0.0036\pm0.0024$} & \multicolumn{1}{p{1.5cm}|}{$-0.0089\pm0.0023$} & \multicolumn{1}{r|}{} & \multicolumn{1}{p{1.5cm}|}{$0.0021\pm0.0023$} & \multicolumn{1}{p{1.5cm}|}{$0.0069\pm0.0028$} & \multicolumn{1}{p{1.5cm}|}{$0.0009\pm0.0018$} \bigstrut\\
		\cline{3-5}\cline{7-9}          &       &       &       &       &       &       &       &  \bigstrut\\
		\cline{3-5}\cline{7-9}          & \multicolumn{1}{c|}{\multirow{2}[4]{*}{Idle}} & \multicolumn{1}{p{1.5cm}|}{$X$} & \multicolumn{1}{p{1.5cm}|}{$Y$} & \multicolumn{1}{p{1.5cm}|}{$Z$} & \multicolumn{1}{r|}{} & \multicolumn{1}{p{1.5cm}|}{$X$} & \multicolumn{1}{p{1.5cm}|}{$Y$} & \multicolumn{1}{p{1.5cm}|}{$Z$} \bigstrut\\
		\cline{3-5}\cline{7-9}          & \multicolumn{1}{c|}{} & \multicolumn{1}{p{1.5cm}|}{$-0.0010\pm0.0034$} & \multicolumn{1}{p{1.5cm}|}{$0.0015\pm0.0029$} & \multicolumn{1}{p{1.5cm}|}{$-0.0031\pm0.0041$} & \multicolumn{1}{r|}{} & \multicolumn{1}{p{1.5cm}|}{$0.0010\pm0.0058$} & \multicolumn{1}{p{1.5cm}|}{$0.0004\pm0.0020$} & \multicolumn{1}{p{1.5cm}|}{$0.0106\pm0.0063$} \bigstrut\\
		\cline{3-5}\cline{7-9}          &       &       &       &       &       &       &       &  \bigstrut\\
		\hline
		\multirow{10}[20]{*}{(0, $\ket{\uparrow}$)} &       &       &       &       &       &       &       &  \bigstrut\\
		\cline{3-5}\cline{7-9}          & \multicolumn{1}{c|}{\multirow{2}[4]{*}{$X_{\frac{\pi}{2}}$}} & \multicolumn{1}{p{1.5cm}|}{$X$} & \multicolumn{1}{p{1.5cm}|}{$Y$} & \multicolumn{1}{p{1.5cm}|}{$Z$} & \multicolumn{1}{r|}{} & \multicolumn{1}{p{1.5cm}|}{$X$} & \multicolumn{1}{p{1.5cm}|}{$Y$} & \multicolumn{1}{p{1.5cm}|}{$Z$} \bigstrut\\
		\cline{3-5}\cline{7-9}          & \multicolumn{1}{c|}{} & \multicolumn{1}{p{1.5cm}|}{$0.0385\pm0.0021$} & \multicolumn{1}{p{1.5cm}|}{$0.0007\pm0.0018$} & \multicolumn{1}{p{1.5cm}|}{$0.0008\pm0.0018$} & \multicolumn{1}{r|}{} & \multicolumn{1}{p{1.5cm}|}{$0.0043\pm0.0027$} & \multicolumn{1}{p{1.5cm}|}{$0.0004\pm0.0015$} & \multicolumn{1}{p{1.5cm}|}{$0.0011\pm0.0020$} \bigstrut\\
		\cline{3-5}\cline{7-9}          &       &       &       &       &       &       &       &  \bigstrut\\
		\cline{3-5}\cline{7-9}          & \multicolumn{1}{c|}{\multirow{2}[4]{*}{$Y_{\frac{\pi}{2}}$}} & \multicolumn{1}{p{1.5cm}|}{$X$} & \multicolumn{1}{p{1.5cm}|}{$Y$} & \multicolumn{1}{p{1.5cm}|}{$Z$} & \multicolumn{1}{r|}{} & \multicolumn{1}{p{1.5cm}|}{$X$} & \multicolumn{1}{p{1.5cm}|}{$Y$} & \multicolumn{1}{p{1.5cm}|}{$Z$} \bigstrut\\
		\cline{3-5}\cline{7-9}          & \multicolumn{1}{c|}{} & \multicolumn{1}{p{1.5cm}|}{$0.0007\pm0.0020$} & \multicolumn{1}{p{1.5cm}|}{$0.0348\pm0.0019$} & \multicolumn{1}{p{1.5cm}|}{$-0.0007\pm0.0020$} & \multicolumn{1}{r|}{} & \multicolumn{1}{p{1.5cm}|}{$0.0012\pm0.0025$} & \multicolumn{1}{p{1.5cm}|}{$0.0031\pm0.0023$} & \multicolumn{1}{p{1.5cm}|}{$0.0030\pm0.0026$} \bigstrut\\
		\cline{3-5}\cline{7-9}          &       &       &       &       &       &       &       &  \bigstrut\\
		\cline{3-5}\cline{7-9}          & \multicolumn{1}{c|}{\multirow{2}[4]{*}{Idle}} & \multicolumn{1}{p{1.5cm}|}{$X$} & \multicolumn{1}{p{1.5cm}|}{$Y$} & \multicolumn{1}{p{1.5cm}|}{$Z$} & \multicolumn{1}{r|}{} & \multicolumn{1}{p{1.5cm}|}{$X$} & \multicolumn{1}{p{1.5cm}|}{$Y$} & \multicolumn{1}{p{1.5cm}|}{$Z$} \bigstrut\\
		\cline{3-5}\cline{7-9}          & \multicolumn{1}{c|}{} & \multicolumn{1}{p{1.5cm}|}{$0.0064\pm0.0042$} & \multicolumn{1}{p{1.5cm}|}{$0.0013\pm0.0021$} & \multicolumn{1}{p{1.5cm}|}{$0.0169\pm0.0032$} & \multicolumn{1}{r|}{} & \multicolumn{1}{p{1.5cm}|}{$0.0009\pm0.0040$} & \multicolumn{1}{p{1.5cm}|}{$0.0010\pm0.0040$} & \multicolumn{1}{p{1.5cm}|}{$0.0103\pm0.0030$} \bigstrut\\
		\cline{3-5}\cline{7-9}          &       &       &       &       &       &       &       &  \bigstrut\\
		\hline
		\multirow{10}[20]{*}{(1, $\ket{\downarrow}$)} &       &       &       &       &       &       &       &  \bigstrut\\
		\cline{3-5}\cline{7-9}          & \multicolumn{1}{c|}{\multirow{2}[4]{*}{$X_{\frac{\pi}{2}}$}} & \multicolumn{1}{p{1.5cm}|}{$X$} & \multicolumn{1}{p{1.5cm}|}{$Y$} & \multicolumn{1}{p{1.5cm}|}{$Z$} & \multicolumn{1}{r|}{} & \multicolumn{1}{p{1.5cm}|}{$X$} & \multicolumn{1}{p{1.5cm}|}{$Y$} & \multicolumn{1}{p{1.5cm}|}{$Z$} \bigstrut\\
		\cline{3-5}\cline{7-9}          & \multicolumn{1}{c|}{} & \multicolumn{1}{p{1.5cm}|}{$0.0062\pm0.0022$} & \multicolumn{1}{p{1.5cm}|}{$0.0007\pm0.0021$} & \multicolumn{1}{p{1.5cm}|}{$0.0007\pm0.0022$} & \multicolumn{1}{r|}{} & \multicolumn{1}{p{1.5cm}|}{$0.0002\pm0.0033$} & \multicolumn{1}{p{1.5cm}|}{$0.0008\pm0.0022$} & \multicolumn{1}{p{1.5cm}|}{$0.0040\pm0.0025$} \bigstrut\\
		\cline{3-5}\cline{7-9}          &       &       &       &       &       &       &       &  \bigstrut\\
		\cline{3-5}\cline{7-9}          & \multicolumn{1}{c|}{\multirow{2}[4]{*}{$Y_{\frac{\pi}{2}}$}} & \multicolumn{1}{p{1.5cm}|}{$X$} & \multicolumn{1}{p{1.5cm}|}{$Y$} & \multicolumn{1}{p{1.5cm}|}{$Z$} & \multicolumn{1}{r|}{} & \multicolumn{1}{p{1.5cm}|}{$X$} & \multicolumn{1}{p{1.5cm}|}{$Y$} & \multicolumn{1}{p{1.5cm}|}{$Z$} \bigstrut\\
		\cline{3-5}\cline{7-9}          & \multicolumn{1}{c|}{} & \multicolumn{1}{p{1.5cm}|}{$0.0007\pm0.0024$} & \multicolumn{1}{p{1.5cm}|}{$0.0039\pm0.0023$} & \multicolumn{1}{p{1.5cm}|}{$-0.0007\pm0.0022$} & \multicolumn{1}{r|}{} & \multicolumn{1}{p{1.5cm}|}{$0.0003\pm0.0025$} & \multicolumn{1}{p{1.5cm}|}{$0.0024\pm0.0032$} & \multicolumn{1}{p{1.5cm}|}{$0.0010\pm0.0032$} \bigstrut\\
		\cline{3-5}\cline{7-9}          &       &       &       &       &       &       &       &  \bigstrut\\
		\cline{3-5}\cline{7-9}          & \multicolumn{1}{c|}{\multirow{2}[4]{*}{Idle}} & \multicolumn{1}{p{1.5cm}|}{$X$} & \multicolumn{1}{p{1.5cm}|}{$Y$} & \multicolumn{1}{p{1.5cm}|}{$Z$} & \multicolumn{1}{r|}{} & \multicolumn{1}{p{1.5cm}|}{$X$} & \multicolumn{1}{p{1.5cm}|}{$Y$} & \multicolumn{1}{p{1.5cm}|}{$Z$} \bigstrut\\
		\cline{3-5}\cline{7-9}          & \multicolumn{1}{c|}{} & \multicolumn{1}{p{1.5cm}|}{$0.0001\pm0.0040$} & \multicolumn{1}{p{1.5cm}|}{$-0.0019\pm0.0027$} & \multicolumn{1}{p{1.5cm}|}{$0.0205\pm0.0035$} & \multicolumn{1}{r|}{} & \multicolumn{1}{p{1.5cm}|}{$0.0010\pm0.0055$} & \multicolumn{1}{p{1.5cm}|}{$0.0004\pm0.0061$} & \multicolumn{1}{p{1.5cm}|}{$0.0093\pm0.0040$} \bigstrut\\
		\cline{3-5}\cline{7-9}          &       &       &       &       &       &       &       &  \bigstrut\\
		\hline
		\hline
	\end{tabular}%
	}
	}	
	\caption{Estimated one-qubit error generator rates for the gate set consisting of single-qubit $X_\frac{\pi}{2}$ and $Y_\frac{\pi}{2}$ rotations together with an idle operation. One-qubit GST is performed using unconditional operations derived from the native conditional two-qubit operations by first initializing the control qubit in either the $\ket{\downarrow}$,  $\ket{\uparrow}$, or $\frac{1}{\sqrt{2}}\left(\ket{\uparrow} + \ket{\downarrow}\right)$ and sequentially applying conditional rotations conditioned on $\ket{\downarrow}$ and $\ket{\uparrow}$ respectively. Thereafter considering only the action on the target qubit is considered. (Table continued on next page).}
	\label{tab:one_qubit_error_generator_rates_GST_unconditional_part_1}%
\end{table}%

\begin{table}[htbp]

    % \addtocounter{table}{-1}
    % \renewcommand{\thetable}{\Roman{table} B}
    % \renewcommand{\theHtable}{\thetable B}% To keep hyperref happy
	{\setstretch{.9}
	{\renewcommand{\arraystretch}{.85}
	\centering
	\begin{tabular}{ccrrrrrrr}
		\hline \hline\\[.05em]
		\multicolumn{1}{p{2.5cm}}{(Target, Control State)} & Gate  & \multicolumn{3}{c}{Hamiltonian} &       & \multicolumn{3}{c}{Stochastic} \bigstrut[b]\\
		\hline
		\multirow{10}[20]{*}{(1, $\frac{1}{\sqrt{2}}\left(\ket{\uparrow} + \ket{\downarrow}\right)$)} &       &       &       &       &       &       &       &  \bigstrut\\
		\cline{3-5}\cline{7-9}          & \multicolumn{1}{c|}{\multirow{2}[4]{*}{$X_{\frac{\pi}{2}}$}} & \multicolumn{1}{p{1.5cm}|}{$X$} & \multicolumn{1}{p{1.5cm}|}{$Y$} & \multicolumn{1}{p{1.5cm}|}{$Z$} & \multicolumn{1}{r|}{} & \multicolumn{1}{p{1.5cm}|}{$X$} & \multicolumn{1}{p{1.5cm}|}{$Y$} & \multicolumn{1}{p{1.5cm}|}{$Z$} \bigstrut\\
		\cline{3-5}\cline{7-9}          & \multicolumn{1}{c|}{} & \multicolumn{1}{p{1.5cm}|}{$0.0016\pm0.0021$} & \multicolumn{1}{p{1.5cm}|}{$0.0023\pm0.0021$} & \multicolumn{1}{p{1.5cm}|}{$0.0023\pm0.0021$} & \multicolumn{1}{r|}{} & \multicolumn{1}{p{1.5cm}|}{$0.0009\pm0.0020$} & \multicolumn{1}{p{1.5cm}|}{$0.0018\pm0.0027$} & \multicolumn{1}{p{1.5cm}|}{$0.0021\pm0.0030$} \bigstrut\\
		\cline{3-5}\cline{7-9}          &       &       &       &       &       &       &       &  \bigstrut\\
		\cline{3-5}\cline{7-9}          & \multicolumn{1}{c|}{\multirow{2}[4]{*}{$Y_{\frac{\pi}{2}}$}} & \multicolumn{1}{p{1.5cm}|}{$X$} & \multicolumn{1}{p{1.5cm}|}{$Y$} & \multicolumn{1}{p{1.5cm}|}{$Z$} & \multicolumn{1}{r|}{} & \multicolumn{1}{p{1.5cm}|}{$X$} & \multicolumn{1}{p{1.5cm}|}{$Y$} & \multicolumn{1}{p{1.5cm}|}{$Z$} \bigstrut\\
		\cline{3-5}\cline{7-9}          & \multicolumn{1}{c|}{} & \multicolumn{1}{p{1.5cm}|}{$0.0023\pm0.0023$} & \multicolumn{1}{p{1.5cm}|}{$0.0041\pm0.0019$} & \multicolumn{1}{p{1.5cm}|}{$-0.0023\pm0.0021$} & \multicolumn{1}{r|}{} & \multicolumn{1}{p{1.5cm}|}{$0.0017\pm0.0028$} & \multicolumn{1}{p{1.5cm}|}{$0.0011\pm0.0029$} & \multicolumn{1}{p{1.5cm}|}{$0\pm0.0023$} \bigstrut\\
		\cline{3-5}\cline{7-9}          &       &       &       &       &       &       &       &  \bigstrut\\
		\cline{3-5}\cline{7-9}          & \multicolumn{1}{c|}{\multirow{2}[4]{*}{Idle}} & \multicolumn{1}{p{1.5cm}|}{$X$} & \multicolumn{1}{p{1.5cm}|}{$Y$} & \multicolumn{1}{p{1.5cm}|}{$Z$} & \multicolumn{1}{r|}{} & \multicolumn{1}{p{1.5cm}|}{$X$} & \multicolumn{1}{p{1.5cm}|}{$Y$} & \multicolumn{1}{p{1.5cm}|}{$Z$} \bigstrut\\
		\cline{3-5}\cline{7-9}          & \multicolumn{1}{c|}{} & \multicolumn{1}{p{1.5cm}|}{$-0.0021\pm0.0049$} & \multicolumn{1}{p{1.5cm}|}{$0.0023\pm0.0035$} & \multicolumn{1}{p{1.5cm}|}{$-0.0016\pm0.0130$} & \multicolumn{1}{r|}{} & \multicolumn{1}{p{1.5cm}|}{$0.0005\pm0.0075$} & \multicolumn{1}{p{1.5cm}|}{$0.0002\pm0.0030$} & \multicolumn{1}{p{1.5cm}|}{$0.0082\pm0.0159$} \bigstrut\\
		\cline{3-5}\cline{7-9}          &       &       &       &       &       &       &       &  \bigstrut\\
		\hline
		\multirow{10}[20]{*}{(1, $\ket{\uparrow}$)} &       &       &       &       &       &       &       &  \bigstrut\\
		\cline{3-5}\cline{7-9}          & \multicolumn{1}{c|}{\multirow{2}[4]{*}{$X_{\frac{\pi}{2}}$}} & \multicolumn{1}{p{1.5cm}|}{$X$} & \multicolumn{1}{p{1.5cm}|}{$Y$} & \multicolumn{1}{p{1.5cm}|}{$Z$} & \multicolumn{1}{r|}{} & \multicolumn{1}{p{1.5cm}|}{$X$} & \multicolumn{1}{p{1.5cm}|}{$Y$} & \multicolumn{1}{p{1.5cm}|}{$Z$} \bigstrut\\
		\cline{3-5}\cline{7-9}          & \multicolumn{1}{c|}{} & \multicolumn{1}{p{1.5cm}|}{$-0.0001\pm0.0020$} & \multicolumn{1}{p{1.5cm}|}{$0.0017\pm0.0022$} & \multicolumn{1}{p{1.5cm}|}{$0.0017\pm0.0023$} & \multicolumn{1}{r|}{} & \multicolumn{1}{p{1.5cm}|}{$0\pm0.0036$} & \multicolumn{1}{p{1.5cm}|}{$0.0016\pm0.0022$} & \multicolumn{1}{p{1.5cm}|}{$0.0016\pm0.0022$} \bigstrut\\
		\cline{3-5}\cline{7-9}          &       &       &       &       &       &       &       &  \bigstrut\\
		\cline{3-5}\cline{7-9}          & \multicolumn{1}{c|}{\multirow{2}[4]{*}{$Y_{\frac{\pi}{2}}$}} & \multicolumn{1}{p{1.5cm}|}{$X$} & \multicolumn{1}{p{1.5cm}|}{$Y$} & \multicolumn{1}{p{1.5cm}|}{$Z$} & \multicolumn{1}{r|}{} & \multicolumn{1}{p{1.5cm}|}{$X$} & \multicolumn{1}{p{1.5cm}|}{$Y$} & \multicolumn{1}{p{1.5cm}|}{$Z$} \bigstrut\\
		\cline{3-5}\cline{7-9}          & \multicolumn{1}{c|}{} & \multicolumn{1}{p{1.5cm}|}{$0.0017\pm0.0026$} & \multicolumn{1}{p{1.5cm}|}{$-0.0005\pm0.0022$} & \multicolumn{1}{p{1.5cm}|}{$-0.0017\pm0.0025$} & \multicolumn{1}{r|}{} & \multicolumn{1}{p{1.5cm}|}{$0.0017\pm0.0030$} & \multicolumn{1}{p{1.5cm}|}{$0.0016\pm0.0034$} & \multicolumn{1}{p{1.5cm}|}{$0\pm0.0021$} \bigstrut\\
		\cline{3-5}\cline{7-9}          &       &       &       &       &       &       &       &  \bigstrut\\
		\cline{3-5}\cline{7-9}          & \multicolumn{1}{c|}{\multirow{2}[4]{*}{Idle}} & \multicolumn{1}{p{1.5cm}|}{$X$} & \multicolumn{1}{p{1.5cm}|}{$Y$} & \multicolumn{1}{p{1.5cm}|}{$Z$} & \multicolumn{1}{r|}{} & \multicolumn{1}{p{1.5cm}|}{$X$} & \multicolumn{1}{p{1.5cm}|}{$Y$} & \multicolumn{1}{p{1.5cm}|}{$Z$} \bigstrut\\
		\cline{3-5}\cline{7-9}          & \multicolumn{1}{c|}{} & \multicolumn{1}{p{1.5cm}|}{$-0.0026\pm0.0034$} & \multicolumn{1}{p{1.5cm}|}{$0.0013\pm0.0056$} & \multicolumn{1}{p{1.5cm}|}{$-0.0101\pm0.0045$} & \multicolumn{1}{r|}{} & \multicolumn{1}{p{1.5cm}|}{$0.0007\pm0.0048$} & \multicolumn{1}{p{1.5cm}|}{$0.0017\pm0.0102$} & \multicolumn{1}{p{1.5cm}|}{$0.0110\pm0.0075$} \bigstrut\\
		\cline{3-5}\cline{7-9}          &       &       &       &       &       &       &       &  \bigstrut\\
		\hline
		\hline
	\end{tabular}%
	}
	}
    \caption{Estimated one-qubit error generator rates for the gate set consisting of single-qubit $X_\frac{\pi}{2}$ and $Y_\frac{\pi}{2}$ rotations together with an idle operation derived from unconditional two-qubit operations. (Table continued from previous page).}
	\label{tab:one_qubit_error_generator_rates_GST_unconditional_part_2}
\end{table}%

\subsection{FOGI Error Rates} \label{sec:suppFOGI}

In the main text, Fig. 3 and 4, we broke down the error budgets for unconditional and conditional single-qubit GST into their contributions due to four specific error mechanisms: dephasing, bitflip errors, overrotation and axis misalignment. These error mechanisms are not in one-to-one correspondence with the error generator rates presented in Supplementary Note \ref{suppG} or the upcoming Supplementary Note \ref{suppH}, though they can be defined in terms of linear combinations of these error generator rates, so we'll properly define these quantities here. In terms of error generator rates these error rates are given in Supplementary Table \ref{tab:fogi_quantities}. These error rates are called first-order gauge invariant (FOGI) quantities, and they correspond to linear combinations of error generator rates that are insensitive to small gauge transformations applied to the representation of the gate set. These FOGI quantities are also constructed so as to have an intuitive and useful operational interpretation in terms of well-known error mechanisms. For more on these FOGI quantities and their construction see \cite{mkadzik2022precision} and its supplementary material.

Note that in the FOGI quantities presented for the idle gate here we are implicitly defining the ideal target operation to be a zero degree rotation about the Z-axis of the Bloch sphere, $Z_{0}$. As we operate in the rotating frame performing an idle at its most basic reduces to tracking the accumulation of phase induced on the state in the lab frame due to the energy splitting between $\ket{0}$ and $\ket{1}$ (i.e. the qubit frequency). In this picture a small Z overrotation or underrotation on the idle means that the actual frequency is slightly different than anticipated resulting in classical timing/phase tracking errors. One last point of note regarding the FOGI quantities for the idle gate is that (restricted to Hamiltonian and stochastic errors are we are here) there are actually six FOGI quantities in total. The two unreported FOGI quantities are called the asymmetry of bitflip, $S_X-S_Y$ and the azimuthal angle of the axis of rotation in the X-Y plane. These two quantities don't affect the infidelity of the idle operation, and are typically of limited operational utility, so we've chosen not to report these here.

\begin{table}[htbp]
	\centering
	\begin{tabular}{cll}
		\hline
		\hline
		\multicolumn{1}{l}{Gate} & \multicolumn{1}{l}{Error Mechanism} & \multicolumn{1}{l}{FOGI Quantity} \bigstrut\\
		\hline
		\multirow{3}[2]{*}{$X_{\frac{\pi}{2}}$} & Overrotation & $H_X$ \bigstrut[t]\\
		& Dephasing & $S_X$ \\
		& Bitflips & $S_Y+S_Z$ \bigstrut[b]\\
		\hline
		\multirow{3}[2]{*}{$Y_{\frac{\pi}{2}}$} & Overrotation & $H_Y$ \bigstrut[t]\\
		& Dephasing & $S_Y$ \\
		& Bitflips & $S_X+S_Z$ \bigstrut[b]\\
		\hline \\[-.75em]
		\multicolumn{1}{l}{$X_{\frac{\pi}{2}}/Y_{\frac{\pi}{2}}$} & Axis Misalignment & $\frac{1}{2}\bigg(H_Y(X_{\frac{\pi}{2}})  + H_Z(X_{\frac{\pi}{2}}) + H_X(Y_{\frac{\pi}{2}}) - H_Z(Y_{\frac{\pi}{2}})    \bigg) $ \bigstrut\\[1em]
		\hline
		\multirow{4}[2]{*}{Idle} & Overrotation & $H_Z$ \bigstrut[t]\\
		& Dephasing & $S_Z$ \\
		& Bitflips & $S_X + S_Y$ \\
		& Axis Misalignment & $\sqrt{H_X^2 + H_Y^2}$ \bigstrut[b]\\
		\hline
	\end{tabular}%
	\caption{First-order gauge invariant (FOGI) error rates and the error mechanisms they correspond to. These FOGI quantities correspond to linear combinations of error generator rates that are, to first order, insensitive to small transformations of the gauge chosen to represent the estimated gate set. $X_{\frac{\pi}{2}}/Y_{\frac{\pi}{2}}$ is meant to note that this quantity is a relational property between the $X_{\frac{\pi}{2}}$ and $Y_{\frac{\pi}{2}}$ gates.}
	\label{tab:fogi_quantities}%
\end{table}%

\section{Two-Qubit GST: Extended Results}\label{suppH}

In this section we present detailed results from the analysis of two-qubit GST using the native conditional two-qubit operations, as discussed in Section V. In Supplementary Table \ref{tab:two_qubit_complete_generator_infidelity_table} we present a complete set of the estimated generator infidelities for the conditional operations, as well as the contributions to this infidelity due to coherent and incoherent noise sources. In Supplementary Tables \ref{tab:two_qubit_GST_error_generator_rates_full_part_1} and \ref{tab:two_qubit_GST_error_generator_rates_full_part_2} we present the full set of estimated Hamiltonian and stochastic error generator rates for the native conditional operations.

	\begin{table}[htbp]
		\centering
		\begin{tabular}{cccc}
			\toprule
			Gate  & Coherent & Incoherent & Total Generator Infidelity \\
			\hline
			$X_{\frac{\pi}{2}}(0, \ket{\downarrow})$ & $7.24 \pm 0.84 \%$ & $9.22 \pm 1.48\%$ & $16.46 \pm 2.14\%$ \\
			$Y_{\frac{\pi}{2}}(0, \ket{\downarrow})$ & $7.27 \pm 0.80\%$ & $9.92 \pm 2.44\%$ & $17.19 \pm 3.38\%$ \\
			$X_{\frac{\pi}{2}}(0, \ket{\uparrow})$ & $9.49 \pm 1.29\%$ & $12.16 \pm 6.77\%$ & $21.64 \pm 7.60\%$ \\
			$Y_{\frac{\pi}{2}}(0, \ket{\uparrow})$ & $8.23 \pm 1.28\%$ & $21.08 \pm 1.76\%$ & $29.31 \pm 2.77\%$ \\
			$X_{\frac{\pi}{2}}(1, \ket{\downarrow})$ & $6.97 \pm 0.82\%$ & $6.92 \pm 3.50\%$ & $13.89 \pm 4.17\%$ \\
			$Y_{\frac{\pi}{2}}(1, \ket{\downarrow})$ & $6.99 \pm 0.84\%$ & $9.94 \pm 3.95\%$ & $16.92 \pm 4.70\%$ \\
			$X_{\frac{\pi}{2}}(1, \ket{\uparrow})$ & $9.17 \pm 0.91\%$ & $9.22 \pm 1.17\%$ & $18.40 \pm 2.18\%$ \\
			$Y_{\frac{\pi}{2}}(1, \ket{\uparrow})$ & $6.98 \pm 0.37\%$ & $21.88 \pm 15.53\%$ & $28.86 \pm 15.04\%$ \\
			Idle  & $0.17 \pm 0.50\%$ & $11.63 \pm 15.82 \%$ & $11.80 \pm 15.73\%$ \\
			\hline \hline
		\end{tabular}%
        \caption{Estimated generator infidelities for native two-qubit conditional rotations and global idling operation as estimated using two-qubit GST \cite{mkadzik2022precision}. The contributions to the total generator infidelity due to coherent and incoherent noise sources have additionally been broken out. In this table we have identified each of the gates with a tuple, the first value of which denotes the target qubit and the second the state of the control qubit being conditioned on.}
		\label{tab:two_qubit_complete_generator_infidelity_table}%
	\end{table}%

% \begin{table}[p]
\begin{table}[H]
	\centering
	{\setstretch{.6}
	{\renewcommand{\arraystretch}{.55}
	\begin{tabular}{p{.5cm}p{.25cm}cccccp{0cm}cccccc}
		\hline
		\hline\\[.1em]
		\multicolumn{1}{c}{Gate} & \multicolumn{6}{c}{Hamiltonian}               &       & \multicolumn{6}{c}{Stochastic} \bigstrut\\
		\hline
		\multirow{8}[12]{=}{\raisebox{-1.25cm}{$\hspace{.5em}X_{\frac{\pi}{2}}$}\\[-1.15cm]\raisebox{-1.25cm}{$(0, \ket{\downarrow})$}} &       &       &       &       &       &       &       &       &       &       &       &       &  \bigstrut[t]\\
		&       &       & \multicolumn{4}{c}{Q2}        &       &       &       & \multicolumn{4}{c}{Q2} \\
		&       &       & \multicolumn{1}{c}{$I$} & \multicolumn{1}{c}{$X$} & \multicolumn{1}{c}{$Y$} & \multicolumn{1}{c}{$Z$} &       &       &       & \multicolumn{1}{c}{$I$} & \multicolumn{1}{c}{$X$} & \multicolumn{1}{c}{$Y$} & \multicolumn{1}{c}{$Z$} \bigstrut[b]\\
		\cline{4-7}\cline{11-14}          & \multicolumn{1}{c}{\multirow{4}[8]{*}{\rotatebox{90}{Q1}}} & \multicolumn{1}{c|}{\rotatebox[origin=r]{90}{$I$}} & \multicolumn{1}{r|}{} & \multicolumn{1}{p{1cm}|}{$ \scriptscriptstyle-0.0071\pm0.0089$} & \multicolumn{1}{p{1cm}|}{$ \scriptscriptstyle-0.0170\pm0.0002$} & \multicolumn{1}{p{1cm}|}{$ \scriptscriptstyle0.0091\pm0.0091$} &       & \multicolumn{1}{c}{\multirow{4}[8]{*}{\rotatebox{90}{Q1}}} & \multicolumn{1}{c|}{\rotatebox[origin=r]{90}{$I$}} & \multicolumn{1}{r|}{} & \multicolumn{1}{p{1cm}|}{$ \scriptscriptstyle0.0015\pm0.0179$} & \multicolumn{1}{p{1cm}|}{$ \scriptscriptstyle0.0023\pm0.0242$} & \multicolumn{1}{p{1cm}|}{$ \scriptscriptstyle0.0515\pm0.0290$} \bigstrut\\
		\cline{4-7}\cline{11-14}          &       & \multicolumn{1}{c|}{\rotatebox[origin=r]{90}{$X$}} & \multicolumn{1}{p{1cm}|}{$ \scriptscriptstyle0.0129\pm0.0106$} & \multicolumn{1}{p{1cm}|}{$ \scriptscriptstyle0.0041\pm0.0103$} & \multicolumn{1}{p{1cm}|}{$ \scriptscriptstyle0.0044\pm0.0213$} & \multicolumn{1}{p{1cm}|}{$ \scriptscriptstyle0.0380\pm0.0165$} &       &       & \multicolumn{1}{c|}{\rotatebox[origin=r]{90}{$X$}} & \multicolumn{1}{p{1cm}|}{$ \scriptscriptstyle0.0031\pm0.0338$} & \multicolumn{1}{p{1cm}|}{$ \scriptscriptstyle0.0010\pm0.0103$} & \multicolumn{1}{p{1cm}|}{$ \scriptscriptstyle0.0014\pm0.0260$} & \multicolumn{1}{p{1cm}|}{$ \scriptscriptstyle0.0012\pm0.0168$} \bigstrut\\
		\cline{4-7}\cline{11-14}          &       & \multicolumn{1}{c|}{\rotatebox[origin=r]{90}{$Y$}} & \multicolumn{1}{p{1cm}|}{$ \scriptscriptstyle-0.1441\pm0.0137$} & \multicolumn{1}{p{1cm}|}{$ \scriptscriptstyle0.0134\pm0.0127$} & \multicolumn{1}{p{1cm}|}{$ \scriptscriptstyle-0.0112\pm0.0171$} & \multicolumn{1}{p{1cm}|}{$ \scriptscriptstyle-0.1203\pm0.0171$} &       &       & \multicolumn{1}{c|}{\rotatebox[origin=r]{90}{$Y$}} & \multicolumn{1}{p{1cm}|}{$ \scriptscriptstyle0.0008\pm0.0076$} & \multicolumn{1}{p{1cm}|}{$ \scriptscriptstyle0.0021\pm0.0136$} & \multicolumn{1}{p{1cm}|}{$ \scriptscriptstyle0.0020\pm0.0042$} & \multicolumn{1}{p{1cm}|}{$ \scriptscriptstyle0.0028\pm0.0188$} \bigstrut\\
		\cline{4-7}\cline{11-14}          &       & \multicolumn{1}{c|}{\rotatebox[origin=r]{90}{$Z$}} & \multicolumn{1}{p{1cm}|}{$ \scriptscriptstyle-0.1314\pm0.0049$} & \multicolumn{1}{p{1cm}|}{$ \scriptscriptstyle-0.0029\pm0.0253$} & \multicolumn{1}{p{1cm}|}{$ \scriptscriptstyle-0.0025\pm0.0148$} & \multicolumn{1}{p{1cm}|}{$ \scriptscriptstyle-0.1325\pm0.0180$} &       &       & \multicolumn{1}{c|}{\rotatebox[origin=r]{90}{$Z$}} & \multicolumn{1}{p{1cm}|}{$ \scriptscriptstyle0.0144\pm0.0235$} & \multicolumn{1}{p{1cm}|}{$ \scriptscriptstyle0.0030\pm0.0094$} & \multicolumn{1}{p{1cm}|}{$ \scriptscriptstyle0.0036\pm0.0189$} & \multicolumn{1}{p{1cm}|}{$ \scriptscriptstyle0.0017\pm0.0091$} \bigstrut\\
		\cline{4-7}\cline{11-14}          &       &       &       &       &       &       &       &       &       &       &       &       &  \bigstrut\\
		\hline
		\multirow{8}[12]{=}{\raisebox{-1.25cm}{$\hspace{.5em}Y_{\frac{\pi}{2}}$}\\[-1.15cm]\raisebox{-1.25cm}{$(0, \ket{\downarrow})$}} &       &       &       &       &       &       &       &       &       &       &       &       &  \bigstrut[t]\\
		&       &       & \multicolumn{4}{c}{Q2}        &       &       &       & \multicolumn{4}{c}{Q2} \\
		&       &       & \multicolumn{1}{c}{$I$} & \multicolumn{1}{c}{$X$} & \multicolumn{1}{c}{$Y$} & \multicolumn{1}{c}{$Z$} &       &       &       & \multicolumn{1}{c}{$I$} & \multicolumn{1}{c}{$X$} & \multicolumn{1}{c}{$Y$} & \multicolumn{1}{c}{$Z$} \bigstrut[b]\\
		\cline{4-7}\cline{11-14}          & \multicolumn{1}{c}{\multirow{4}[8]{*}{\rotatebox{90}{Q1}}} & \multicolumn{1}{c|}{\rotatebox[origin=r]{90}{$I$}} & \multicolumn{1}{r|}{} & \multicolumn{1}{p{1cm}|}{$ \scriptscriptstyle0.0018\pm0.0089$} & \multicolumn{1}{p{1cm}|}{$ \scriptscriptstyle-0.0028\pm0.0047$} & \multicolumn{1}{p{1cm}|}{$ \scriptscriptstyle-0.0046\pm0.0200$} &       & \multicolumn{1}{c}{\multirow{4}[8]{*}{\rotatebox{90}{Q1}}} & \multicolumn{1}{c|}{\rotatebox[origin=r]{90}{$I$}} & \multicolumn{1}{r|}{} & \multicolumn{1}{p{1cm}|}{$ \scriptscriptstyle0.0013\pm0.0060$} & \multicolumn{1}{p{1cm}|}{$ \scriptscriptstyle0.0062\pm0.0208$} & \multicolumn{1}{p{1cm}|}{$ \scriptscriptstyle0.0501\pm0.0424$} \bigstrut\\
		\cline{4-7}\cline{11-14}          &       & \multicolumn{1}{c|}{\rotatebox[origin=r]{90}{$X$}} & \multicolumn{1}{p{1cm}|}{$ \scriptscriptstyle0.1361\pm0.0189$} & \multicolumn{1}{p{1cm}|}{$ \scriptscriptstyle0.0005\pm0.0038$} & \multicolumn{1}{p{1cm}|}{$ \scriptscriptstyle0.0081\pm0.0089$} & \multicolumn{1}{p{1cm}|}{$ \scriptscriptstyle0.1378\pm0.0108$} &       &       & \multicolumn{1}{c|}{\rotatebox[origin=r]{90}{$X$}} & \multicolumn{1}{p{1cm}|}{$ \scriptscriptstyle0.0017\pm0.0047$} & \multicolumn{1}{p{1cm}|}{$ \scriptscriptstyle0.0036\pm0.0388$} & \multicolumn{1}{p{1cm}|}{$ \scriptscriptstyle0.0044\pm0.0196$} & \multicolumn{1}{p{1cm}|}{$ \scriptscriptstyle0.0024\pm0.0117$} \bigstrut\\
		\cline{4-7}\cline{11-14}          &       & \multicolumn{1}{c|}{\rotatebox[origin=r]{90}{$Y$}} & \multicolumn{1}{p{1cm}|}{$ \scriptscriptstyle0.0254\pm0.0096$} & \multicolumn{1}{p{1cm}|}{$ \scriptscriptstyle-0.0156\pm0.0119$} & \multicolumn{1}{p{1cm}|}{$ \scriptscriptstyle-0.0087\pm0.0211$} & \multicolumn{1}{p{1cm}|}{$ \scriptscriptstyle0.0216\pm0.0206$} &       &       & \multicolumn{1}{c|}{\rotatebox[origin=r]{90}{$Y$}} & \multicolumn{1}{p{1cm}|}{$ \scriptscriptstyle0.0032\pm0.0097$} & \multicolumn{1}{p{1cm}|}{$ \scriptscriptstyle0.0026\pm0.0130$} & \multicolumn{1}{p{1cm}|}{$ \scriptscriptstyle0.0024\pm0.0111$} & \multicolumn{1}{p{1cm}|}{$ \scriptscriptstyle0.0016\pm0.0051$} \bigstrut\\
		\cline{4-7}\cline{11-14}          &       & \multicolumn{1}{c|}{\rotatebox[origin=r]{90}{$Z$}} & \multicolumn{1}{p{1cm}|}{$ \scriptscriptstyle-0.1351\pm0.0097$} & \multicolumn{1}{p{1cm}|}{$ \scriptscriptstyle0.0023\pm0.0231$} & \multicolumn{1}{p{1cm}|}{$ \scriptscriptstyle-0.0047\pm0.0400$} & \multicolumn{1}{p{1cm}|}{$ \scriptscriptstyle-0.1240\pm0.0158$} &       &       & \multicolumn{1}{c|}{\rotatebox[origin=r]{90}{$Z$}} & \multicolumn{1}{p{1cm}|}{$ \scriptscriptstyle0.0095\pm0.0482$} & \multicolumn{1}{p{1cm}|}{$ \scriptscriptstyle0.0037\pm0.0178$} & \multicolumn{1}{p{1cm}|}{$ \scriptscriptstyle0.0016\pm0.0138$} & \multicolumn{1}{p{1cm}|}{$ \scriptscriptstyle0.0049\pm0.0132$} \bigstrut\\
		\cline{4-7}\cline{11-14}          &       &       &       &       &       &       &       &       &       &       &       &       &  \bigstrut\\
		\hline
		\multirow{8}[12]{=}{\raisebox{-1.25cm}{$\hspace{.5em}X_{\frac{\pi}{2}}$}\\[-1.15cm]\raisebox{-1.25cm}{$(0, \ket{\uparrow})$}} &       &       &       &       &       &       &       &       &       &       &       &       &  \bigstrut[t]\\
		&       &       & \multicolumn{4}{c}{Q2}        &       &       &       & \multicolumn{4}{c}{Q2} \\
		&       &       & \multicolumn{1}{c}{$I$} & \multicolumn{1}{c}{$X$} & \multicolumn{1}{c}{$Y$} & \multicolumn{1}{c}{$Z$} &       &       &       & \multicolumn{1}{c}{$I$} & \multicolumn{1}{c}{$X$} & \multicolumn{1}{c}{$Y$} & \multicolumn{1}{c}{$Z$} \bigstrut[b]\\
		\cline{4-7}\cline{11-14}          & \multicolumn{1}{c}{\multirow{4}[8]{*}{\rotatebox{90}{Q1}}} & \multicolumn{1}{c|}{\rotatebox[origin=r]{90}{$I$}} & \multicolumn{1}{r|}{} & \multicolumn{1}{p{1cm}|}{$ \scriptscriptstyle-0.0184\pm0.0087$} & \multicolumn{1}{p{1cm}|}{$ \scriptscriptstyle-0.0128\pm0.0159$} & \multicolumn{1}{p{1cm}|}{$ \scriptscriptstyle-0.0073\pm0.0112$} &       & \multicolumn{1}{c}{\multirow{4}[8]{*}{\rotatebox{90}{Q1}}} & \multicolumn{1}{c|}{\rotatebox[origin=r]{90}{$I$}} & \multicolumn{1}{r|}{} & \multicolumn{1}{p{1cm}|}{$ \scriptscriptstyle0.0045\pm0.0082$} & \multicolumn{1}{p{1cm}|}{$ \scriptscriptstyle0.0021\pm0.0211$} & \multicolumn{1}{p{1cm}|}{$ \scriptscriptstyle0.0610\pm0.0255$} \bigstrut\\
		\cline{4-7}\cline{11-14}          &       & \multicolumn{1}{c|}{\rotatebox[origin=r]{90}{$X$}} & \multicolumn{1}{p{1cm}|}{$ \scriptscriptstyle-0.0400\pm0.0260$} & \multicolumn{1}{p{1cm}|}{$ \scriptscriptstyle-0.0118\pm0.0178$} & \multicolumn{1}{p{1cm}|}{$ \scriptscriptstyle-0.0148\pm0.0273$} & \multicolumn{1}{p{1cm}|}{$ \scriptscriptstyle0.0412\pm0.0380$} &       &       & \multicolumn{1}{c|}{\rotatebox[origin=r]{90}{$X$}} & \multicolumn{1}{p{1cm}|}{$ \scriptscriptstyle0.0054\pm0.0229$} & \multicolumn{1}{p{1cm}|}{$ \scriptscriptstyle0.0062\pm0.0194$} & \multicolumn{1}{p{1cm}|}{$ \scriptscriptstyle0.0030\pm0.0088$} & \multicolumn{1}{p{1cm}|}{$ \scriptscriptstyle0.0051\pm0.0497$} \bigstrut\\
		\cline{4-7}\cline{11-14}          &       & \multicolumn{1}{c|}{\rotatebox[origin=r]{90}{$Y$}} & \multicolumn{1}{p{1cm}|}{$ \scriptscriptstyle0.1369\pm0.0187$} & \multicolumn{1}{p{1cm}|}{$ \scriptscriptstyle-0.0031\pm0.0170$} & \multicolumn{1}{p{1cm}|}{$ \scriptscriptstyle0.0164\pm0.0029$} & \multicolumn{1}{p{1cm}|}{$ \scriptscriptstyle-0.1587\pm0.0261$} &       &       & \multicolumn{1}{c|}{\rotatebox[origin=r]{90}{$Y$}} & \multicolumn{1}{p{1cm}|}{$ \scriptscriptstyle0.0075\pm0.0271$} & \multicolumn{1}{p{1cm}|}{$ \scriptscriptstyle0.0022\pm0.0166$} & \multicolumn{1}{p{1cm}|}{$ \scriptscriptstyle0.0034\pm0.0335$} & \multicolumn{1}{p{1cm}|}{$ \scriptscriptstyle0.0057\pm0.0231$} \bigstrut\\
		\cline{4-7}\cline{11-14}          &       & \multicolumn{1}{c|}{\rotatebox[origin=r]{90}{$Z$}} & \multicolumn{1}{p{1cm}|}{$ \scriptscriptstyle0.1514\pm0.0128$} & \multicolumn{1}{p{1cm}|}{$ \scriptscriptstyle0.0205\pm0.0154$} & \multicolumn{1}{p{1cm}|}{$ \scriptscriptstyle0.0078\pm0.0144$} & \multicolumn{1}{p{1cm}|}{$ \scriptscriptstyle-0.1520\pm0.0111$} &       &       & \multicolumn{1}{c|}{\rotatebox[origin=r]{90}{$Z$}} & \multicolumn{1}{p{1cm}|}{$ \scriptscriptstyle0.0040\pm0.0220$} & \multicolumn{1}{p{1cm}|}{$ \scriptscriptstyle0.0029\pm0.0028$} & \multicolumn{1}{p{1cm}|}{$ \scriptscriptstyle0.0031\pm0.0480$} & \multicolumn{1}{p{1cm}|}{$ \scriptscriptstyle0.0053\pm0.0246$} \bigstrut\\
		\cline{4-7}\cline{11-14}          &       &       &       &       &       &       &       &       &       &       &       &       &  \bigstrut\\
		\hline
		\multirow{8}[12]{=}{\raisebox{-1.25cm}{$\hspace{.5em}Y_{\frac{\pi}{2}}$}\\[-1.15cm]\raisebox{-1.25cm}{$(0, \ket{\uparrow})$}} &       &       &       &       &       &       &       &       &       &       &       &       &  \bigstrut[t]\\
		&       &       & \multicolumn{4}{c}{Q2}        &       &       &       & \multicolumn{4}{c}{Q2} \\
		&       &       & \multicolumn{1}{c}{$I$} & \multicolumn{1}{c}{$X$} & \multicolumn{1}{c}{$Y$} & \multicolumn{1}{c}{$Z$} &       &       &       & \multicolumn{1}{c}{$I$} & \multicolumn{1}{c}{$X$} & \multicolumn{1}{c}{$Y$} & \multicolumn{1}{c}{$Z$} \bigstrut[b]\\
		\cline{4-7}\cline{11-14}          & \multicolumn{1}{c}{\multirow{4}[8]{*}{\rotatebox{90}{Q1}}} & \multicolumn{1}{c|}{\rotatebox[origin=r]{90}{$I$}} & \multicolumn{1}{r|}{} & \multicolumn{1}{p{1cm}|}{$ \scriptscriptstyle0.0014\pm0.0198$} & \multicolumn{1}{p{1cm}|}{$ \scriptscriptstyle0.0028\pm0.0239$} & \multicolumn{1}{p{1cm}|}{$ \scriptscriptstyle-0.0100\pm0.0488$} &       & \multicolumn{1}{c}{\multirow{4}[8]{*}{\rotatebox{90}{Q1}}} & \multicolumn{1}{c|}{\rotatebox[origin=r]{90}{$I$}} & \multicolumn{1}{r|}{} & \multicolumn{1}{p{1cm}|}{$ \scriptscriptstyle0.0044\pm0.0169$} & \multicolumn{1}{p{1cm}|}{$ \scriptscriptstyle0.0017\pm0.0043$} & \multicolumn{1}{p{1cm}|}{$ \scriptscriptstyle0.1611\pm0.0385$} \bigstrut\\
		\cline{4-7}\cline{11-14}          &       & \multicolumn{1}{c|}{\rotatebox[origin=r]{90}{$X$}} & \multicolumn{1}{p{1cm}|}{$ \scriptscriptstyle-0.1288\pm0.0219$} & \multicolumn{1}{p{1cm}|}{$ \scriptscriptstyle0.0055\pm0.0250$} & \multicolumn{1}{p{1cm}|}{$ \scriptscriptstyle-0.0138\pm0.0268$} & \multicolumn{1}{p{1cm}|}{$ \scriptscriptstyle0.1266\pm0.0395$} &       &       & \multicolumn{1}{c|}{\rotatebox[origin=r]{90}{$X$}} & \multicolumn{1}{p{1cm}|}{$ \scriptscriptstyle0.0028\pm0.0228$} & \multicolumn{1}{p{1cm}|}{$ \scriptscriptstyle0.0019\pm0.0101$} & \multicolumn{1}{p{1cm}|}{$ \scriptscriptstyle0.0009\pm0.0064$} & \multicolumn{1}{p{1cm}|}{$ \scriptscriptstyle0.0011\pm0.0148$} \bigstrut\\
		\cline{4-7}\cline{11-14}          &       & \multicolumn{1}{c|}{\rotatebox[origin=r]{90}{$Y$}} & \multicolumn{1}{p{1cm}|}{$ \scriptscriptstyle-0.0581\pm0.0013$} & \multicolumn{1}{p{1cm}|}{$ \scriptscriptstyle0.0148\pm0.0099$} & \multicolumn{1}{p{1cm}|}{$ \scriptscriptstyle-0.0015\pm0.0270$} & \multicolumn{1}{p{1cm}|}{$ \scriptscriptstyle0.0369\pm0.0216$} &       &       & \multicolumn{1}{c|}{\rotatebox[origin=r]{90}{$Y$}} & \multicolumn{1}{p{1cm}|}{$ \scriptscriptstyle0.0040\pm0.0112$} & \multicolumn{1}{p{1cm}|}{$ \scriptscriptstyle0.0020\pm0.0357$} & \multicolumn{1}{p{1cm}|}{$ \scriptscriptstyle0.0006\pm0.0311$} & \multicolumn{1}{p{1cm}|}{$ \scriptscriptstyle0.0080\pm0.0037$} \bigstrut\\
		\cline{4-7}\cline{11-14}          &       & \multicolumn{1}{c|}{\rotatebox[origin=r]{90}{$Z$}} & \multicolumn{1}{p{1cm}|}{$ \scriptscriptstyle0.1429\pm0.0450$} & \multicolumn{1}{p{1cm}|}{$ \scriptscriptstyle0.0168\pm0.0186$} & \multicolumn{1}{p{1cm}|}{$ \scriptscriptstyle-0.0158\pm0.0165$} & \multicolumn{1}{p{1cm}|}{$ \scriptscriptstyle-0.1530\pm0.0230$} &       &       & \multicolumn{1}{c|}{\rotatebox[origin=r]{90}{$Z$}} & \multicolumn{1}{p{1cm}|}{$ \scriptscriptstyle0.0124\pm0.0286$} & \multicolumn{1}{p{1cm}|}{$ \scriptscriptstyle0.0012\pm0.0249$} & \multicolumn{1}{p{1cm}|}{$ \scriptscriptstyle0.0034\pm0.0140$} & \multicolumn{1}{p{1cm}|}{$ \scriptscriptstyle0.0054\pm0.0287$} \bigstrut\\
		\cline{4-7}\cline{11-14}          &       &       &       &       &       &       &       &       &       &       &       &       &  \bigstrut\\
		\hline
		\multirow{8}[12]{=}{\raisebox{-1.25cm}{$\hspace{.5em}X_{\frac{\pi}{2}}$}\\[-1.15cm]\raisebox{-1.25cm}{$(1, \ket{\downarrow})$}} &       &       &       &       &       &       &       &       &       &       &       &       &  \bigstrut[t]\\
		&       &       & \multicolumn{4}{c}{Q2}        &       &       &       & \multicolumn{4}{c}{Q2} \\
		&       &       & \multicolumn{1}{c}{$I$} & \multicolumn{1}{c}{$X$} & \multicolumn{1}{c}{$Y$} & \multicolumn{1}{c}{$Z$} &       &       &       & \multicolumn{1}{c}{$I$} & \multicolumn{1}{c}{$X$} & \multicolumn{1}{c}{$Y$} & \multicolumn{1}{c}{$Z$} \bigstrut[b]\\
		\cline{4-7}\cline{11-14}          & \multicolumn{1}{c}{\multirow{4}[8]{*}{\rotatebox{90}{Q1}}} & \multicolumn{1}{c|}{\rotatebox[origin=r]{90}{$I$}} & \multicolumn{1}{r|}{} & \multicolumn{1}{p{1cm}|}{$ \scriptscriptstyle0.0063\pm0.0211$} & \multicolumn{1}{p{1cm}|}{$ \scriptscriptstyle0.1235\pm0.0018$} & \multicolumn{1}{p{1cm}|}{$ \scriptscriptstyle0.1321\pm0.0174$} &       & \multicolumn{1}{c}{\multirow{4}[8]{*}{\rotatebox{90}{Q1}}} & \multicolumn{1}{c|}{\rotatebox[origin=r]{90}{$I$}} & \multicolumn{1}{r|}{} & \multicolumn{1}{p{1cm}|}{$ \scriptscriptstyle0.0006\pm0.0286$} & \multicolumn{1}{p{1cm}|}{$ \scriptscriptstyle0.0022\pm0.0104$} & \multicolumn{1}{p{1cm}|}{$ \scriptscriptstyle0.0079\pm0.0192$} \bigstrut\\
		\cline{4-7}\cline{11-14}          &       & \multicolumn{1}{c|}{\rotatebox[origin=r]{90}{$X$}} & \multicolumn{1}{p{1cm}|}{$ \scriptscriptstyle0.0043\pm0.0184$} & \multicolumn{1}{p{1cm}|}{$ \scriptscriptstyle-0.0021\pm0.0047$} & \multicolumn{1}{p{1cm}|}{$ \scriptscriptstyle0.0034\pm0.0101$} & \multicolumn{1}{p{1cm}|}{$ \scriptscriptstyle-0.0154\pm0.0133$} &       &       & \multicolumn{1}{c|}{\rotatebox[origin=r]{90}{$X$}} & \multicolumn{1}{p{1cm}|}{$ \scriptscriptstyle0.0006\pm0.0190$} & \multicolumn{1}{p{1cm}|}{$ \scriptscriptstyle0.0019\pm0.0178$} & \multicolumn{1}{p{1cm}|}{$ \scriptscriptstyle0.0019\pm0.0177$} & \multicolumn{1}{p{1cm}|}{$ \scriptscriptstyle0.0016\pm0.0171$} \bigstrut\\
		\cline{4-7}\cline{11-14}          &       & \multicolumn{1}{c|}{\rotatebox[origin=r]{90}{$Y$}} & \multicolumn{1}{p{1cm}|}{$ \scriptscriptstyle-0.0010\pm0.0173$} & \multicolumn{1}{p{1cm}|}{$ \scriptscriptstyle0.0003\pm0.0075$} & \multicolumn{1}{p{1cm}|}{$ \scriptscriptstyle-0.0024\pm0.0155$} & \multicolumn{1}{p{1cm}|}{$ \scriptscriptstyle-0.0075\pm0.0197$} &       &       & \multicolumn{1}{c|}{\rotatebox[origin=r]{90}{$Y$}} & \multicolumn{1}{p{1cm}|}{$ \scriptscriptstyle0.0007\pm0.0271$} & \multicolumn{1}{p{1cm}|}{$ \scriptscriptstyle0.0016\pm0.0167$} & \multicolumn{1}{p{1cm}|}{$ \scriptscriptstyle0.0003\pm0.0222$} & \multicolumn{1}{p{1cm}|}{$ \scriptscriptstyle0.0005\pm0.0161$} \bigstrut\\
		\cline{4-7}\cline{11-14}          &       & \multicolumn{1}{c|}{\rotatebox[origin=r]{90}{$Z$}} & \multicolumn{1}{p{1cm}|}{$ \scriptscriptstyle-0.0040\pm0.0124$} & \multicolumn{1}{p{1cm}|}{$ \scriptscriptstyle0.0081\pm0.0140$} & \multicolumn{1}{p{1cm}|}{$ \scriptscriptstyle0.1466\pm0.0061$} & \multicolumn{1}{p{1cm}|}{$ \scriptscriptstyle0.1229\pm0.0278$} &       &       & \multicolumn{1}{c|}{\rotatebox[origin=r]{90}{$Z$}} & \multicolumn{1}{p{1cm}|}{$ \scriptscriptstyle0.0442\pm0.0285$} & \multicolumn{1}{p{1cm}|}{$ \scriptscriptstyle0.0013\pm0.0196$} & \multicolumn{1}{p{1cm}|}{$ \scriptscriptstyle0.0005\pm0.0181$} & \multicolumn{1}{p{1cm}|}{$ \scriptscriptstyle0.0033\pm0.0094$} \bigstrut\\
		\cline{4-7}\cline{11-14}          &       &       &       &       &       &       &       &       &       &       &       &       &  \bigstrut\\
		\hline
		\multirow{8}[12]{=}{\raisebox{-1.25cm}{$\hspace{.5em}Y_{\frac{\pi}{2}}$}\\[-1.15cm]\raisebox{-1.25cm}{$(1, \ket{\downarrow})$}} &       &       &       &       &       &       &       &       &       &       &       &       &  \bigstrut[t]\\
		&       &       & \multicolumn{4}{c}{Q2}        &       &       &       & \multicolumn{4}{c}{Q2} \\
		&       &       & \multicolumn{1}{c}{$I$} & \multicolumn{1}{c}{$X$} & \multicolumn{1}{c}{$Y$} & \multicolumn{1}{c}{$Z$} &       &       &       & \multicolumn{1}{c}{$I$} & \multicolumn{1}{c}{$X$} & \multicolumn{1}{c}{$Y$} & \multicolumn{1}{c}{$Z$} \bigstrut[b]\\
		\cline{4-7}\cline{11-14}          & \multicolumn{1}{c}{\multirow{4}[8]{*}{\rotatebox{90}{Q1}}} & \multicolumn{1}{c|}{\rotatebox[origin=r]{90}{$I$}} & \multicolumn{1}{r|}{} & \multicolumn{1}{p{1cm}|}{$ \scriptscriptstyle-0.1255\pm0.0197$} & \multicolumn{1}{p{1cm}|}{$ \scriptscriptstyle0\pm0.0057$} & \multicolumn{1}{p{1cm}|}{$ \scriptscriptstyle0.1290\pm0.0287$} &       & \multicolumn{1}{c}{\multirow{4}[8]{*}{\rotatebox{90}{Q1}}} & \multicolumn{1}{c|}{\rotatebox[origin=r]{90}{$I$}} & \multicolumn{1}{r|}{} & \multicolumn{1}{p{1cm}|}{$ \scriptscriptstyle0.0054\pm0.0098$} & \multicolumn{1}{p{1cm}|}{$ \scriptscriptstyle0.0027\pm0.0091$} & \multicolumn{1}{p{1cm}|}{$ \scriptscriptstyle0.0095\pm0.0179$} \bigstrut\\
		\cline{4-7}\cline{11-14}          &       & \multicolumn{1}{c|}{\rotatebox[origin=r]{90}{$X$}} & \multicolumn{1}{p{1cm}|}{$ \scriptscriptstyle-0.0025\pm0.0270$} & \multicolumn{1}{p{1cm}|}{$ \scriptscriptstyle-0.0128\pm0.0137$} & \multicolumn{1}{p{1cm}|}{$ \scriptscriptstyle0.0009\pm0.0103$} & \multicolumn{1}{p{1cm}|}{$ \scriptscriptstyle-0.0120\pm0.0127$} &       &       & \multicolumn{1}{c|}{\rotatebox[origin=r]{90}{$X$}} & \multicolumn{1}{p{1cm}|}{$ \scriptscriptstyle0.0009\pm0.0185$} & \multicolumn{1}{p{1cm}|}{$ \scriptscriptstyle0.0048\pm0.0199$} & \multicolumn{1}{p{1cm}|}{$ \scriptscriptstyle0.0023\pm0.0166$} & \multicolumn{1}{p{1cm}|}{$ \scriptscriptstyle0.0022\pm0.0196$} \bigstrut\\
		\cline{4-7}\cline{11-14}          &       & \multicolumn{1}{c|}{\rotatebox[origin=r]{90}{$Y$}} & \multicolumn{1}{p{1cm}|}{$ \scriptscriptstyle0.0015\pm0.0198$} & \multicolumn{1}{p{1cm}|}{$ \scriptscriptstyle0.0058\pm0.0143$} & \multicolumn{1}{p{1cm}|}{$ \scriptscriptstyle0.0073\pm0.0196$} & \multicolumn{1}{p{1cm}|}{$ \scriptscriptstyle0.0034\pm0.0178$} &       &       & \multicolumn{1}{c|}{\rotatebox[origin=r]{90}{$Y$}} & \multicolumn{1}{p{1cm}|}{$ \scriptscriptstyle0.0019\pm0.0141$} & \multicolumn{1}{p{1cm}|}{$ \scriptscriptstyle0.0024\pm0.0162$} & \multicolumn{1}{p{1cm}|}{$ \scriptscriptstyle0.0020\pm0.0252$} & \multicolumn{1}{p{1cm}|}{$ \scriptscriptstyle0.0016\pm0.0273$} \bigstrut\\
		\cline{4-7}\cline{11-14}          &       & \multicolumn{1}{c|}{\rotatebox[origin=r]{90}{$Z$}} & \multicolumn{1}{p{1cm}|}{$ \scriptscriptstyle-0.0154\pm0.0377$} & \multicolumn{1}{p{1cm}|}{$ \scriptscriptstyle-0.1400\pm0.0074$} & \multicolumn{1}{p{1cm}|}{$ \scriptscriptstyle0.0114\pm0.0070$} & \multicolumn{1}{p{1cm}|}{$ \scriptscriptstyle0.1306\pm0.0142$} &       &       & \multicolumn{1}{c|}{\rotatebox[origin=r]{90}{$Z$}} & \multicolumn{1}{p{1cm}|}{$ \scriptscriptstyle0.0531\pm0.0306$} & \multicolumn{1}{p{1cm}|}{$ \scriptscriptstyle0.0054\pm0.0215$} & \multicolumn{1}{p{1cm}|}{$ \scriptscriptstyle0.0026\pm0.0198$} & \multicolumn{1}{p{1cm}|}{$ \scriptscriptstyle0.0027\pm0.0068$} \bigstrut\\
		\cline{4-7}\cline{11-14}          &       &       &       &       &       &       &       &       &       &       &       &       &  \bigstrut\\
		\hline
		\hline
		\end{tabular}}}
    \caption{Complete set of estimated Hamiltonian and Stochastic error generator rates for native two-qubit gate set as estimated using two-qubit GST. (Table continued on next page.)}
	\label{tab:two_qubit_GST_error_generator_rates_full_part_1}%
\end{table}

\begin{table}[H]
        % \addtocounter{table}{-1}
        % \renewcommand{\thetable}{\Roman{table} B}
        % \renewcommand{\theHtable}{\thetable B}% To keep hyperref happy
		\centering
		{\setstretch{.6}
		{\renewcommand{\arraystretch}{.55}
		\begin{tabular}{p{.5cm}p{.25cm}cccccp{0cm}cccccc}
		\hline
		\hline\\[.1em]
		\multicolumn{1}{l}{Gate} & \multicolumn{6}{c}{Hamiltonian}               &       & \multicolumn{6}{c}{Stochastic} \bigstrut\\
		\hline
		\multirow{8}[12]{=}{\raisebox{-1.25cm}{$\hspace{.5em}X_{\frac{\pi}{2}}$}\\[-1.15cm]\raisebox{-1.25cm}{$(1, \ket{\uparrow})$}} &       &       &       &       &       &       &       &       &       &       &       &       &  \bigstrut[t]\\
		&       &       & \multicolumn{4}{c}{Q2}        &       &       &       & \multicolumn{4}{c}{Q2} \\
		&       &       & \multicolumn{1}{c}{$I$} & \multicolumn{1}{c}{$X$} & \multicolumn{1}{c}{$Y$} & \multicolumn{1}{c}{$Z$} &       &       &       & \multicolumn{1}{c}{$I$} & \multicolumn{1}{c}{$X$} & \multicolumn{1}{c}{$Y$} & \multicolumn{1}{c}{$Z$} \bigstrut[b]\\
		\cline{4-7}\cline{11-14}          & \multicolumn{1}{c}{\multirow{4}[8]{*}{\rotatebox{90}{Q1}}} & \multicolumn{1}{c|}{\rotatebox[origin=r]{90}{$I$}} & \multicolumn{1}{r|}{} & \multicolumn{1}{p{1cm}|}{$ \scriptscriptstyle-0.0101\pm0.0205$} & \multicolumn{1}{p{1cm}|}{$ \scriptscriptstyle-0.1522\pm0.0205$} & \multicolumn{1}{p{1cm}|}{$ \scriptscriptstyle-0.1576\pm0.0191$} &       & \multicolumn{1}{c}{\multirow{4}[8]{*}{\rotatebox{90}{Q1}}} & \multicolumn{1}{c|}{\rotatebox[origin=r]{90}{$I$}} & \multicolumn{1}{r|}{} & \multicolumn{1}{p{1cm}|}{$ \scriptscriptstyle0.0022\pm0.0059$} & \multicolumn{1}{p{1cm}|}{$ \scriptscriptstyle0.0122\pm0.0250$} & \multicolumn{1}{p{1cm}|}{$ \scriptscriptstyle0.0031\pm0.0262$} \bigstrut\\
		\cline{4-7}\cline{11-14}          &       & \multicolumn{1}{c|}{\rotatebox[origin=r]{90}{$X$}} & \multicolumn{1}{p{1cm}|}{$ \scriptscriptstyle0.0109\pm0.0261$} & \multicolumn{1}{p{1cm}|}{$ \scriptscriptstyle-0.0090\pm0.0097$} & \multicolumn{1}{p{1cm}|}{$ \scriptscriptstyle-0.0002\pm0.0173$} & \multicolumn{1}{p{1cm}|}{$ \scriptscriptstyle0.0057\pm0.0187$} &       &       & \multicolumn{1}{c|}{\rotatebox[origin=r]{90}{$X$}} & \multicolumn{1}{p{1cm}|}{$ \scriptscriptstyle0.0016\pm0.0152$} & \multicolumn{1}{p{1cm}|}{$ \scriptscriptstyle0.0008\pm0.0103$} & \multicolumn{1}{p{1cm}|}{$ \scriptscriptstyle0.0029\pm0.0285$} & \multicolumn{1}{p{1cm}|}{$ \scriptscriptstyle0.0027\pm0.0266$} \bigstrut\\
		\cline{4-7}\cline{11-14}          &       & \multicolumn{1}{c|}{\rotatebox[origin=r]{90}{$Y$}} & \multicolumn{1}{p{1cm}|}{$ \scriptscriptstyle-0.0004\pm0.0191$} & \multicolumn{1}{p{1cm}|}{$ \scriptscriptstyle-0.0063\pm0.0065$} & \multicolumn{1}{p{1cm}|}{$ \scriptscriptstyle0.0037\pm0.0143$} & \multicolumn{1}{p{1cm}|}{$ \scriptscriptstyle0.0022\pm0.0239$} &       &       & \multicolumn{1}{c|}{\rotatebox[origin=r]{90}{$Y$}} & \multicolumn{1}{p{1cm}|}{$ \scriptscriptstyle0.0013\pm0.0216$} & \multicolumn{1}{p{1cm}|}{$ \scriptscriptstyle0.0013\pm0.0197$} & \multicolumn{1}{p{1cm}|}{$ \scriptscriptstyle0.0025\pm0.0347$} & \multicolumn{1}{p{1cm}|}{$ \scriptscriptstyle0.0022\pm0.0219$} \bigstrut\\
		\cline{4-7}\cline{11-14}          &       & \multicolumn{1}{c|}{\rotatebox[origin=r]{90}{$Z$}} & \multicolumn{1}{p{1cm}|}{$ \scriptscriptstyle-0.0098\pm0.0246$} & \multicolumn{1}{p{1cm}|}{$ \scriptscriptstyle-0.0126\pm0.0180$} & \multicolumn{1}{p{1cm}|}{$ \scriptscriptstyle0.1402\pm0.0150$} & \multicolumn{1}{p{1cm}|}{$ \scriptscriptstyle0.1532\pm0.0084$} &       &       & \multicolumn{1}{c|}{\rotatebox[origin=r]{90}{$Z$}} & \multicolumn{1}{p{1cm}|}{$ \scriptscriptstyle0.0509\pm0.0355$} & \multicolumn{1}{p{1cm}|}{$ \scriptscriptstyle0.0027\pm0.0180$} & \multicolumn{1}{p{1cm}|}{$ \scriptscriptstyle0.0006\pm0.0143$} & \multicolumn{1}{p{1cm}|}{$ \scriptscriptstyle0.0051\pm0.0216$} \bigstrut\\
		\cline{4-7}\cline{11-14}          &       &       &       &       &       &       &       &       &       &       &       &       &  \bigstrut\\
		\hline
		\multirow{8}[12]{=}{\raisebox{-1.25cm}{$\hspace{.5em}Y_{\frac{\pi}{2}}$}\\[-1.15cm]\raisebox{-1.25cm}{$(1, \ket{\uparrow})$}} &       &       &       &       &       &       &       &       &       &       &       &       &  \bigstrut[t]\\
		&       &       & \multicolumn{4}{c}{Q2}        &       &       &       & \multicolumn{4}{c}{Q2} \\
		&       &       & \multicolumn{1}{c}{$I$} & \multicolumn{1}{c}{$X$} & \multicolumn{1}{c}{$Y$} & \multicolumn{1}{c}{$Z$} &       &       &       & \multicolumn{1}{c}{$I$} & \multicolumn{1}{c}{$X$} & \multicolumn{1}{c}{$Y$} & \multicolumn{1}{c}{$Z$} \bigstrut[b]\\
		\cline{4-7}\cline{11-14}          & \multicolumn{1}{c}{\multirow{4}[8]{*}{\rotatebox{90}{Q1}}} & \multicolumn{1}{c|}{\rotatebox[origin=r]{90}{$I$}} & \multicolumn{1}{r|}{} & \multicolumn{1}{p{1cm}|}{$ \scriptscriptstyle0.1474\pm0.0291$} & \multicolumn{1}{p{1cm}|}{$ \scriptscriptstyle-0.0061\pm0.0317$} & \multicolumn{1}{p{1cm}|}{$ \scriptscriptstyle-0.1096\pm0.0433$} &       & \multicolumn{1}{c}{\multirow{4}[8]{*}{\rotatebox{90}{Q1}}} & \multicolumn{1}{c|}{\rotatebox[origin=r]{90}{$I$}} & \multicolumn{1}{r|}{} & \multicolumn{1}{p{1cm}|}{$ \scriptscriptstyle0.0070\pm0.0200$} & \multicolumn{1}{p{1cm}|}{$ \scriptscriptstyle0.0011\pm0.0327$} & \multicolumn{1}{p{1cm}|}{$ \scriptscriptstyle0.0092\pm0.0261$} \bigstrut\\
		\cline{4-7}\cline{11-14}          &       & \multicolumn{1}{c|}{\rotatebox[origin=r]{90}{$X$}} & \multicolumn{1}{p{1cm}|}{$ \scriptscriptstyle-0.0122\pm0.0110$} & \multicolumn{1}{p{1cm}|}{$ \scriptscriptstyle-0.0044\pm0.0074$} & \multicolumn{1}{p{1cm}|}{$ \scriptscriptstyle0.0028\pm0.0042$} & \multicolumn{1}{p{1cm}|}{$ \scriptscriptstyle0\pm0.0054$} &       &       & \multicolumn{1}{c|}{\rotatebox[origin=r]{90}{$X$}} & \multicolumn{1}{p{1cm}|}{$ \scriptscriptstyle0.0025\pm0.0453$} & \multicolumn{1}{p{1cm}|}{$ \scriptscriptstyle0.0015\pm0.0514$} & \multicolumn{1}{p{1cm}|}{$ \scriptscriptstyle0.0025\pm0.0737$} & \multicolumn{1}{p{1cm}|}{$ \scriptscriptstyle0.0021\pm0.0166$} \bigstrut\\
		\cline{4-7}\cline{11-14}          &       & \multicolumn{1}{c|}{\rotatebox[origin=r]{90}{$Y$}} & \multicolumn{1}{p{1cm}|}{$ \scriptscriptstyle0.0082\pm0.0136$} & \multicolumn{1}{p{1cm}|}{$ \scriptscriptstyle0.0039\pm0.0108$} & \multicolumn{1}{p{1cm}|}{$ \scriptscriptstyle-0.0234\pm0.0040$} & \multicolumn{1}{p{1cm}|}{$ \scriptscriptstyle0.0115\pm0.0214$} &       &       & \multicolumn{1}{c|}{\rotatebox[origin=r]{90}{$Y$}} & \multicolumn{1}{p{1cm}|}{$ \scriptscriptstyle0.0054\pm0.0186$} & \multicolumn{1}{p{1cm}|}{$ \scriptscriptstyle0.0022\pm0.0150$} & \multicolumn{1}{p{1cm}|}{$ \scriptscriptstyle0.0047\pm0.0171$} & \multicolumn{1}{p{1cm}|}{$ \scriptscriptstyle0.0054\pm0.0073$} \bigstrut\\
		\cline{4-7}\cline{11-14}          &       & \multicolumn{1}{c|}{\rotatebox[origin=r]{90}{$Z$}} & \multicolumn{1}{p{1cm}|}{$ \scriptscriptstyle-0.0016\pm0.0346$} & \multicolumn{1}{p{1cm}|}{$ \scriptscriptstyle-0.1272\pm0.0501$} & \multicolumn{1}{p{1cm}|}{$ \scriptscriptstyle-0.0073\pm0.0325$} & \multicolumn{1}{p{1cm}|}{$ \scriptscriptstyle0.1373\pm0.0211$} &       &       & \multicolumn{1}{c|}{\rotatebox[origin=r]{90}{$Z$}} & \multicolumn{1}{p{1cm}|}{$ \scriptscriptstyle0.1509\pm0.0470$} & \multicolumn{1}{p{1cm}|}{$ \scriptscriptstyle0.0045\pm0.0178$} & \multicolumn{1}{p{1cm}|}{$ \scriptscriptstyle0.0042\pm0.0104$} & \multicolumn{1}{p{1cm}|}{$ \scriptscriptstyle0.0154\pm0.0404$} \bigstrut\\
		\cline{4-7}\cline{11-14}          &       &       &       &       &       &       &       &       &       &       &       &       &  \bigstrut\\
		\hline
		\multirow{8}[12]{=}{\raisebox{-1.25cm}{Idle}} &       &       &       &       &       &       &       &       &       &       &       &       &  \bigstrut[t]\\
		&       &       & \multicolumn{4}{c}{Q2}        &       &       &       & \multicolumn{4}{c}{Q2} \\
		&       &       & \multicolumn{1}{c}{$I$} & \multicolumn{1}{c}{$X$} & \multicolumn{1}{c}{$Y$} & \multicolumn{1}{c}{$Z$} &       &       &       & \multicolumn{1}{c}{$I$} & \multicolumn{1}{c}{$X$} & \multicolumn{1}{c}{$Y$} & \multicolumn{1}{c}{$Z$} \bigstrut[b]\\
		\cline{4-7}\cline{11-14}          & \multicolumn{1}{c}{\multirow{4}[8]{*}{\rotatebox{90}{Q1}}} & \multicolumn{1}{c|}{\rotatebox[origin=r]{90}{$I$}} & \multicolumn{1}{r|}{} & \multicolumn{1}{p{1cm}|}{$ \scriptscriptstyle-0.0149\pm0.0106$} & \multicolumn{1}{p{1cm}|}{$ \scriptscriptstyle-0.0085\pm0.0323$} & \multicolumn{1}{p{1cm}|}{$ \scriptscriptstyle0.0112\pm0.0651$} &       & \multicolumn{1}{c}{\multirow{4}[8]{*}{\rotatebox{90}{Q1}}} & \multicolumn{1}{c|}{\rotatebox[origin=r]{90}{$I$}} & \multicolumn{1}{r|}{} & \multicolumn{1}{p{1cm}|}{$ \scriptscriptstyle0.0079\pm0.0401$} & \multicolumn{1}{p{1cm}|}{$ \scriptscriptstyle0.0022\pm0.0378$} & \multicolumn{1}{p{1cm}|}{$ \scriptscriptstyle0.0114\pm0.0769$} \bigstrut\\
		\cline{4-7}\cline{11-14}          &       & \multicolumn{1}{c|}{\rotatebox[origin=r]{90}{$X$}} & \multicolumn{1}{p{1cm}|}{$ \scriptscriptstyle-0.0172\pm0.0578$} & \multicolumn{1}{p{1cm}|}{$ \scriptscriptstyle0.0135\pm0.0232$} & \multicolumn{1}{p{1cm}|}{$ \scriptscriptstyle0.0098\pm0.0415$} & \multicolumn{1}{p{1cm}|}{$ \scriptscriptstyle0.0034\pm0.0431$} &       &       & \multicolumn{1}{c|}{\rotatebox[origin=r]{90}{$X$}} & \multicolumn{1}{p{1cm}|}{$ \scriptscriptstyle0.0043\pm0.0614$} & \multicolumn{1}{p{1cm}|}{$ \scriptscriptstyle0.0123\pm0.0482$} & \multicolumn{1}{p{1cm}|}{$ \scriptscriptstyle0.0045\pm0.0646$} & \multicolumn{1}{p{1cm}|}{$ \scriptscriptstyle0.0050\pm0.0498$} \bigstrut\\
		\cline{4-7}\cline{11-14}          &       & \multicolumn{1}{c|}{\rotatebox[origin=r]{90}{$Y$}} & \multicolumn{1}{p{1cm}|}{$ \scriptscriptstyle0.0012\pm0.0381$} & \multicolumn{1}{p{1cm}|}{$ \scriptscriptstyle0.0109\pm0.0230$} & \multicolumn{1}{p{1cm}|}{$ \scriptscriptstyle0.0039\pm0.0444$} & \multicolumn{1}{p{1cm}|}{$ \scriptscriptstyle-0.0134\pm0.0757$} &       &       & \multicolumn{1}{c|}{\rotatebox[origin=r]{90}{$Y$}} & \multicolumn{1}{p{1cm}|}{$ \scriptscriptstyle0.0071\pm0.0363$} & \multicolumn{1}{p{1cm}|}{$ \scriptscriptstyle0.0039\pm0.1249$} & \multicolumn{1}{p{1cm}|}{$ \scriptscriptstyle0.0100\pm0.0555$} & \multicolumn{1}{p{1cm}|}{$ \scriptscriptstyle0.0160\pm0.0304$} \bigstrut\\
		\cline{4-7}\cline{11-14}          &       & \multicolumn{1}{c|}{\rotatebox[origin=r]{90}{$Z$}} & \multicolumn{1}{p{1cm}|}{$ \scriptscriptstyle-0.0129\pm0.0780$} & \multicolumn{1}{p{1cm}|}{$ \scriptscriptstyle-0.0099\pm0.0237$} & \multicolumn{1}{p{1cm}|}{$ \scriptscriptstyle-0.0019\pm0.0351$} & \multicolumn{1}{p{1cm}|}{$ \scriptscriptstyle0.0111\pm0.0741$} &       &       & \multicolumn{1}{c|}{\rotatebox[origin=r]{90}{$Z$}} & \multicolumn{1}{p{1cm}|}{$ \scriptscriptstyle0.0198\pm0.1040$} & \multicolumn{1}{p{1cm}|}{$ \scriptscriptstyle0.0055\pm0.0132$} & \multicolumn{1}{p{1cm}|}{$ \scriptscriptstyle0.0036\pm0.0294$} & \multicolumn{1}{p{1cm}|}{$ \scriptscriptstyle0.0028\pm0.0445$} \bigstrut\\
		\cline{4-7}\cline{11-14}          &       &       &       &       &       &       &       &       &       &       &       &       &  \bigstrut\\
		\hline
		\hline
	\end{tabular}}}
    \caption{Complete set of estimated Hamiltonian and Stochastic error generator rates for native two-qubit gate set as estimated using two-qubit GST. (Continued from previous page.)}
	\label{tab:two_qubit_GST_error_generator_rates_full_part_2}%
\end{table}%

\section{Phase reversal tomography}\label{suppI}

\subsection{Finding a physical density matrix}
The full measured density matrices, reconstructed using phase reversal tomography, both with and without SPAM extraction, are as follows:

\begin{align}
&\hat{\rho}_{\text{SPAM}} = \\
&\begin{pmatrix}
    0.460  & 0.000 & 0.000 & 0.332  + 0.109j \\
    0.000 & 0.116  & 0.000 & 0.000 \\
    0.000 & 0.000 & 0.097 & 0.000 \\
    0.332  - 0.109j & 0.000 & 0.000 & 0.327
\end{pmatrix}
\end{align}

\begin{align}
&\hat{\rho}_{\text{no SPAM}} = \\
&\begin{pmatrix}
    0.431 & 0.000 & 0.000 & 0.464 + 0.154j \\
    0.000 & -0.023 & 0.000 & 0.000 \\
    0.000 & 0.000 & 0.050 & 0.000 \\
    0.464 - 0.154j & 0.000 & 0.000 & 0.485
\end{pmatrix}
\end{align}

Measurement imperfections can result in the estimated density matrix not representing a physical density matrix. To find the nearest physical density matrix to the estimated matrix, $\hat{\rho}$, we therefore used a Nelder-Mead optimisation algorithm to optimise the parameters $t_{n}$ in the following lower triangular matrix 

\begin{equation}
T = \left( \begin{array}{cccc}
t_{1} & 0 & 0 & 0\\
0 & t_{5} & 0 & 0\\
0 & 0 & t_{6} & 0\\
t_{2}+it_{3} & 0 & 0 & t_{4}
\end{array} \right),
\end{equation}

The cost function to be minimised was given by the overlap between $\hat{\sigma}$ and the measured density matrix $\hat{\rho}$, where 

\begin{equation}
    \hat{\sigma} = \frac{T^{\dagger}(t)T(t)}{Tr(T^{\dagger}(t)T(t))}.
\end{equation}

\begin{figure}[!h]
\begin{center}
\includegraphics[width=1\textwidth]{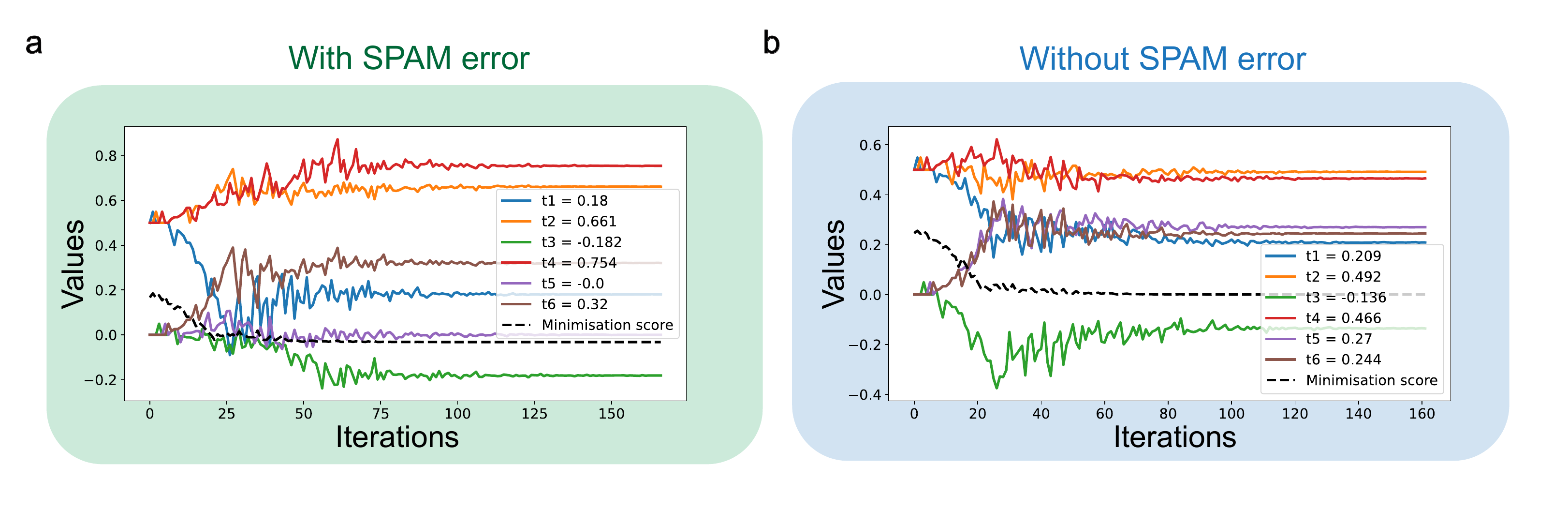}
\caption[Nelder-Mead estimation of density matrix elements.]{\textbf{Nelder-Mead estimation of density matrix elements.} \textbf{a.} Plot showing iterations of the Nelder-Mead optimisation algorithm to find the nearest physical density matrix to the measured matrix. The values $t_n$ are seen to converge to their optimal value. \textbf{b.} Plot of the same Nelder-Mead algorithm run on the density matrix with SPAM error extracted.}
\label{fig:optimisation}
\end{center}
\end{figure}

Supplementary Fig.~\ref{fig:optimisation} shows the individual iterations of the optimisation algorithm for both the density matrix with (Supplementary Fig.~\ref{fig:optimisation}, a) and without (Supplementary Fig.~\ref{fig:optimisation}, b) SPAM extracted. The $t_n$ values converge to optimal values within 175 iterations of the optimisation algorithm in both cases, rendering a final overlap between the measured and estimated density matrices of 100.0$\%$ with SPAM error and 99.79$\%$ for the density matrix with SPAM extracted. 

\begin{figure}[!h]
    \centering
    \includegraphics[width=1\textwidth]{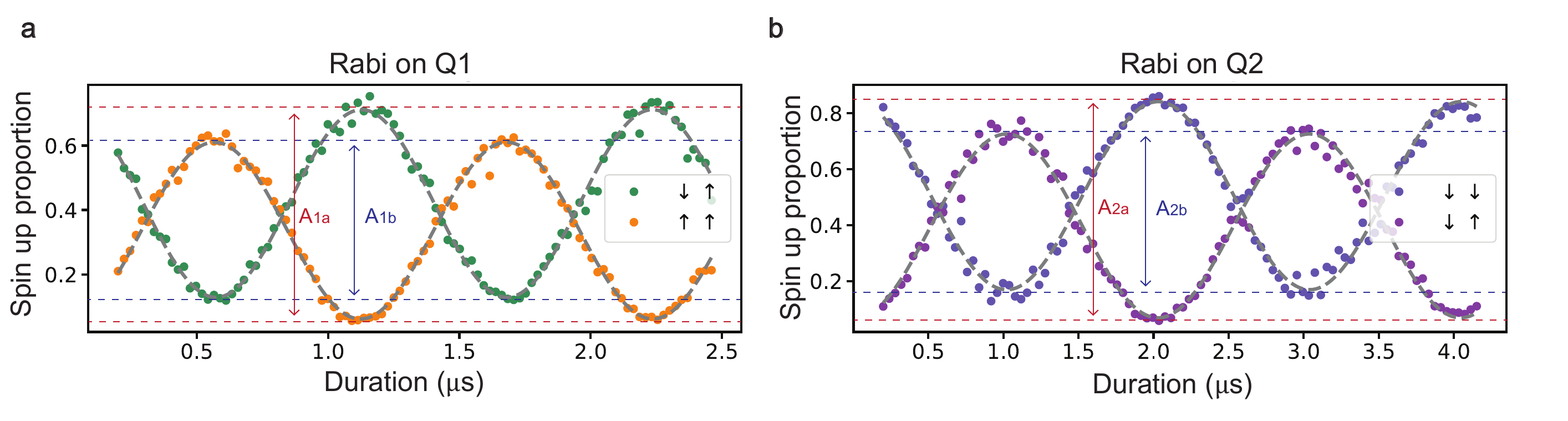}
    \caption[Phase reversal tomography SPAM extraction.]{\textbf{Phase reversal tomography SPAM extraction.} \textbf{a.} Rabi oscillation with two-qubit readout performed on Q1. $A_{1a}$ represents the amplitude of the oscillation of the $\ket{\downarrow \uparrow}$ state while $A_{1b}$ represents the amplitude of the oscillation of the $\ket{\uparrow \uparrow}$ state. These amplitudes provide a scaling factor for the phase reversal oscillations and direct Bell state measurement, allowing for the extraction of SPAM error from the results. \textbf{b.} Rabi oscillation with two-qubit readout performed on Q2. $A_{2a}$ represents the amplitude of the oscillation of the $\ket{\downarrow \downarrow}$ state while $A_{2b}$ represents the amplitude of the oscillation of the $\ket{\downarrow \uparrow}$ state.}
    \label{fig:spam_removal}
\end{figure}

\subsection{SPAM removal}

Reconstructing the density matrix of the prepared Bell state using phase reversal tomography involves performing both one and two-qubit readout. To extract SPAM from the single qubit readout we performed a Rabi oscillation on both electrons individually, and used the amplitude of these Rabi oscillations as a scaling factor for the phase reversal oscillations. In order to extract the SPAM from two-qubit readout, a Rabi was first performed on Q2, between the states $\ket{\downarrow \downarrow}$ and $\widetilde{\ket{\downarrow \uparrow}}$, followed by a Rabi oscillation performed on Q1, between the states $\widetilde{\ket{\downarrow \uparrow}}$ and $\ket{\uparrow \uparrow}$. Both Rabi oscillations were read out using two-qubit readout, by combining the single-shot readout results of both electrons to obtain the probabilities $P(\ket{\downarrow \downarrow}, P(\ket{\downarrow \uparrow}), P(\uparrow \downarrow), P(\uparrow \uparrow))$. The amplitude of these Rabi oscillations are dictated by a combination of initialisation, $I_{\psi}$ and readout, $R_{\psi}$, error, where $\psi = \{\downarrow \downarrow, \downarrow \uparrow, \uparrow \downarrow, \uparrow \uparrow \}$ denotes the state being initialised or read out. The amplitudes of each of the four Rabi oscillations, as shown in Supplementary Fig.~\ref{fig:spam_removal}, can therefore be written as the following:

\begin{align}
    A_{1a} = I_{\downarrow \uparrow} \times R_{\downarrow \uparrow},\\
    A_{1b} = I_{\downarrow \uparrow} \times R_{\uparrow \uparrow},\\
    A_{2a} = I_{\downarrow \downarrow} \times R_{\downarrow \downarrow},\\
    A_{2b} = I_{\downarrow \downarrow} \times R_{\downarrow \uparrow},\\
\end{align}

In order to extract the SPAM from our phase reversal measurement, we require the information $I_{\downarrow \downarrow} \times R_{\downarrow \downarrow}$ and $I_{\downarrow \downarrow} \times R_{\uparrow \uparrow}$. The value of $I_{\downarrow \downarrow} \times R_{\downarrow \downarrow}$ is obtained directly from the amplitude $A_{2a}$ (see Supplementary Fig.~\ref{fig:spam_removal}), while $I_{\downarrow \downarrow} \times R_{\uparrow \uparrow}$ can be obtained with the following expression:

\begin{align}
    I_{\downarrow \downarrow} \times R_{\uparrow \uparrow} &= \frac{I_{\downarrow \uparrow} \times R_{\uparrow \uparrow} \times I_{\downarrow \downarrow} \times R_{\downarrow \uparrow}}{I_{\downarrow \uparrow} \times R_{\downarrow \uparrow}}\\
    &= \frac{A_{1b}\times A_{2b}}{A_{1a}}
\end{align}

These values were then used as a scaling factor for the Bell state measurements.

\begin{figure}[!h]
    \centering
    \includegraphics[width=1\textwidth]{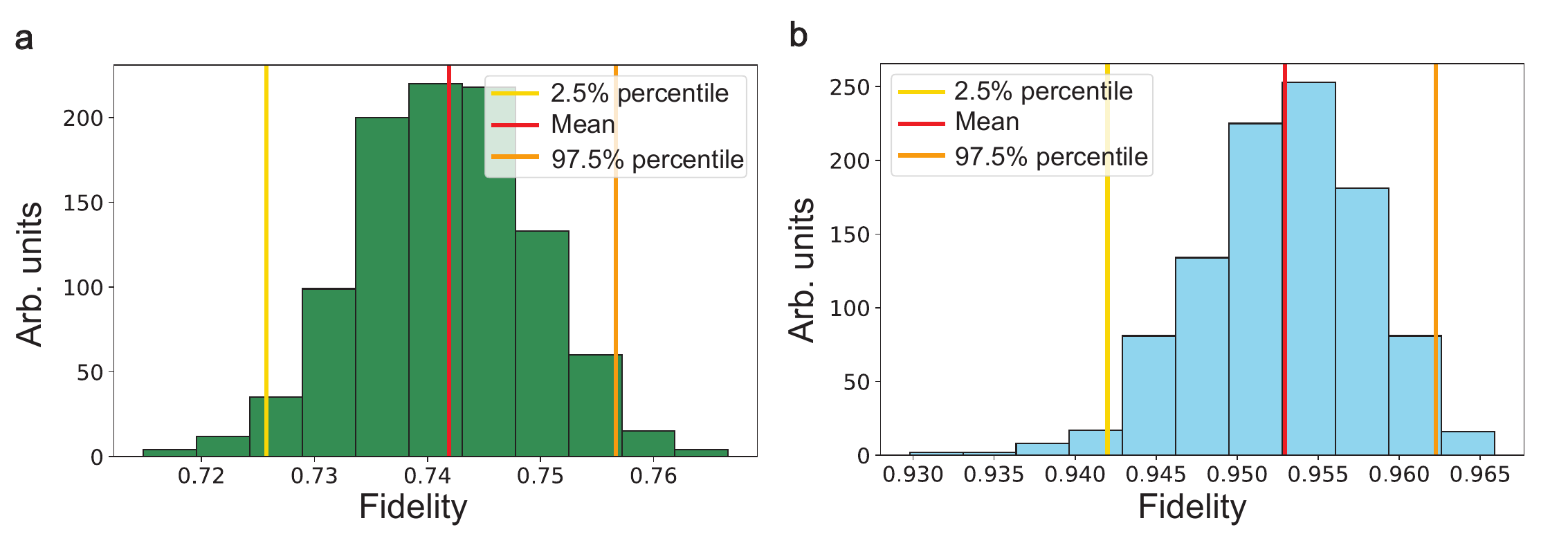}
    \caption[Non-parametric bootstrapping error bar estimations.]{\textbf{Non-parametric bootstrapping error bar estimations.} Bell state fidelity distribution histogram, obtained by resampling 10 measured Bell state fidelity values, 1000 times. This was performed for both the SPAM included dataset (\textbf{a.}) and SPAM extracted dataset  (\textbf{b.}). }
    \label{fig:fidelity_hist_spam}
\end{figure}

\subsection{Error bars}
The error bars were calculated by repeating the phase reversal tomography experiment 10 times, resulting in 10 distinct Bell state fidelities. A non-parametric bootstrapping method was then used, whereby 10 new fidelity values were chosen at random from the list of 10 measured fidelities. The mean of these 10 resampled fidelity values was then calculated. This process was then repeated 1000 times. The histogram of the 1000 average fidelity values is shown in Supplementary Fig.~\ref{fig:fidelity_hist_spam}, for the case of both the SPAM included (Supplementary Fig.~\ref{fig:fidelity_hist_spam}, a) and SPAM extracted (Supplementary Fig.~\ref{fig:fidelity_hist_spam}, b) results. The error bars were then calculated by finding the 95$\%$ confidence interval of the resulting fidelity distribution, which represents 2$\sigma$. The mean of the 10 originally measured fidelity values is then taken as the final fidelity value reported.

\section{Concurrence}
To determine the degree of entanglement present between the electrons of the J-coupled system we calculated a commonly used entanglement metric known as concurrence, $C$. The definition of concurrence, which currently only exists for pairs of qubits, is given by \cite{wootters2001entanglement}

\begin{equation}
    C = |\braket{\Phi|\tilde{\Phi}}|,
\end{equation}

where $\ket{\Phi}$ is the state of a pair of qubits, written in the basis $\{ \ket{00}, \ket{01}, \ket{10}, \ket{11} \}$. $\ket{\tilde{\Phi}} = (\hat{\sigma_{y}} \otimes \hat{\sigma_{y}})\ket{\Phi^{*}}$ is the `spin-flip' operation where $\ket{\Phi^{*}}$ is the complex conjugate of $\ket{\Phi}$ and $\sigma_y$ is the Pauli-y matrix. When applied to a state, the spin-flip operator will take each qubit to the orthogonal state on the Bloch sphere. Physically the concurrence therefore represents the overlap between a state $\ket{\Phi}$ and a state that is  diametrically opposite to $\ket{\Phi}$ on the Bloch sphere.\\

Experimentally, the most straightforward method of calculating the concurrence is by reconstructing the density matrix of the physical system and calculating the quantity

\begin{equation}
    \hat{R} = \sqrt{\sqrt{\hat{\rho}}\tilde{\rho}\sqrt{\hat{\rho}}}.
\end{equation}

where $\hat{\rho}$ is the measured density matrix and $\tilde{\rho}$ is the result of applying the spin flip operation to $\hat{\rho}$, given by

\begin{equation}
    \tilde{\rho} = (\hat{\sigma}_{y} \otimes \hat{\sigma}_{y})\hat{\rho}^*(\hat{\sigma}_{y} \otimes \hat{\sigma}_{y}),
\end{equation}

where $\hat{\rho}^*$ is the complex conjugate transpose of the measured density matrix. The value of concurrence is then given by 

\begin{equation}
    C= max\{0,\lambda_1 - \lambda_2 - \lambda_3 - \lambda_4\},
\end{equation}

where $\lambda_{i}$ are the non-negative eigenvalues of $\hat{R}$ in decreasing order. 
\begin{figure}[!h]
    \centering
    \includegraphics[width=0.85\textwidth]{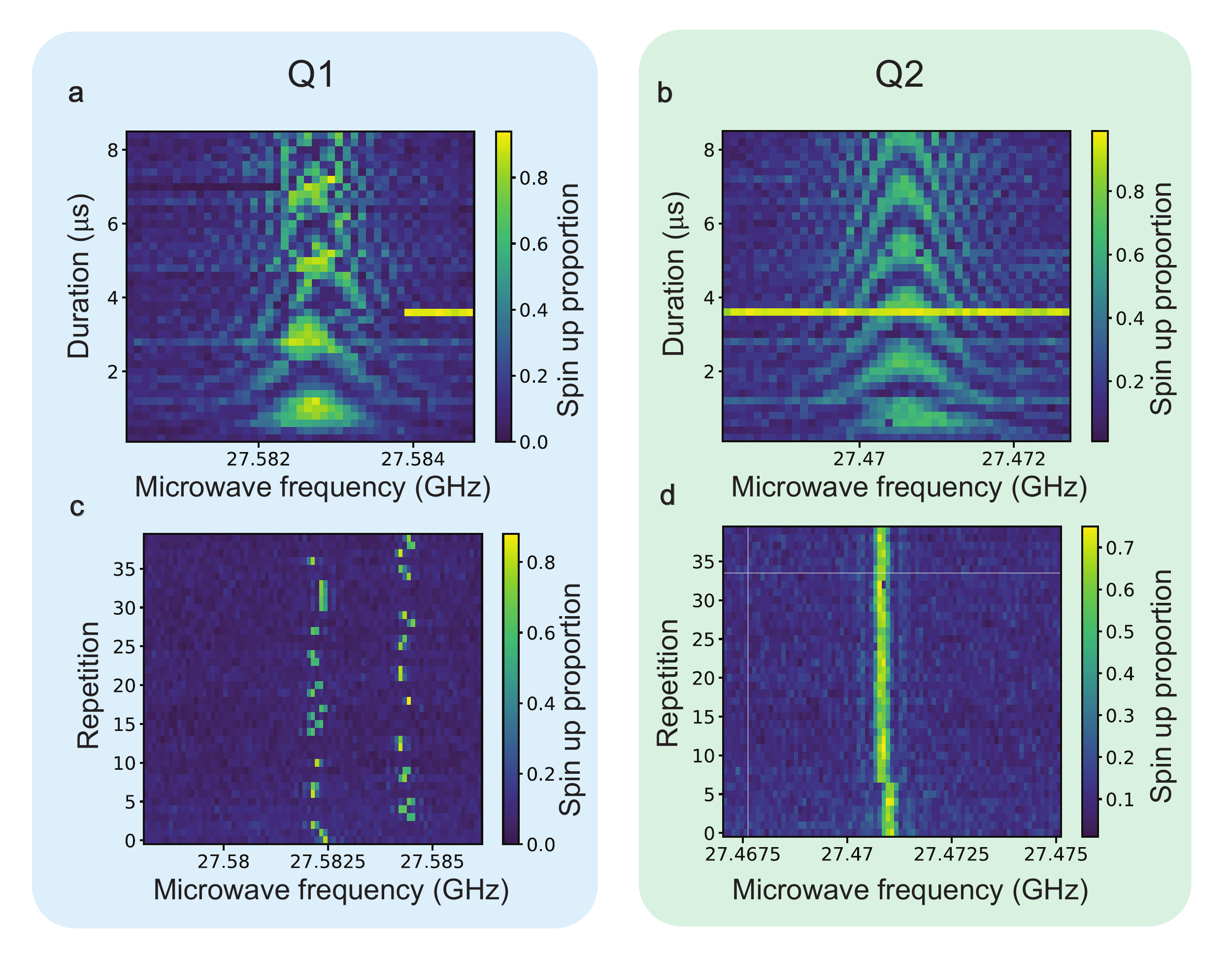}
    \caption[Electron frequency stability.]{\textbf{Electron frequency stability.} \textbf{a.}(\textbf{b.}) Rabi chevrons performed on electron Q1(Q2) showing the greater stability of Q2. An automated retuning algorithm was run after the completion of every frequency sweep, in order to stay at the correct gate tuning position for gate readout. The yellow horizontal lines present in the Rabi chevrons are due to the device readout position drifting out of tune during the measurement, resulting in a spin up proportion of 1 being erroneously measured. As the retuning algorithm is only repeated once every frequency sweep, the device hence remains out of tune until the end of the frequency sweep. \textbf{c.} Repeated ESR frequency spectrum of Q1 taken over a time period of a few hours. The jumps observed in the resonance frequency of Q1 are as a result of two $^{29}$Si nuclei that are hyperfine coupled to Q1 with a hyperfine coupling of 200 kHz and 2 MHz respectively. \textbf{d.} Repeated ESR frequency spectrum of Q2 taken over a time period of a few hours, exhibiting a much more stable ESR frequency as a result of nuclear freezing.}
    \label{fig:29si_electron_jumps}
\end{figure}

\section{$^{29}$Si spin bath}

All experiments were carried out on a device fabricated on an enriched $^{28}$Si substrate, with 800 ppm residual $^{29}$Si, which possess a nuclear spin of $I = \frac{1}{2}$. By repeatedly probing the ESR spectrum of the electrons, we can determine the hyperfine coupling strength between the donor electrons and the residual $^{29}$Si nuclear spins which occasionally cause shifts in electron resonance frequency. Supplementary Fig.~\ref{fig:29si_electron_jumps} shows a Rabi chevrons (a-b) and low-power ESR frequency spectra (c-d) performed on electrons Q1 and Q2. The resonance frequency of Q2 is highly stable as a result of the nuclear freezing effect \cite{madzik2020controllable}, discussed in Section II. The spectrum of Q1 instead shows shifts in frequency of 2 MHz and 200 kHz, corresponding to the hyperfine coupling between Q1 and two $^{29}$Si nuclear spins. Through separate Ramsey experiments we further identified a third $^{29}$Si with 20 kHz coupling to Q1, not visible here due to the power broadening introduced by the Rabi drive.

%References bbl for supplementary
\providecommand{\noopsort}[1]{}\providecommand{\singleletter}[1]{#1}%
%

%Uncomment if wishing to use references.bib file
% \bibliography{references}

\end{document}